\documentclass[11pt]{article}

\usepackage[letterpaper,margin=1in]{geometry}
\usepackage[T1]{fontenc}
\usepackage{microtype}
\usepackage{xcolor}

\usepackage{amsmath,amssymb,amsthm,mathtools}
\usepackage{algorithm,algpseudocode}
\floatstyle{boxed}
\restylefloat{algorithm}
\usepackage{booktabs,array,tabularx}
\usepackage{caption}

\usepackage[hidelinks,pagebackref=true]{hyperref}
\usepackage[capitalize,nameinlink,noabbrev]{cleveref}
\usepackage{authblk}

\let\cites\cite
\let\manuscriptcitation\cite
\newcommand{\citeat}[2]{\manuscriptcitation[#2]{#1}}

\renewcommand*{\backref}[1]{}
\renewcommand*{\backrefalt}[4]{%
  \ifcase #1%
  \or (Cited on page~#2.)%
  \else (Cited on pages~#2.)%
  \fi}

\newtheorem{theorem}{Theorem}
\newtheorem*{mainresultrestated}{Theorem~\ref{thm:main-dyadic-intro}}
\newtheorem{lemma}[theorem]{Lemma}
\newtheorem{proposition}[theorem]{Proposition}
\newtheorem{corollary}[theorem]{Corollary}
\theoremstyle{definition}
\newtheorem{definition}[theorem]{Definition}

\newtheorem{claim}[theorem]{Claim}

\theoremstyle{remark}
\newtheorem{remark}[theorem]{Remark}

\crefname{theorem}{theorem}{theorems}
\Crefname{theorem}{Theorem}{Theorems}
\crefname{lemma}{lemma}{lemmas}
\Crefname{lemma}{Lemma}{Lemmas}
\crefname{proposition}{proposition}{propositions}
\Crefname{proposition}{Proposition}{Propositions}
\crefname{corollary}{corollary}{corollaries}
\Crefname{corollary}{Corollary}{Corollaries}
\crefname{definition}{definition}{definitions}
\Crefname{definition}{Definition}{Definitions}
\crefname{example}{example}{examples}
\Crefname{example}{Example}{Examples}
\crefname{claim}{claim}{claims}
\Crefname{claim}{Claim}{Claims}
\crefname{conjecture}{conjecture}{conjectures}
\Crefname{conjecture}{Conjecture}{Conjectures}
\crefname{remark}{remark}{remarks}
\Crefname{remark}{Remark}{Remarks}

\newcommand{\C}{\mathbb C}
\newcommand{\Z}{\mathbb Z}
\newcommand{\E}{\operatorname{\mathbb{E}}}
\newcommand{\Prob}{\operatorname{Pr}}
\newcommand{\supp}{\operatorname{supp}}

\newcommand{\HT}{\mathcal H}
\newcommand{\iu}{\mathrm{i}}
\newcommand{\norm}[1]{%
  \mathopen{}\mathclose{%
    \mathchoice
      {\left\lVert#1\right\rVert}
      {\lVert#1\rVert}
      {\lVert#1\rVert}
      {\lVert#1\rVert}%
  }%
}
\newcommand{\ceil}[1]{\left\lceil#1\right\rceil}
\newcommand{\conj}[1]{\overline{#1}}
\newcommand{\ind}{\mathbf 1}
\newcommand{\eps}{\varepsilon}

\let\tilde\widetilde

\newcommand{\X}{X}
\newcommand{\Xhat}{\hat{\X}}
\newcommand{\Xtilde}{\tilde{\X}}
\newcommand{\R}{R}
\newcommand{\Rhat}{\hat{\R}}

\newcommand{\papertitle}{A Nonadaptive Sparse Fourier Transform
  in Every Dimension}
\title{\papertitle}

\author{\textsc{Mahdi Cheraghchi}\thanks{%
  Emails: \{mahdich, jylyu\}@umich.edu.
  This research was partially supported by the National Science
  Foundation under Grant No.\ CCF-2236931.}}

\author{\textsc{Jingyi Lyu}\protect\footnotemark[1]}

\affil{Department of EECS\\
  University of Michigan, Ann Arbor MI, USA}

\date{}

\hypersetup{
  pdftitle={A Nonadaptive Sparse Fourier Transform in Every Dimension},
  pdfauthor={Mahdi Cheraghchi; Jingyi Lyu},
  pdfsubject={Sparse Fourier transforms and sparse recovery},
  pdfkeywords={sparse Fourier transform, sparse recovery,
    nonadaptive sampling, high-dimensional Fourier transform},
  pdfdisplaydoctitle=true
}

\begin{document}
\pagenumbering{roman}
\maketitle
\begin{abstract}
\normalsize
Let $G=(\Z/2^a\Z)^d$, of order $N=2^{ad}$.  Given point-query access to any
complex signal $\X$ with Fourier transform $\Xhat$ over $G$, we provide a
randomized algorithm that
constructs a $k$-sparse estimate $\Xtilde$
via non-adaptive queries such that, with large probability,
\[
 \norm{\Xtilde-\Xhat}_2^2 \leq
 C \cdot \norm{\HT_k(\Xhat)-\Xhat}_2^2,
\]
for an absolute constant $C>0$,
where $\HT_k(\cdot)$ denotes the best $k$-sparse approximation.
The constant $C$ can be improved to any desired
$1+\eps$ by either supplying a constant-factor estimate on the $k$-sparse
tail norm or allowing an arbitrarily small
additive error term $\delta \|\Xhat\|_2^2$ on the right-hand side.
With a supplied tail estimate, we can
also satisfy an $\ell_\infty/\ell_2$ heavy hitters guarantee.
For fixed $\delta$, the number of samples is
$O\left(\frac{k}{\eps^2}(\log N)\log(k/\eps)\right)$,
and the recovery algorithm requires a polynomial in $(k/\eps)\log N$
exact operations.
The algorithm naturally extends to
$G=(\Z/b^a\Z)^d$ for any fixed integer $b>1$.
\end{abstract}

\tableofcontents
\clearpage
\pagenumbering{arabic}

\section{Introduction}
\label{sec:intro}

We consider the $d$-dimensional Discrete Fourier Transform over the group
$G=(\Z/Q\Z)^d$ of order $N=Q^d$, for any dimension $d \geq 1$.
Let $\X\colon G \to \C$ be an $N$-dimensional complex signal.
The Fourier character $\chi_f$ for group element $f \in G$ is given by
\[
  \chi_f(t)=\exp\!\left(\frac{2\pi\iu\langle f,t\rangle}{Q}\right),
\]
where $\iu = \sqrt{-1}$ denotes the imaginary unit.
The Fourier coefficient $\Xhat(f)$ corresponding to any frequency $f \in G$
is then given by the (normalized) complex inner product of the signal and
$\chi_f$; namely,
\[
  \Xhat(f):=\frac1N\sum_{t\in G}\X(t)\conj{\chi_f(t)}.
\]
With this normalization, the inverse transform becomes
\begin{equation}
\label{eqn:fourier-normalization}
  \X(t)=\sum_{f\in G}\Xhat(f)\chi_f(t).
\end{equation}
We prove our results for $Q=2^a$, where $a\geq1$, and extend them to
$Q=b^a$ for any fixed integer $b\geq2$.  The bounds are uniform in $a$
and $d$; their constants may depend on $b$.  Thus ``every dimension''
refers to this family of groups.  We do not claim the same bounds for
large-prime side lengths or for arbitrary finite abelian groups.

The textbook Fast Fourier Transform (FFT) algorithm can compute
$\Xhat\colon G \to \C$ in nearly linear time $O(N \log N)$
in the signal length.  In many applications,
however, the signal is known to exhibit sparsity
in the frequency domain.  Applications range from audiovisual
compression (e.g., JPEG, MPEG), medical imaging (e.g., fMRI),
and astronomy (e.g., Event Horizon Telescope) to theoretical areas
such as computational learning theory, complexity theory, and
cryptography.
In such cases, it is possible to recover the sparse spectrum
exponentially more efficiently (both in terms of the number
of samples needed and the algorithmic runtime).  Sparse Fourier Transforms
aim to provide such improvements, both in theory and practice.

Sparse Fourier Transforms are closely related to compressed sensing.
In compressed sensing, the goal is to robustly recover a sparse (or
approximately sparse) signal using as few linear measurements as
possible.  Namely, for a $k$-sparse (or nearly $k$-sparse)
spectrum $\Xhat \in \C^N$, the goal is to design a measurement scheme
$M\in \C^{n\times N}$ such that from a sketch $M \cdot \Xhat \in \C^n$,
one can recover $\Xhat$ either exactly or approximately.
For randomized nonadaptive $\ell_2/\ell_2$ recovery from linear
measurements, a standard lower-bound scale is
$n=\Omega(k\log(N/k))$.  Price and Woodruff~\citeat{ref:PW11}{Thm.~4.4}
establish this bound for fixed approximation factor $1+\eps_0$, where
$0<\eps_0<1$, and constant failure probability below $1/2$, in the
regime $k\log N=o(N)$.  Their decoder receives the sketch without a
supplied tail estimate, and its output need not be $k$-sparse.
Since the Fourier transform is unitary up to a scalar normalization,
the question of
recovering a sparse spectrum by sampling the signal is equivalent
to recovering a sparse signal by querying its spectrum (with the
roles of the signal and its transform swapped, up to normalization).
The sampling matrix $M$ is then a submatrix of the Fourier transform
matrix over the group $G$.  For sparse Fourier transforms, we also seek
recovery in polynomial (ideally nearly linear) time in the number
of samples $n$.

Borrowing the terminology of compressed sensing, we adopt the
$\ell_q/\ell_p$ benchmark for assessing the robustness guarantees
of sparse Fourier transform (cf.\ \cite{ref:FR13}).
Let $\HT_k(\Xhat)$ be the best $k$-sparse approximation of $\Xhat$; i.e.,
what is obtained from $\Xhat$ by zeroing out all but the top $k$
entries of $\Xhat$ in magnitude.  Let
\[ \sigma_k(\Xhat)_p
  :=\norm{\Xhat-\HT_k(\Xhat)}_p\]
  be the $\ell_p$ norm of the signal's $k$-sparse tail.
Then, an $\ell_q/\ell_p$ recovery guarantee (assuming $0 < p\leq q$)
requires
the reconstruction $\Xtilde$ of the spectrum $\Xhat$
to satisfy
\begin{equation}
\label{ellqellp}
  \norm{\Xhat-\Xtilde}_q
  \le C\,k^{1/q-1/p}\,\sigma_k(\Xhat)_p,
\end{equation}
where, ideally, one wishes $C=1+\eps$ for a small parameter $\eps>0$.
Under this normalization, increasing either $p$ or $q$ strengthens the
guarantee in the sense of efficient reductions, up to constant changes
in sparsity and approximation factor~\cite{ref:MN16}.  These reductions
do not in general preserve a factor $1+\eps$.
This clean guarantee is often relaxed in the literature to accommodate
the limitations of the techniques.  Common relaxations include an
additional additive error term on the right-hand side, bounds on the
dynamic range of the signal entries, and bounds on the SNR
(signal-to-noise ratio), such as the ratio between the signal norm and
its $k$-sparse tail norm.
Of special note is the $\ell_\infty/\ell_2$
case, which captures a tail-relative \emph{heavy hitters} guarantee:
the coordinatewise error is $O(\sigma_k(\Xhat)_2/\sqrt{k})$.
This scale is measured relative to the tail norm.

In this work, we focus on the important case $p=q=2$, which is
the standard convention in the sparse Fourier transform literature.
It is known that for this case, sublinear sample complexity is
only achievable when the use of randomness in sampling is allowed
\cite{ref:CDD09}.  This setting is often called nonuniform or
``for each'' in the literature (cf.\ \cite{ref:MN16}).
Our focus is moreover on the nonadaptive query model: all query
locations are chosen in advance using randomness, and the queries are
performed independently of earlier outcomes.

\subsection{Main result and related work}

Our main result is the following.

\begin{theorem}[Nonadaptive sparse Fourier
recovery]\label{thm:main-dyadic-intro}
Fix a constant $c\geq1$.
Let $a,d\geq1$ be integers, $G=(\Z/2^a\Z)^d$, $N=|G|=2^{ad}$, $\eps \in
(0,1/2]$, and
$2 \leq k < N$.
There is a randomized algorithm which, given point-query access to any fixed
$\X \in \C^G$ and a tail estimate $B$ satisfying
\[
  \sigma_k(\Xhat)_2 \leq B \leq c\,\sigma_k(\Xhat)_2,
\]
uses non-adaptive queries, computes an estimate $\Xtilde^{\rm pre}$
with $O_c(k/\eps^2)$ nonzero coordinates, and returns the $k$-sparse vector
$\Xtilde=\HT_k(\Xtilde^{\rm pre})$.  With probability at least $9/10$,
both
\[
  \norm{\Xhat-\Xtilde^{\rm pre}}_\infty
  \leq\frac{\eps}{\sqrt{k}}\,\sigma_k(\Xhat)_2
\]
and
\begin{equation}
\label{eqn:main-error}
  \norm{\Xhat-\Xtilde}_2^2
  \leq(1+\eps)\sigma_k(\Xhat)_2^2,
\end{equation}
where $\Xhat \in \C^G$ denotes the Fourier transform of $\X$ over $G$
and $\sigma_k(\Xhat)_2$ denotes its best $k$-sparse approximation error in
$\ell_2$.
Moreover, the algorithm makes at most
\begin{equation}
  \min\!\left\{N,
    O\left(\frac{k}{\eps^2} (\log N) \log(k/\eps)\right)\right\}
\end{equation}
queries.
The runtime of the algorithm (assuming exact operations) is at most
\[
  O\!\left(
    \frac{k^{7/2}}{\eps^7}(\log N)^{9/4}
    \log\!\left(\frac{k\log N}{\eps}\right)
  \right) = \mathsf{poly}((k/\eps) \log N).
\]
The hidden constants may depend on $c$.
\end{theorem}

The coordinatewise estimate and its hard thresholding are obtained on
the same success event, as shown in \Cref{sec:resources}.
We can remove the supplied tail estimate at the cost of an absolute
constant approximation factor.  The following result is proved as
\Cref{thm:tail-free} in \Cref{sec:tail-free}.
In this result and the next, let $G=(\Z/2^a\Z)^d$ and $N=2^{ad}$
for integers $a,d\geq1$.

\begin{theorem}[Recovery without a tail estimate]
\label{thm:tail-free-intro}
Let $2\leq k<N$.  There are an absolute constant $C$ and a randomized
algorithm which, given point-query access to any fixed $\X\in\C^G$,
returns a $k$-sparse vector $\Xtilde$ such that, with probability at
least $9/10$,
\[
  \norm{\Xhat-\Xtilde}_2^2\leq C\,\sigma_k(\Xhat)_2^2.
\]
No tail estimate or signal norm is supplied.  The algorithm uses at most
$\min\{N,O(k\log N\log k)\}$ nonadaptive point queries and
$\mathsf{poly}(k\log N)$ exact operations on every execution.
In particular, if $\Xhat$ is $k$-sparse, it is recovered exactly with
probability at least $9/10$.
\end{theorem}

Alternatively, we retain the factor $1+\eps$ without a supplied tail
estimate or signal norm by allowing a relative additive error.
\Cref{thm:relative-additive} in \Cref{sec:relative-additive} proves the
following result and gives the full polynomial running-time bound.

\begin{theorem}[Recovery with relative additive error]
\label{thm:relative-additive-intro}
Let $2\leq k<N$, $0<\eps\leq1/2$, and $0<\delta\leq1/4$.
There is a randomized algorithm which, given point-query access to any
fixed $\X\in\C^G$, returns a $k$-sparse vector $\Xtilde$ such that,
with probability at least $9/10$,
\[
  \norm{\Xhat-\Xtilde}_2^2
  \leq(1+\eps)\sigma_k(\Xhat)_2^2+\delta\norm{\Xhat}_2^2.
\]
No tail estimate or signal norm is supplied.  The algorithm uses at most
\[
  \min\!\left\{N,O\!\left(\frac{k}{\eps^2}
    \left((\log N)\log\frac{k}{\eps}
      +\log\log\frac1{\eps\delta}\right)\right)\right\}
\]
nonadaptive point queries and
$\mathsf{poly}((k/\eps)\log N,\log(1/\delta))$ exact operations on
every execution.  All hidden constants are absolute.
\end{theorem}

For fixed $\delta$, the sample bound in
\Cref{thm:relative-additive-intro} has the same asymptotic form as in
\Cref{thm:main-dyadic-intro}.

\Cref{thm:main-dyadic-intro} uses $O(k\log N\log k)$ samples for
constant $\eps$ in every dimension $d$, and at the same time provides
polynomial-time recovery in the number of samples
(i.e., $\mathsf{poly}((k/\eps)\log N)$).
For fixed $0<c_0<1$ and $k\leq N^{1-c_0}$, this sample bound is within
an $O(\log k)$ factor of the standard lower-bound scale above, since
$\log(N/k)\geq c_0\log N$; the cited lower bound applies in this regime
for sufficiently large $N$.

The inverse-quadratic dependence on $\eps$ is also known in the
standard linear-sketch model.  Price and
Woodruff~\citeat{ref:PW11}{Sec.~7} prove an
$\Omega(k/\eps^2)$ measurement bound for outputs with at most $k$
nonzero coordinates, when $k>1$, $\eps$ and the constant failure
probability are sufficiently small, and $N\geq Ck/\eps^2$ for a
sufficiently large absolute constant $C$.
The output restriction matters: their Theorem~3.1 allows $2k$ nonzero
coordinates and uses $O((k/\eps)\log N)$ measurements.
These results use unsquared norm error; replacing their accuracy
parameter by $\sqrt{1+\eps}-1=\Theta(\eps)$ gives the same dependence
under our squared-error convention.
These comparisons concern the ordinary linear-sketch model; they do
not establish optimality under the supplied-tail promise of
\Cref{thm:main-dyadic-intro}.

For Fourier heavy hitters, the closest prior result uniform in $d$
is due to Nakos, Song, and Wang~\cite{ref:NSW19}.  It uses
$O(k\log k\log N)$ samples and $O(N\log^3N\log k)$ time, with a
supplied tail estimate and a polynomial-range assumption.
Our bounds replace the dependence on $N$ in decoding by a polynomial
in $k$, $1/\eps$, and $\log N$, and remove the range assumption in the
exact-arithmetic model of \Cref{sec:model}.

Bounds for randomly subsampled Fourier matrices were developed by
Cand\`es and Tao~\cite{ref:CT06}, Rudelson and Vershynin~\cite{ref:RV08},
Cheraghchi, Guruswami, and Velingker~\cite{ref:CGV13}, and
Bourgain~\cite{ref:Bou14}.
Haviv and Regev~\cite{ref:HR16} obtain the restricted isometry property
(RIP) with $O(k\log^2k\log N)$ samples, which, combined with
CoSaMP~\citeat{ref:NT09}{Thm.~2.2 in the full version}, gives
nonadaptive, for-each constant-factor $\ell_2/\ell_2$ recovery over any
finite abelian group in $\mathsf{poly}(N)$ exact operations, without a
supplied tail estimate or range bound.\footnote{The usual uniform
RIP guarantee is $\ell_2/\ell_1$.  For the for-each consequence, write
$\Xhat=\HT_k(\Xhat)+z$ and let $A$ be the normalized sampled Fourier
matrix.  Then $\E\norm{Az}_2^2=\norm z_2^2$.  Markov's inequality
bounds the measurement error $Az$ by a constant times $\norm z_2$ with
constant probability; the deterministic noise-robust guarantee for
CoSaMP recovers the $k$-sparse head to that accuracy.}
Our constant-factor theorem saves one $\log k$ factor in this sample
bound and decodes in $\mathsf{poly}(k\log N)$ operations.
With a supplied tail estimate, \Cref{thm:main-dyadic-intro} additionally
gives $1+\eps$ error and a coordinatewise guarantee for the intermediate
estimate; without that input, \Cref{thm:relative-additive-intro} gives
$1+\eps$ with a relative additive error.

Historically, one of the first examples of sparse Fourier transforms
is the celebrated Goldreich-Levin theorem \cite{ref:GL89}.
This theorem applies to the sparse Hadamard transform, which is
the extremal case when $a=1$ (i.e., $G=\mathbb{F}_2^d$).
The original result from \cite{ref:GL89} provides an algorithm
for this case in time $\mathsf{poly}(k \log N)$; see also the exposition
in~\cite{ref:Gol01}.
To be more precise, this result provides a global heavy hitters algorithm,
by which we mean recovering all frequencies $f$ satisfying
$|\Xhat(f)|^2\geq\eta\norm{\Xhat}_2^2$ for a prescribed $\eta>0$.
Here and throughout, Fourier mass means a sum of squared magnitudes.  This
guarantee
is weaker than the $\ell_\infty/\ell_2$ requirement in \eqref{ellqellp}
and, unlike the latter, does not automatically imply an $\ell_2/\ell_2$
guarantee (or vice versa).

More recently, Amrollahi, Zandieh, Kapralov, and
Krause~\citeat{ref:AZKK19}{Thm.~3} gave an $\ell_2/\ell_2$ algorithm
for arbitrary real signals on the Boolean cube, with a $1+\eps$
squared-error guarantee and $O(k)$-sparse output.  For fixed $\eps$,
setting their degree parameter to the cube dimension gives
$O(k\log N(\log\log N)^2)$ samples and
$\mathsf{poly}(k,\log N)$ decoding time, with no supplied tail estimate
or coefficient-range assumption.  Their shifted linear hashing
construction permits the query locations to be fixed in advance:
residuals affect the subtraction from acquired measurements, rather
than their locations.  Thus robust nonadaptive hashing is already
available at the Boolean endpoint.  SparseFHT~\cite{ref:SHV15} and
SPRIGHT~\cite{ref:LBPR15} give other fast sparse Walsh--Hadamard
algorithms, under random-support and noise assumptions that differ
from arbitrary-tail recovery.
Error-correcting identification also appears in the sparse-recovery
scheme of Gilbert, Li, Porat, and Strauss~\cite{ref:GLPS10}, whose
measurements are freely chosen linear sketches.

Also in the Walsh--Hadamard setting, Cheraghchi and Indyk
\cite{ref:CI16} gave a deterministic, nonadaptive sparse recovery
algorithm for real-valued signals.  Their recovery guarantee is
$\ell_1/\ell_1$, rather than the $\ell_2/\ell_2$ guarantee considered here.

In the other extreme of $d=1$, a long line of research provides
increasingly better guarantees to the point of practical performance
\cites{ref:GGIMS02,ref:AGS03,ref:GMS05,ref:ZGSD05,ref:IGS07,
ref:Iwen10,ref:HIKP12practical,ref:HIKP12,ref:SP14}.
Notably,
Hassanieh, Indyk, Katabi, and Price developed algorithms that outperform
standard FFT implementations in sufficiently sparse regimes
\cite{ref:HIKP12practical}, and obtained an $O(k\log N)$-time algorithm
for exactly sparse spectra and nearly linear dependence on $k$ for
approximately sparse spectra, up to logarithmic factors
\cite{ref:HIKP12}.

These algorithms use flat frequency filters: a window is nearly
constant on a passband and small outside a slightly larger band,
allowing a bucket to retain nearby Fourier coefficients while controlling
leakage~\citeat{ref:HIKP12}{Def.~2.3}.
The constructions flatten a Gaussian or Dolph--Chebyshev window by
convolution with a boxcar; the theoretical Gaussian construction is
also discretized, periodized, and
truncated~\citeat{ref:HIKP12practical}{Def.~3.3 and Claim~3.4},
\citeat{ref:HIKP12}{Lem.~7.3}.
Tensor products extend such filters to higher dimensions, but a direct
use of sharp filters can turn an $O(k\log N)$ sample cost into
$O(k\log^dN)$; coarser filters reduce this
cost~\citeat{ref:IK14}{Sec.~3}, \cite{ref:Kap16}.

For fixed dimension $d$, Indyk and Kapralov obtained
$\ell_2/\ell_2$ recovery using $O_d(k\log N)$ samples
and $O(N\log^3 N)$ time
\cite{ref:IK14}.  Kapralov subsequently achieved
$O_d(k\log N\log\log N)$ samples and $O_d(k\log^{d+3}N)$ time
\cite{ref:Kap16}.  Here the accuracy is constant, and the hidden
constants depend (exponentially) on $d$; these results also use a supplied
tail-norm estimate
and polynomial-range assumptions.  For arbitrary dimension $d$,
Nakos, Song, and Wang obtained a nonadaptive $\ell_\infty/\ell_2$
guarantee using $O(k\log k\log N)$ samples, with no additional
dimension dependence, but with $O(N\log^3 N\log k)$ running time
and a supplied tail-norm estimate and polynomial-range assumptions
\cite{ref:NSW19}.

Other Fourier recovery models allow different side information.
Akavia's deterministic Fourier approximation algorithm~\cite{ref:Aka10}
has polynomial dependence on a supplied Fourier $\ell_1$ bound and a
coefficient threshold, rather than the tail-relative parameters used
here.  For high-dimensional trigonometric polynomials on the continuous
torus, K\"ammerer, Potts, and Volkmer~\cite{ref:KPV17} use sampling on
multiple rank-1 lattices.  Their algorithm starts from a prescribed
frequency candidate set and a threshold below the nonzero coefficient
magnitudes; later lattices depend on the projected support already
identified.  These are different guarantees and sampling models from
our nonadaptive point-query recovery of arbitrary discrete signals.

A different approach, based on adaptive aliasing
filters, gives dimension-independent recovery in
$\mathsf{poly}(k,\log N)$ time for exactly sparse spectra
\cite{ref:KVZ19}.  Bringmann, Kapralov, Makarov, Nakos, Yagudin, and Zandieh
subsequently improved the
exact-sparse running time to nearly quadratic in $k$ and developed
a robust variant under a head-tail separation assumption
\cite{ref:BKM23}.  These latter algorithms use adaptive sampling.
Their FFT computation tree is the same frequency partition tree as our
subgroup chain, up to coordinate order and child labels.  On the
Boolean cube it is the Goldreich--Levin/Kushilevitz--Mansour prefix
tree~\cite{ref:GL89,ref:KM93}; at $d=1$ it is residue-class decimation.
Our contribution lies in the robust node test, its nonadaptive
implementation, and the traversal and cost analysis.
The resulting guarantee is uniform throughout the family
$(\Z/2^a\Z)^d$, including the regime where both $a$ and $d$ grow.
The Boolean endpoint already has the arbitrary-tail algorithm above.
In constant dimension, the filter-based algorithms have linear dependence on
$k$ in their running times,
with the dimension and logarithmic factors shown in \Cref{tab:comparison}.

\Cref{tab:comparison} compares selected quantitative bounds, with their
accuracy dependence shown and constant failure probability.

\begin{table}[p]
\centering
\setlength{\tabcolsep}{4pt}
\renewcommand{\arraystretch}{1.15}
\setlength{\extrarowheight}{2pt}
\caption{Selected sparse Fourier bounds.  Where two bounds are given,
samples are listed first and running time second; ``both'' gives a
common bound.  An asterisk
marks a row whose sample bound is inherited from running time.
Randomized guarantees
have constant failure probability.  Except for the RIP row, sides are
powers of two; the Boolean cube has side two and $d=\log_2N$.
Dimension dependence is explicit, and
$\alpha>0$ is fixed.  For Bringmann et al., $\widetilde O$ hides
polynomial factors in $\log N$.}
\label{tab:comparison}
\begin{tabularx}{\textwidth}{|>{\raggedright\arraybackslash}p{0.15\textwidth}|
  >{\centering\arraybackslash}p{0.43\textwidth}|
  >{\raggedright\arraybackslash}X|}
\hline
Source & Samples / running time & Setting and promises \\
\hline
\cite{ref:GL89}
& Both: $\mathsf{poly}(k,\log N)$
& Boolean cube, Boolean input;
  global heavy hitters $|\Xhat(f)|\geq k^{-1/2}$. \\
\hline
\citeat{ref:CI16}{Thm.~1}
& Both$^*$: $k^{1+\alpha}(\log N)^{2/\alpha+O(1)}$
& Boolean cube; deterministic, nonadaptive; $\ell_1/\ell_1$;
  real input, $\log^{O(1)}N$ bits. \\
\hline
\citeat{ref:HIKP12}{Thm.~4.9}
& Both$^*$: $\begin{gathered}
  O\!\bigl((k/\eps)\log(N/k)\\
  {}\cdot\log(N/\rho)\bigr)
  \end{gathered}$
& $d=1$; nonadaptive; additive norm error $\rho\norm{\Xhat}_2$;
  polynomial SNR. \\
\hline
\citeat{ref:IK14}{Thm.~6.3}
& $\begin{gathered}
  2^{O(d\log(d+1))}(k/\eps)\log N\\
  O(N\log^3N)
  \end{gathered}$
& Nonadaptive; supplied tail-norm estimate, polynomial SNR. \\
\hline
\citeat{ref:Kap16}{Thm.~3.5}
& $\begin{gathered}
  2^{O(d^2)}k\log N(\log\log N+\eps^{-1})\\
  2^{O(d^2)}(k/\eps)\log^{d+3}N
  \end{gathered}$
& Supplied tail-norm estimate, polynomial range. \\
\hline
\citeat{ref:Kap17}{Thm.~1.2}
& $\begin{gathered}
  O((k/\eps)\log N)\\
  O((k/\eps)\log^{17/4}N)
  \end{gathered}$
& $d=1$; supplied tail-norm estimate, polynomial SNR;
  $N^{-1}<\eps<1$; fixed runtime tradeoff. \\
\hline
\cite{ref:HR16}, \cite{ref:NT09}
& $\begin{gathered}
  O(k\log^2k\log N)\\
  \mathsf{poly}(N)
  \end{gathered}$
& Fourier RIP and CoSaMP; any finite abelian group; nonadaptive;
  constant-factor for-each $\ell_2/\ell_2$. \\
\hline
\citeat{ref:NSW19}{Thm.~1.1}
& $\begin{gathered}
  O(k\log N\log k)\\
  O(N\log^3N\log k)
  \end{gathered}$
& All $d$; nonadaptive; $\ell_\infty/\ell_2$;
  supplied tail-norm estimate, polynomial range. \\
\hline
\citeat{ref:KVZ19}{Thm.~3}
& Both: $O(k^3\log^2k\log^2N)$
& All $d$; deterministic, adaptive; exactly $k$-sparse. \\
\hline
\citeat{ref:BKM23}{Thm.~1}
& Both$^*$: $\widetilde O\!\left(k^2 2^{8\sqrt{\log k\log\log N}}\right)$
& All $d$; deterministic, adaptive; exactly sparse. \\
\hline
\citeat{ref:BKM23}{Thm.~4}
& $\begin{gathered}
  \widetilde O\!\left(k^2/\eps+
    k^2 2^{\Theta(\sqrt{\log k\log\log N})}\right)\\
  \widetilde O(k^3/\eps)
  \end{gathered}$
& All $d$; adaptive; disjoint head $w$ and tail $\eta$,
  $|w_f|\geq3\norm{\eta}_2$. \\
\hline
This paper, \Cref{thm:main-dyadic-intro}
& $\begin{gathered}
  O\!\left((k/\eps^2)\log N\log(k/\eps)\right)\\
  O\!\left(k^{7/2}\eps^{-7}\log^{9/4}N\cdot{}
  \right.\\[-2pt]
  \left.{}\log(k\log N/\eps)\right)
  \end{gathered}$
& All $d$; nonadaptive; arbitrary tail;
  constant-factor tail-norm estimate. \\
\hline
\end{tabularx}
\end{table}

\paragraph{Limitations and open questions.}
The constants in the tests and query allocation are not optimized.
The algorithm uses the full-FFT branch whenever its planned sample
count reaches $N$; practical performance is outside this analysis.
For fixed accuracy, our displayed operation bound is sublinear when
$k^{7/2}(\log N)^{9/4}\log(k\log N)=o(N)$, corresponding to
sparsity below about
$N^{2/7}$ up to logarithmic factors.
The filter-based algorithms of \cite{ref:Kap16,ref:Kap17} have linear
dependence on $k$ in their time bounds under their promises, compared
with our $k^{7/2}$ dependence.  Our advantages there are the absence of
range assumptions and the variants without a supplied tail estimate.
The sample bound leaves a $\log k$ gap from the usual lower-bound scale
at constant accuracy.  Removing that gap, improving the accuracy
dependence with $O(k)$-sparse output, and extending the results to groups
with large prime factors remain natural questions.

\subsection{Overview of techniques and sketch of the results}
\label{sec:overview}

We give an overview of our techniques and all the technical results used
to obtain the recovery guarantees, so that the argument can be followed
before the detailed proofs.  Let $\X\in\C^G$, where
$G=(\Z/2^a\Z)^d$ has order $N=2^{ad}$.
We seek a $k$-sparse approximation $\Xtilde$ to the spectrum $\Xhat$
from nonadaptive point queries to $\X$.
As is common in the sparse recovery literature, the decoder proceeds by
iterative peeling.  In each round, it locates
significant frequencies of the current residual, adds them to a growing
support, and refits the coefficients on the entire support by least squares.
It computes the needed residual values by subtracting the fitted signal
from the stored samples and, if another round is needed, searches at a
lower threshold.  All point queries are made before decoding.
The main difficulties are to
search without enumerating $G$, to implement the search with Fourier
samples, and to control its total cost without a coefficient-range
assumption.  All success guarantees below are for each fixed signal,
over the sampling randomness; the recovery theorems succeed with
probability at least $9/10$.

We first sketch how the main components fit together, before discussing
each in detail.  Exact coset projections let us test frequency buckets
using point queries.  A heavy coefficient
forces its projection to have large average magnitude despite
cancellation from other coefficients.  A multiscale test detects this
using only second moments, and a node potential lets heavy paths survive
occasional errors while controlling the total search size.  Least
squares bounds the residual norm simultaneously for all supports inside
one fixed set of significant frequencies, so the bound remains valid
for the adaptively chosen supports on a successful run.  In
\Cref{sec:schedule}, we charge the sum of the locator cost parameters to
still-unidentified significant frequencies and to the tail allowance.
Geometric threshold decrease
bounds each significant frequency's total charge; the tail contributions
also sum to a constant multiple of the recovery sparsity.  This controls
the total cost parameter without a coefficient-range assumption and
lets us allocate all samples in advance as in \Cref{sec:tapes}.
The decoder reads them sequentially: conditional on the entries read so
far, the next unread
blocks, including their locations, have the same distribution as fresh
samples.

Without a supplied tail estimate, the decoder of \Cref{sec:tail-free}
saves every fit instead of trying to recognize the right stopping scale.
Its empirical residual norm estimate need not always bound the true
residual.  On the
least-squares event, however, it does so until a sufficiently good fit
has been saved, provided the preceding locator calls succeed.  The same
cost accounting forces a good fit to appear before the support or work
limits stop the run.  Independent samples compare candidates by
estimating differences of squared errors.  Sparsity bounds the variance
of these estimates.  Selecting a candidate by these comparisons gives
constant-factor
recovery without a tail estimate.  For the relative additive guarantee,
the reduction in \Cref{sec:relative-additive} uses a preliminary sample
to identify a range of scales, unless zero is already a good
approximation.  We reuse the recovery query sets at
those scales and apply the same candidate selection rule.

\subsubsection{Coset projections and the search tree}

The algebraic starting point is that a frequency bucket can be accessed
by averaging the signal over a subgroup in the time domain.  For a
subgroup $H\leq G$, let
$H^\perp=\{z\in G:\chi_h(z)=1\text{ for every }h\in H\}$ be its
annihilator.  Character orthogonality gives
\begin{equation}
\label{eqn:overview-projection}
  \sum_{f\in v+H}\Xhat(f)\chi_f(t)
  =\E_{z\in H^\perp}\X(t+z)\conj{\chi_v(z)}.
\end{equation}
Thus the average on a time coset $t+H^\perp$ isolates the frequency
coset $v+H$.  At $t=0$ it gives the sum of that bucket's coefficients;
varying $t$ changes their phases.  This coset projection identity is the
algebraic ingredient used by Cheraghchi and Indyk~\cite{ref:CI16} in
their sparse Walsh--Hadamard transform.  We retain that principle,
while developing a different recovery mechanism for arbitrary complex
tails and an $\ell_2/\ell_2$ guarantee.

To organize the buckets, write each coordinate of a frequency in binary,
read its bits from least to most significant, and concatenate the
coordinates in a fixed order.  Let $H_j$ contain the frequencies whose
first $j$ bits are zero.  More explicitly, for $j=\ell a+c<ad$, with
$0\leq c<a$, set
\[
  H_j=\{0\}^{\ell}\times(2^c\Z/2^a\Z)
       \times(\Z/2^a\Z)^{d-\ell-1},
  \qquad H_{ad}=\{0\}^d.
\]
A coset of $H_j$ fixes the first $j$ bits; its two children fix the
next bit.  These prefixes label cyclic cosets, without identifying
cyclic addition with bitwise addition.

For example, write $Z=\Z/8\Z$ and take $G=Z^2$.  The subgroup chain
and its annihilators are
\[
\begin{array}{c|c|c}
 j&H_j&K_j=H_j^\perp\\ \hline
 0&Z\times Z&\{0\}\times\{0\}\\
 1&2Z\times Z&4Z\times\{0\}\\
 2&4Z\times Z&2Z\times\{0\}\\
 3&\{0\}\times Z&Z\times\{0\}\\
 4&\{0\}\times2Z&Z\times4Z\\
 5&\{0\}\times4Z&Z\times2Z\\
 6&\{0\}\times\{0\}&Z\times Z.
\end{array}
\]
At depth two, a bucket fixes the first coordinate modulo four and
leaves the second free.  Its projection is evaluated by averaging over
shifts of $K_2=\{0,2,4,6\}\times\{0\}$.  On the Boolean cube the
annihilator simply exchanges fixed and free coordinates; the table
illustrates the divisibility condition that replaces this picture for
larger cyclic groups.

The resulting binary tree has depth $\log N$, $N$ leaves, and $2N-1$ nodes.
The $2^j$ buckets at depth $j$ partition $G$.  The root represents the
whole spectrum and has projected signal $\X(t)$, while leaf $f$ has
projected signal $\Xhat(f)\chi_f(t)$.  Only a pruned part of this tree
will be explored.  \Cref{sec:projections} defines this tree formally;
\Cref{prop:orthogonality} proves its projection identity, and
\Cref{prop:statistics} gives the statistics used to guide the search.

\subsubsection{Nonadaptive samples and least-squares residuals}

We first solve a coordinatewise recovery problem.  Given a sparsity
parameter $2\leq s<N$ and an accuracy scale $\tau\geq0$ satisfying
\[
  \sigma_s(\Xhat)_2^2\leq s\tau^2,
\]
we seek an estimate supported on fewer than $17s$ frequencies with
coordinatewise error at most $\tau/2$; this is
\Cref{thm:upper-tail}.  In the main theorem, $0<\eps\leq1/2$ specifies
the desired relative squared-error factor $1+\eps$.  We take
$s=O(k/\eps^2)$, with the hidden constant depending only on the fixed
quality of the supplied tail estimate, and choose $\tau$ from that
estimate.  We describe the positive-scale case first and return to
$\tau=0$ below.

We use the query sets of \Cref{sec:tapes-def}, all chosen and queried
before decoding.
As in \Cref{sec:model}, each is a multiset with fixed indices.
\begin{itemize}
\item \emph{Least-squares queries} fit the coefficients on the current
  support and estimate the residual norm.  The same set serves every round.
\item \emph{Tree search queries} test frequency buckets.  Their structure
  is described below; all buckets at a given depth share the same queries.
\item \emph{Verification queries} estimate coefficients at retained leaves
  and filter the candidate frequencies using coefficient estimates.
\item \emph{Final coefficient estimation queries} refine the coefficients
  after support identification.  In the constant-factor variant without a
  tail estimate, this independent final set instead provides
  \emph{candidate selection queries}.
\end{itemize}
These sets are mutually independent; tree search samples have the
within-block dependencies described next.
At depth $j$, choose an \emph{anchor} $t$ uniformly in $G$, choose
\emph{offsets} $z$ uniformly in $K_j$, and query $\X(t+z)$.  The
anchor specifies where to evaluate a projection; offsets approximate
the average in \eqref{eqn:overview-projection}.  To accommodate several
possible distributions of projected values, use power-of-two sampling
resolutions.
For a total search budget $K=O(s)$, rounded up to a power of two $K'$,
resolution $w\in\{1,2,4,\ldots,K'\}$ receives $O(K/w)$ anchors and
$O(w)$ offsets per anchor.  Each resolution costs $O(K)$ samples, so
the structured samples cost
\[
  O(K\log K\log N)=O(s\log s\log N).
\]
All buckets at the same depth use these same samples, with different
character weights.  Exploring another bucket costs computation but
requires no further queries.  We will justify the budget and the
adaptive use of this fixed collection after describing the decoding procedure.

The decoder maintains a growing support $S$, initially empty, and a
least-squares fit $C_S$ supported on $S$.  A search round locates
frequencies in the residual spectrum $\Rhat=\Xhat-C_S$ and adds
them to $S$.  Residual time values are available from the original
samples through
\[
  \R(t)=\X(t)-\sum_{f\in S}C_S(f)\chi_f(t).
\]
After enlarging $S$, we refit its coefficients using one independent
query set of $n_{\rm ls}=O(s\log s)$ uniform points $t_b$.  Define
\[
  A_{bf}=\frac{\chi_f(t_b)}{\sqrt{n_{\rm ls}}},\qquad
  \mathbf{b}_b=\frac{\X(t_b)}{\sqrt{n_{\rm ls}}},\qquad
  C_S=\arg\min_{c\in\C^S}\norm{\mathbf{b}-A_Sc}_2.
\]
We extend $C_S$ by zero outside $S$.  The fit minimizes the squared discrepancy
at the observed
points, and its empirical residual $z_S=\mathbf{b}-A_SC_S$ also provides
information about the true Fourier residual.

Since $S$ depends on earlier answers, we need a residual norm estimate
valid simultaneously for every support the successful decoder can reach.  The
fixed set
\[
  \Omega=\{f:|\Xhat(f)|>\tau/4\}
\]
has fewer than $17s$ elements: outside the largest $s$ coefficients,
the tail promise permits fewer than $16s$ coefficients above
$\tau/4$.  Matrix Chernoff bounds make $A_\Omega$ an approximate
isometry, and a separate second-moment bound controls the sampled
contribution outside $\Omega$.  On their common event,
\Cref{lem:ls} gives, simultaneously for every $S\subseteq\Omega$,
a computable residual norm estimate
\begin{equation}
\label{eqn:overview-certificate}
\begin{gathered}
  M_S=2\norm{z_S}_2+C_{\rm ls}\sqrt{s}\,\tau,\qquad
  \norm{\Xhat-C_S}_2\leq M_S,\\
  M_S^2\leq O\!\left(\norm{\Xhat_{\Omega\setminus S}}_2^2
                       +s\tau^2\right),
\end{gathered}
\end{equation}
where $C_{\rm ls}=3b_{\rm ls}$ and $b_{\rm ls}=\sqrt{255}/2$, as in
\Cref{lem:ls}.  Thus the same fitting
samples can be reused throughout.  The embedding is required on this
one fixed frequency set, rather than on every sparse support.  Although
the empirical fitting error decreases as $S$ grows, our analysis does
not assume that the true Fourier error decreases in every round.

\subsubsection{Locating heavy coefficients with multiscale tests}

In one round, fix the residual $\Rhat$, its norm bound $M$, and a
magnitude threshold $0<T\leq M$.  The locator seeks every frequency
with $|\Rhat(f)|\geq T$.  Its effective sparsity parameter is
\[
  \kappa=M^2/T^2:
\]
there can be at most $\kappa$ such frequencies, even though the
residual need not be sparse.  We search for them by breadth-first
traversal of the frequency search tree.

For a bucket $v+H_j$, suppressing the depth in the notation, put
\[
  h_v(t)=\sum_{f\in v+H_j}\Rhat(f)\chi_f(t),\qquad
  F_v=\E_t|h_v(t)|,\qquad
  W_v=\E_t|h_v(t)|^2
      =\sum_{f\in v+H_j}|\Rhat(f)|^2.
\]
Fourier inversion and Cauchy--Schwarz give the bounds of
\Cref{prop:statistics}:
\begin{equation}
\label{eqn:overview-sandwich}
  |\Rhat(f)|\leq F_v\leq\sqrt{W_v}
  \qquad(f\in v+H_j).
\end{equation}
Consequently, every ancestor of a $T$-heavy frequency has $F_v\geq T$.
At any one depth, however, only $O(\kappa)$ buckets can have energy
comparable to $T^2$.  Our test uses these two facts: it returns positive with
high probability when $F_v\geq T$, and returns negative with high probability
when $W_v\leq\theta T^2$, for a small absolute constant $\theta$.
Either answer is allowed between these two promises.
The projection has no leakage from outside the coset, and a heavy
coefficient need not be isolated from other large coefficients in its
bucket.  The cost is that the variance of an offset sample is controlled
by the whole residual energy, rather than the bucket energy.  This
leads to the multiscale allocation below, shared by every bucket at a
depth, instead of the flat filters and random frequency permutations
used in \cite{ref:HIKP12,ref:IK14,ref:Kap16}.

Random anchors are needed because the Fourier coefficients in a bucket
can cancel at a particular time.  Inequality
\eqref{eqn:overview-sandwich} guarantees enough average magnitude when
the bucket contains a heavy coefficient.  Nevertheless, this magnitude
may be distributed very unevenly: $|h_v(t)|$ could be about $T$ at
most anchors, or about $\kappa T$ on only a $1/\kappa$ fraction of
them.  These are distributions of projected values over time, not
distributions of coefficient magnitudes.  The first case needs a few
accurate evaluations; the second needs many coarse evaluations.
The geometric allocation handles both, along with the intermediate
possibilities.

Specifically, round $\kappa$ up to a power of two $\kappa'$.  At
resolution $w\in\{1,2,4,\ldots,\kappa'\}$, use
$\Theta(\kappa'/w)$ anchors.  At each anchor $t$, take three
independent averages of $\Theta(w)$ observations
\[
  V_v(t,z)=\R(t+z)\conj{\chi_v(z)},\qquad z\text{ uniform in }K_j,
\]
and take their median separately in the real and imaginary coordinates.
Call the resulting estimate $\lambda_{w,t}$.  One repetition returns positive
if some anchor at some resolution satisfies
\[
  |\lambda_{w,t}|\geq\frac{\kappa'T}{16w}.
\]
Five independent repetitions followed by a majority vote make both
error probabilities sufficiently small absolute constants.

The three group means are what allow second moments to suffice.
Conditional on $t$, write $\nu(t)=\E_z|V_v(t,z)|^2$.
By \Cref{lem:median}, the chance that the median misses $h_v(t)$ by a
constant fraction of its testing threshold is at most
\[
  \min\!\left\{1,
    C\left(\frac{\nu(t)w}{\kappa'^2T^2}\right)^2\right\}.
\]
This is the conditional estimate in \eqref{eqn:conditional-noise}.
We sum the truncated square over the power-of-two sampling resolutions
before taking expectation over $t$.  The bound of \Cref{clm:dyadic-sum}
\[
  \sum_{w\in\{1,2,4,\ldots,\kappa'\}}
    \frac{\kappa'}w
    \min\!\left\{1,\left(\frac{wu}{\kappa'}\right)^2\right\}
  \leq4u\qquad(u\geq0),
\]
where the sum is over powers of two, reduces the total estimation-error bound
to $\E_t\nu(t)\leq M^2$.  Taking a second moment of $\nu$ instead
would lose this advantage.  A truncation of $|h_v(t)|$ supplies a
detectable anchor when $F_v\geq T$; when $W_v$ is small, its
quadratic tail bound controls false positives.  Each resolution uses
$O(\kappa)$ samples, giving $O(\kappa\log(2\kappa))$ samples per
depth.  \Cref{lem:multiscale} proves this decision guarantee without
estimating $F_v$ numerically or requiring a fourth moment of $\R$.

\subsubsection{Potentials, supermartingales, and leaf verification}

Pruning a node immediately after one negative test would require every
test on a heavy path to be correct.  Amplifying all those tests would
cost an additional logarithmic factor in samples.  The traversal of
\Cref{sec:potential-traversal} instead maintains a positive node
potential $\phi(v)$, starting at
\[
  \phi_{\max}=\left(\frac{3\kappa\log N}{\gamma}\right)^{1/4}
\]
at the root, where $\gamma$ is the locator's failure probability.
A positive answer multiplies the potential by $2^{1/4}$, capped at
$\phi_{\max}$; a negative answer divides it by four.  Retain a node if its
potential is at least one.  This rule lets a heavy path survive occasional
negative answers, while the cap bounds the potential with which a branch
enters a low-energy region.

Tolerating local errors during a search has a classical precedent in
noisy binary search~\cite{ref:FRPU94}.  That work uses backtracking
under independent comparison errors.  Here the nodes at one depth
share samples, and the test has different completeness and soundness
promises, in terms of $F_v$ and $W_v$ respectively.

This use of prefix search is related to the group-testing and
$\ell_2/\ell_2$ identification procedures of Cheraghchi and
Nakos~\citeat{ref:CN20}{Sections~6 and~10}.  Their compressed-sensing
construction can assign independent Gaussian measurement weights and
use Gaussian anticoncentration to detect significant buckets.  Our
point queries impose the Fourier weights, so their statistical test
is unavailable here.  Our multiscale bucket test instead supplies the
conditional moment bounds for heavy-path survival and a potential drift
bound for the search size.

Price, Scarlett, and Tan~\citeat{ref:PST23}{Appendix~A} use
Fuss--Catalan tree counts in their analysis of prefix-tree algorithms
for sparsity-constrained nonadaptive group testing.

Let $\mathcal F_j$ record the query batches through depth $j$.
The residual is fixed, and the next depth's batch is independent of
$\mathcal F_j$.  On a fixed heavy path, let $A_j\in\{2^{1/4},1/4\}$
be the multiplier before capping.  The test guarantee implies
\[
  \E[A_j^{-4}\mid\mathcal F_{j-1}]<\tfrac34.
\]
For any fixed starting depth $a$, the inverse products
$Z_{a,j}=\prod_{i=a+1}^j A_i^{-4}$, with $Z_{a,a}=1$, form a
nonnegative supermartingale.  Ville's maximal inequality (\Cref{lem:ville})
extends Markov's
inequality to the running maximum of a nonnegative supermartingale:
when the process starts at one, the probability that it ever exceeds
$\lambda>0$ is at most $1/\lambda$.  Taking $\lambda=\phi_{\max}^4$
bounds the crossing probability by $\phi_{\max}^{-4}$.
A pruned heavy path must cross this threshold
on the interval after its last visit to the cap.  A union bound over
all starting depths and the at most $\kappa$ heavy paths gives failure
probability at most $\kappa(\log N)\phi_{\max}^{-4}=\gamma/3$.
This argument uses a separate process for each deterministic starting
depth, so it does not treat the last visit to the cap as a stopping time.

For the search-size bound, let $\mathcal V_j$ be the retained nodes at
depth $j$ and sum their potentials:
\[
  \Phi_j=\sum_{v\in\mathcal V_j}\phi(v).
\]
Every retained node contributes at least one.  At a low-energy child,
the conditional mean multiplier is less than $1/3$.  Since each retained
parent has at most two such children, their total expected contribution
is at most $2\Phi_j/3$.  There are at most $\kappa/\theta$ high-energy
nodes at the next depth, and each contributes at most $\phi_{\max}$.
Consequently,
\[
  \E[\Phi_{j+1}\mid\mathcal F_j]
  \leq\frac23\Phi_j+\frac{\kappa\phi_{\max}}\theta.
\]
Thus the potential contracts, apart from a bounded contribution at each
new depth.  We reserve that contribution for the remaining depths and
add the potential already spent on the search.  The resulting process
\[
  \Psi_j=\Phi_j+\frac13\sum_{i<j}\Phi_i
    +\frac{\kappa\phi_{\max}}\theta(\log N-j)
\]
is a nonnegative supermartingale.  It is at least one third of the
cumulative number of retained nodes through depth $j$.  Ville's
inequality therefore bounds the probability that the search ever exceeds
its prescribed size limit.  This gives the second use of the maximal
inequality: inverse products protect heavy paths, while accumulated
node potentials control the total search.

All tests at one depth share their samples and may be correlated.
Conditioning on whole depth batches preserves the fresh randomness
needed for the path estimates.  The search-size argument uses linearity
of conditional expectation, so it requires no independence between
siblings.  \Cref{sec:potential-traversal} gives both arguments in full.

Finally, retained leaves undergo independent verification.  A uniform
point $t$ gives an unbiased coefficient estimate
$\R(t)\conj{\chi_f(t)}$, with variance at most $M^2$.  Shared
median-of-means samples estimate every candidate to accuracy below
$T/4$.  Keeping estimates of magnitude at least $3T/4$ retains all
coefficients of magnitude at least $T$ and only coefficients of
magnitude greater than $T/2$.  One shared verification query set costs
$O(\kappa\log(\kappa (\log N)/\gamma))$ queries for the entire candidate
list, in addition to the $O(\kappa (\log N)\log(2\kappa))$ tree search queries.
This completes the locator of \Cref{prop:locator}.  Independent
verification is needed because a retained leaf need not be heavy.

\subsubsection{Thresholds, total search cost, and nonadaptive queries}

We now assemble the rounds as in \Cref{sec:recovery-algo}.
In round $i$, let $S_i$ be the current
support, $M_i=M_{S_i}$ its residual norm estimate, and $T_i$ the search
threshold.  The locator's cost is governed by
$\kappa_i=M_i^2/T_i^2$.  Outside $S_i$, the residual coefficients
equal the original coefficients, so a successful call finds all
as-yet-unfitted frequencies of magnitude at least $T_i$.  Soundness
keeps every newly added frequency inside $\Omega$ as long as
$T_i\geq\tau/2$, preserving the simultaneous least-squares guarantee.

Start with $T_1=M_\varnothing$.  After each search and refit, set
\[
  T_{i+1}=\max\{\tau/2,\min\{T_i/2,M_{i+1}\}\}.
\]
Stop after processing the floor $\tau/2$.  Before that final step,
the threshold at least halves, and it can drop further when the new
residual norm estimate is small.  Neither the number of rounds nor their
total cost is analyzed through the initial-to-final threshold ratio.

Instead, account separately for each unidentified frequency's contribution to
the residual norm estimate.  Work on the common least-squares event and fix an
arbitrary finite initial segment through a proposed round whose preceding
locator calls all succeeded.  For $f\in\Omega$, consider the last
round $i$ of this initial segment in which $f$ is still unidentified.  Then
$|\Xhat(f)|<2T_i$: either the threshold equals the current norm
bound, or completeness at the preceding threshold supplies this
inequality.  Earlier thresholds are geometrically larger, apart from
the possible final floor step.  The geometric-series argument in
\Cref{lem:charging} gives, over this initial segment,
\[
  \sum_{i:f\notin S_i}\frac{|\Xhat(f)|^2}{T_i^2}=O(1),
  \qquad \sum_iT_i^{-2}=O(\tau^{-2}).
\]
Summing \eqref{eqn:overview-certificate} now gives
\begin{equation}
\label{eqn:overview-charge}
  \sum_i\kappa_i
  \leq O\!\left(
       \sum_{f\in\Omega}\sum_{i:f\notin S_i}
         \frac{|\Xhat(f)|^2}{T_i^2}
       +s\tau^2\sum_iT_i^{-2}\right)
  =O(s).
\end{equation}
The bound is uniform over such finite initial segments
(\Cref{lem:charging}).  Since each accepted round has $\kappa_i\geq1$, a
successful run must terminate within $O(s)$ rounds.  This
is the source of independence from coefficient range: we sum the
actual round costs rather than bound every round using the largest
possible $\kappa_i$.

The same bound on the total cost parameter justifies the fixed query
allocation of \Cref{sec:tapes-def}.
Store the samples as ordered \emph{query sets}, one for each depth,
repetition, and resolution, together with the fitting and verification
query sets.  A round reads the next unused blocks at the resolutions it
needs.  A budget check prevents the total cost parameter from
exceeding $K=O(s)$ on any execution; the cost bound shows that this
check never interrupts a successful run.  \Cref{lem:capacity} proves that
the allocated queries suffice on every execution.  Although all queries
have already been made, each decoding decision uses only entries read so far,
including
their locations.  Unread blocks retain their fresh conditional
distribution.  \Cref{lem:simulation} proves the equivalence with fresh
sampling.  Taking the failure probability per call
$\gamma=\Theta(1/s)$ permits a union bound over all calls.

The bound on the total cost parameter also accounts for the verification
queries:
\[
  \sum_i O\!\left(\kappa_i\log(\kappa_i (\log N)/\gamma)\right)
  =O\!\left(s\log(s\log N)\right).
\]
The least-squares queries and the final coefficient estimation queries each use
$O(s\log s)$ samples.  Since $\log\log N\leq (\log N)$, these contributions
are absorbed by the $O(s\log N\log s)$ tree search samples.
The complete query set therefore uses $O(s\log s\log N)$
samples, as proved in \Cref{lem:Q-upper}, without multiplying the
node-sampling cost by the number of candidate leaves.

The running-time bound requires the additional accounting in
\Cref{prop:shared-runtime}.  A locator call at ratio $\kappa_i$ has a
search-tree size limit
$D_i=O(\kappa_i^{5/4}(\log N)^{5/4}\gamma^{-5/4})$.
One factor $\gamma^{-1}$ comes from the maximal first-moment bound on
the accumulated search; the remaining $\gamma^{-1/4}$ comes from the
potential cap.
Since $\sum_i\kappa_i=O(s)$ and $\gamma^{-1}=O(s)$,
$\sum_i\kappa_iD_i=O(s^{7/2}(\log N)^{5/4})$.
Each character evaluation takes $O(\log N)$ exact operations, and node
tests and verification process
$O(\kappa_i\log(s\log N))$ samples per tested node or leaf.
Residual values are computed once per sample occurrence and stored for reuse.
Least squares is updated by adding columns as $S$ grows: only $O(s)$
columns are ever added, so repeated fits reuse a Cholesky factor
instead of rebuilding the entire system.  Their total cost, bounded in
\eqref{eqn:column-cost}, is
$O(s^2(\log N)\log s+s^3\log s)$.
These costs fit within
\[
  O\!\left(s^{7/2}(\log N)^{9/4}\log(s\log N)\right).
\]
The cubic least-squares term shows that a sharper traversal estimate
alone would not give a near-quadratic bound for this implementation.
The support and work limits make this an operation bound on every
execution, including failures.  If the planned query count reaches
$N$, we instead query the full signal and use an FFT.  All operation
bounds use the exact-arithmetic model of \Cref{sec:model}.

\subsubsection{The supplied tail estimate and the exactly sparse case}

At the threshold floor, the identified support contains every
coefficient of magnitude at least $\tau/2$.  The remaining Fourier
energy and the squared least-squares residual norm bound are then $O(s\tau^2)$.
An independent query set of $O(s\log s)$ final coefficient estimation samples
estimates the
residual coefficients on the support to accuracy $\tau/4$.
Adding them to $C_S$ gives a
vector $\Xtilde^{\rm pre}$ with fewer than $17s$ nonzero entries and
$\norm{\Xhat-\Xtilde^{\rm pre}}_\infty\leq\tau/2$
(\Cref{lem:cleanup,thm:upper-tail}).  Thus the intermediate result
provides coordinatewise recovery, from which we obtain the
$\ell_2/\ell_2$ conclusion by hard thresholding, which keeps the
$k$ largest-magnitude coordinates.

Fix a constant $c\geq1$, write $\sigma=\sigma_k(\Xhat)_2$, and
suppose the supplied estimate satisfies $\sigma\leq B\leq c\sigma$.
For $B>0$, use the parameters of \eqref{eqn:wrapper-parameters}:
\[
  s=\ceil{25c^2k/\eps^2},\qquad
  \tau=\frac{\eps B}{5c\sqrt{k}}.
\]
These choices ensure $\sigma_s(\Xhat)_2^2\leq s\tau^2$ and make
the coordinate error small enough for the desired relative error.
If $s\geq N$, use the full FFT; otherwise apply the preceding
algorithm and return $\Xtilde=\HT_k(\Xtilde^{\rm pre})$.  The stability
estimate
of \Cref{lem:hard-thresholding} gives
\[
  \norm{\Xhat-\Xtilde}_2^2
  \leq\sigma^2+4\sqrt{k}\,\sigma\tau+5k\tau^2
  \leq(1+\eps)\sigma^2.
\]
The estimate $B$ therefore determines both the tail-norm bound used in the
residual norm estimate and the stopping accuracy.  Substituting
$s=O(k/\eps^2)$ into the preceding bounds yields
\Cref{thm:main-dyadic-intro}.

A constant-factor estimate of the tail cannot generally be obtained
first from sublinear point queries.  For example, a unit spike at a
uniformly random time-domain location has $|\Xhat(f)|=1/N$ for every
$f$, and hence a nonzero $k$-tail for $k<N$.  With $m$ nonadaptive
queries, its location is hit with probability at most $m/N$; until a
hit, the answers agree with the zero signal, whose tail is zero.
Thus $o(N)$ queries cannot give a multiplicative tail estimate for
every signal with a fixed success probability greater than $1/2$.
This explains the need for the separate constructions without $B$;
it does not rule out other tail-free recovery guarantees.

When $B=0$, the promise implies exact $k$-sparsity.  Set $s=k$ and
$\tau=0$, so that $M_S=2\norm{z_S}_2$ and there is no positive
threshold floor.  Test $M_S=0$ initially and after each refit, before
forming a ratio with the next threshold.  On the least-squares event, if all
preceding locator calls succeeded,
this equality implies that
the entire residual vanishes.  The per-frequency cost contributions remain
summable, while the tail term in \eqref{eqn:overview-charge}
disappears.  The total cost parameter is still $O(k)$, forcing finite
termination
at an exact fit without a lower bound on nonzero coefficients.
No final coefficient estimation step is needed in this case.
This is the zero-scale case of \Cref{cor:discovery,thm:upper-tail}.

\Cref{alg:main-recovery} summarizes the algorithm for
\Cref{thm:main-dyadic-intro}.  The routine $\operatorname{Fit}$ is the
least-squares procedure of \Cref{lem:ls}; it returns $C_S$ and
$M=2\norm{z_S}_2+3b_{\rm ls}\sqrt{s}\tau$, where
$b_{\rm ls}=\sqrt{255}/2$.  The locator of \Cref{prop:locator} accesses
the current residual through the stored values
$\R(t)=\X(t)-\sum_{f\in S}C_S(f)\chi_f(t)$.
An abort returns the zero vector.  All queries precede decoding, and
unread samples, including their locations, do not influence its decisions.

\begin{algorithm}[t]
\caption{Sparse Fourier recovery from a supplied tail estimate}
\label{alg:main-recovery}
\begin{algorithmic}[1]
\Require Point-query access to $\X$ and $k,\eps,B,c$ as in
  \Cref{thm:main-dyadic-intro}.
\State If $B=0$, set $(s,\tau)\gets(k,0)$; otherwise set
  $(s,\tau)\gets(\ceil{25c^2k/\eps^2},\eps B/(5c\sqrt{k}))$.
\State $K\gets31000s$, $\gamma\gets1/(30K)$, $T_{\min}\gets\tau/2$.
\State Set $Q_{\rm plan}\gets N$ if $s\geq N$; otherwise compute
  $Q_{\rm plan}\gets Q_{\rm total}$ from \eqref{eqn:Q-upper}.
\If{$Q_{\rm plan}\geq N$}
  \State Query all of $G$ and compute $\Xhat$ by FFT.
  \State Set $\Xtilde^{\rm pre}(f)\gets\Xhat(f)$ if $s\geq N$ or
    $|\Xhat(f)|\geq\tau/2$, and zero otherwise.
  \State \Return $\Xtilde\gets\HT_k(\Xtilde^{\rm pre})$.
\EndIf
\State Generate and query all four collections of \Cref{sec:tapes-def}.
\State Read the least-squares queries $U$; set $S\gets\varnothing$,
  $\kappa_{\rm tot}\gets0$.
\State $(C_S,M)\gets\operatorname{Fit}(U,S;\sqrt{s}\tau)$.
\State If $M=0$, return $\Xtilde\gets\HT_k(C_S)$; otherwise set $T\gets M$.
\While{true}
  \State Abort if $T\leq0$; set $\kappa\gets M^2/T^2$.
  \State Abort if $\kappa<1$, $T<T_{\min}$, or
    $\kappa_{\rm tot}+\kappa>K$; set
    $\kappa_{\rm tot}\gets\kappa_{\rm tot}+\kappa$.
  \State Run the locator on $\R$ with $(M,T,\gamma)$, reading its next
    prescribed batch; let $S_{\rm new}$ be its output.
  \State Abort if the locator aborts or $|S\cup S_{\rm new}|\geq17s$.
  \State $S\gets S\cup S_{\rm new}$;
    $(C_S,M)\gets\operatorname{Fit}(U,S;\sqrt{s}\tau)$.
  \State Abort if the fit is singular;
    if $M=0$, return $\Xtilde\gets\HT_k(C_S)$.
  \State If $T=T_{\min}$, leave the loop.
  \State $T\gets\max\{T_{\min},\min\{T/2,M\}\}$.
\EndWhile
\State Estimate $\Xtilde^{\rm pre}$ on $S$ as in \Cref{sec:cleanup}, using
  the independent final coefficient estimation queries;
  set $\Xtilde^{\rm pre}(f)=0$ outside $S$.
\State \Return $\Xtilde\gets\HT_k(\Xtilde^{\rm pre})$.
\end{algorithmic}
\end{algorithm}

The test $T=T_{\min}$ follows the search and refit, so the final positive
threshold is processed before termination.  When $B=0$, the algorithm
instead stops when $M=0$.  The variants without a supplied tail estimate
modify the stopping rule and final step, as described next.

\subsubsection{Recovery without a supplied tail estimate}

There are two ways to remove $B$, depending on the desired guarantee.
To retain the factor $1+\eps$, \Cref{sec:relative-additive} allows
an additive error $\delta\norm{\Xhat}_2^2$.  Using the pilot estimate of
\Cref{lem:additive-pilot}, preliminary scale-estimation queries either
give an interval containing a suitable tail-norm bound or the zero vector
already satisfies the
required error guarantee.  We run the coordinatewise recovery algorithm using
the corresponding
accuracy thresholds for a geometric list of candidate tail bounds, include
zero among the candidates, and select a candidate using independent
samples.  The recovery query sets have lengths independent of the numerical
scale, so they can be reused for different guesses, restarting each
decoding run from the beginning of the stored query answers.  Conditional on
the pilot, one suitable
scale is fixed independently of those query sets; the proof requires
success at that scale, without requiring simultaneous success at all
guesses.  The pilot therefore changes decoding thresholds but never
query locations.  \Cref{thm:relative-additive} obtains, for
$2\leq k<N$, $0<\eps\leq1/2$, and $0<\delta\leq1/4$,
\[
  \norm{\Xhat-\Xtilde}_2^2
  \leq(1+\eps)\sigma_k(\Xhat)_2^2
       +\delta\norm{\Xhat}_2^2.
\]
For fixed $\delta$, the sample bound has the same asymptotic order
as in the theorem with a supplied tail-norm estimate; the full dependence on
$\delta$
and the additional decoding work are stated there.  This pilot is
not a universal sublinear estimator of arbitrary signal energy, and
the additive term remains when the spectrum is exactly sparse.

For a constant-factor approximation, we can remove $B$ with no
additive term.  In \Cref{sec:tail-free}, we use the empirical residual norm
estimate
$M_S=2\norm{z_S}_2$, omit the threshold floor, and save every
least-squares fit, including the initial zero fit.  This empirical
quantity need not always bound the true residual.  On a common
high-probability least-squares event, however, the next residual norm
estimate is an upper bound as long as all saved fits have error larger
than a suitable constant times the unknown tail norm and all preceding
locator calls succeeded.  The same cost accounting argument shows that a good
fit must be saved before a support or work limit can stop this part
of the run.  The algorithm need not recognize when it reaches that
fit.  It produces $O(k)$ saved fits, and keeping the largest $k$
coefficients of each preserves a constant-factor good candidate
(\Cref{lem:tail-free-candidates}).

The selection rule of \Cref{lem:tail-free-selection} addresses the same
second-moment issue as bucket testing.  Direct empirical estimates of
squared residual norms
can have uncontrolled variance.  Instead, compare differences of
squared errors.  For two $k$-sparse candidates $\Xtilde^{(i)},\Xtilde^{(j)}$,
let $\X^{(i)}$
and $D_{ij}$ be the time signals with spectra $\Xtilde^{(i)}$ and
$\Xtilde^{(j)}-\Xtilde^{(i)}$,
respectively.  Parseval gives
\[
\begin{split}
  \norm{\Xhat-\Xtilde^{(j)}}_2^2-\norm{\Xhat-\Xtilde^{(i)}}_2^2
  ={}&\norm{\Xtilde^{(j)}-\Xtilde^{(i)}}_2^2\\
     &-2\mathrm{Re}\,\E_t[(\X(t)-\X^{(i)}(t))\conj{D_{ij}(t)}].
\end{split}
\]
The common signal-energy term cancels.  Moreover, $\Xtilde^{(j)}-\Xtilde^{(i)}$
is
$2k$-sparse, so
$|D_{ij}(t)|\leq\sqrt{2k}\norm{\Xtilde^{(j)}-\Xtilde^{(i)}}_2$.
By \Cref{lem:sparse-correlation}, this bounds the comparison variance
using only the second moment of the residual; see
\eqref{eqn:tail-free-variance}.

To describe the selection rule, discard duplicate candidates and
return immediately if only one remains.  Let $n_{\rm c}\geq2$ be the number
remaining and write $e_i=\norm{\Xhat-\Xtilde^{(i)}}_2$ for analysis.
With $O(k\log n_{\rm c})$ shared samples, median-of-means estimates
$\widehat a_{ij}$ of $a_{ij}=e_j^2-e_i^2$ satisfy, with high
probability,
\[
  |\widehat a_{ij}-a_{ij}|
  \leq\tfrac14\norm{\Xtilde^{(j)}-\Xtilde^{(i)}}_2e_i
  \qquad\text{for all }i\neq j.
\]
A negative value of $\widehat a_{ij}$ estimates a smaller squared error for
$\Xtilde^{(j)}$ than for $\Xtilde^{(i)}$.
By \eqref{eqn:tail-free-pair-error}, the estimation error in $\widehat a_{ij}$
is at
most a multiple of $\norm{\Xtilde^{(j)}-\Xtilde^{(i)}}_2$.
We therefore normalize by this distance and choose a candidate minimizing
\[
  J_i=\max_{j\neq i}
       \frac{-\widehat a_{ij}}{\norm{\Xtilde^{(j)}-\Xtilde^{(i)}}_2}.
\]
This comparison score is the largest estimated improvement offered by a
competitor, normalized by the distance between the candidates.
If $i_*$ minimizes $e_i$ and $e_*=e_{i_*}$, then
$J_{i_*}\leq e_*/4$, because no competitor has smaller true error.
For any other candidate, comparison with $i_*$ and
$\norm{\Xtilde^{(i)}-\Xtilde^{(i_*)}}_2\leq e_i+e_*$ give
$J_i\geq3e_i/4-e_*$.  The minimizing comparison score therefore yields
$e_i\leq5e_*/3$, without using the unknown errors in the algorithm
(\Cref{lem:tail-free-selection}).

For the relative-additive reduction, use
$O(k\eta^{-2}\log n_{\rm c})$ samples to replace $1/4$ in the comparison
error by any $0<\eta<1$.  The candidate selection procedure of
\Cref{lem:additive-selection} minimizes the same score $J_i$ defined
above, with the unclipped numerator $-\widehat a_{ij}$.
The best candidate's score is at most $\eta e_*$,
whereas another candidate's score is at least
$(1-\eta)e_i-e_*$.  Thus the selected error is at most
$(1+\eta)e_* /(1-\eta)$, giving the required factor arbitrarily
close to one.

Combining the candidate construction and selection proves
\Cref{thm:tail-free}: no signal norm or tail estimate is supplied,
the squared error is at most $C\sigma_k(\Xhat)_2^2$ for an absolute
constant $C$, and the bounds are $O(k\log k\log N)$ queries and
$O(k^{7/2}(\log N)^{9/4}\log(k\log N))$ exact operations, with
the usual full-FFT case.  When the tail is zero, the same
guarantee gives exact recovery even when exact sparsity is not known
in advance.

\subsubsection{Other fixed bases}

All results, including those without a supplied tail estimate, extend
to $G=(\Z/b^a\Z)^d$ for every
fixed integer $b\geq2$, including composite $b$
(\Cref{sec:fixed-base}).  The subgroup chain gives a $b$-ary tree,
and character orthogonality gives the same projection identity.
We decrease the potential multiplier following a negative test and amplify node
tests by
constants depending on $b$, so that false-path decay beats the
$b$-fold branching, giving the locator of \Cref{prop:fixed-base-locator}.
The statistical resolutions can still be powers of two.
The least-squares and cost accounting arguments then apply with the same
dependence on $k$, $\eps$, and $\log N$; only the hidden constants
may additionally depend on the fixed base $b$.

\subsection{Organization}

The rest of the article is organized as follows.
In \Cref{sec:prelim}, we fix notation and specify the computational model.
\Cref{sec:projections} develops the subgroup projections and average-magnitude
statistic used in the search, and \Cref{sec:locator} constructs the
multiscale bucket test and error-tolerant locator.
In \Cref{sec:least-squares}, we use least squares to bound the residual norm.
We combine these ingredients into an adaptive
coordinatewise recovery algorithm in \Cref{sec:schedule}.
In \Cref{sec:tapes}, we implement its sampling nonadaptively using fixed
query sets, and we bound the sample and operation costs of a class of
procedures that also contains the variants below.
In \Cref{sec:resources}, we complete these bounds for the recovery
algorithm and prove the main theorem with a $1+\eps$ guarantee from a
supplied tail estimate.
\Cref{sec:tail-free,sec:relative-additive} remove the supplied tail
estimate, respectively with a constant approximation factor and with
a relative additive error.
Finally, \Cref{sec:fixed-base} provides an overview
of how the results extend to nonbinary
bases (i.e., when $G=(\Z/b^a\Z)^d$, where $b\geq2$ is any fixed integer and
$a,d\geq1$).

\section{Preliminaries}
\label{sec:prelim}

\subsection{Notation}

Throughout, $a,d\geq1$ are integers, $G=(\Z/2^a\Z)^d$, $N=|G|=2^{ad}$.
We write $\iu=\sqrt{-1}$ for the imaginary unit.
For $f,t\in G$, write
$\langle f,t\rangle=\sum_{i=1}^d f_it_i\in\Z/2^a\Z$, so that the
character $\chi_f(t)=\exp(2\pi\iu\langle f,t\rangle/2^a)$ is well defined.

For a positive integer $n$, write $[n]=\{1,\ldots,n\}$.
All logarithms are to base two.
For an integer $e\geq0$, write $\mathcal{W}_{2^e}=\{1,2,4,\ldots,2^e\}$ for the
set of \emph{power-of-two sampling resolutions} up to $2^e$.

For $z\in\C$, write $\conj z$, $\mathrm{Re}(z)$, and $\mathrm{Im}(z)$ for its
complex conjugate, real part, and imaginary part, respectively.
For $\Xhat\in\C^G$, write $\supp(\Xhat)=\{f\in G:\Xhat(f)\neq0\}$,
$\norm{\Xhat}_2$ for the Euclidean norm, and $\norm{\Xhat}_\infty=\max_{f\in
G}|\Xhat(f)|$.

For $S\subseteq G$, write $\C^S$ for the space of vectors whose coordinates are
indexed by the frequencies in $S$.
We regard every $c\in\C^S$ also as a vector in $\C^G$ by setting $c(f)=0$ for
$f\notin S$.
For $\Xhat\in\C^G$, we write $\Xhat_S\in\C^S$ for the restriction of $\Xhat$ to
$S$;
by the convention above, $\Xhat_S$ is also the vector that agrees with $\Xhat$
on $S$ and vanishes outside $S$, so that $\Xhat=\Xhat_S+\Xhat_{G\setminus S}$.

Fix an order on $G$.
For $\Xhat\in\C^G$ and $1\leq k<N$, let $\HT_k(\Xhat)$ be the vector obtained
from $\Xhat$ by keeping $k$ coordinates of largest magnitude, with ties broken
by this order, and setting the other coordinates to zero.
We call these $k$ coordinates the \emph{selected coordinate set} of
$\HT_k(\Xhat)$; it has exactly $k$ elements even if $\Xhat$ has fewer than $k$
nonzero coordinates.
By definition, $\norm{\Xhat-\HT_k(\Xhat)}_2=\sigma_k(\Xhat)_2$.

\subsection{Fourier analysis on \texorpdfstring{$G$}{G}}
\label{sec:fourier}

The bilinearity of the inner product implies
\[
  \chi_f(t+t')=\chi_f(t)\chi_f(t'),\qquad
  \chi_f(t)\chi_g(t)=\chi_{f+g}(t).
\]
The characters are orthonormal under uniform expectation: for $f,g\in G$,
\begin{equation}
\label{eqn:char-orthogonality}
  \E_t\bigl[\chi_f(t)\conj{\chi_g(t)}\bigr]=
  \begin{cases}1,&f=g,\\0,&f\neq g.\end{cases}
\end{equation}

Let $\X\colon G\to\C$ be a signal with spectrum $\Xhat$, so that
\eqref{eqn:fourier-normalization} holds.
By \eqref{eqn:char-orthogonality}, for every $f \in G$ and $S \subseteq G$,
\begin{equation}
\label{eqn:parseval}
  \Xhat(f)=\E_t\bigl[\X(t)\conj{\chi_f(t)}\bigr],\qquad
  \E_t\Bigl[\Bigl|\sum_{f\in
  S}\Xhat(f)\chi_f(t)\Bigr|^2\Bigr]=\norm{\Xhat_S}_2^2 .
\end{equation}
In particular, for $t$ uniform on $G$, the single sample
$\X(t)\conj{\chi_f(t)}$ has mean $\Xhat(f)$.
Furthermore, since $|\chi_f(t)|=1$, taking $S=G$ shows that the variance of
this single sample is
\begin{equation}
\label{eqn:one-sample}
  \E_t\bigl[|\X(t)\conj{\chi_f(t)}-\Xhat(f)|^2\bigr] =
  \norm{\Xhat}_2^2-|\Xhat(f)|^2,
\end{equation}
which is bounded by the squared $\ell_2$ norm of the whole spectrum $\Xhat$.
These identities hold for every signal; besides the input $\X$, we apply them
to residual signals $\R$ with spectrum $\Rhat$.

We also need to estimate correlations with a signal whose spectrum is
sparse.  Sparsity gives a pointwise bound on that signal, so a
second-moment bound on the other signal suffices.

\begin{lemma}[Sparse correlation]
\label{lem:sparse-correlation}
Let $U,V:G\to\C$ have spectra $\widehat U,\widehat V$, and suppose
$|\supp(\widehat U)|\leq q$ for an integer $q\geq1$.  For $t$ uniform
on $G$, the random variable $\xi(t)=V(t)\conj{U(t)}$ satisfies
\[
  \E_t[\xi(t)]
  =\sum_{f\in G}\widehat V(f)\conj{\widehat U(f)},
  \qquad
  \E_t[|\xi(t)|^2]
  \leq q\norm{\widehat U}_2^2\norm{\widehat V}_2^2.
\]
\end{lemma}
\begin{proof}
Expanding $U$ and $V$ in the character basis gives the first identity by
\eqref{eqn:char-orthogonality}.  For the second, Cauchy--Schwarz gives
\[
  |U(t)|
  \leq\sum_{f\in\supp(\widehat U)}|\widehat U(f)|
  \leq\sqrt q\norm{\widehat U}_2
\]
at every $t$.  Therefore Parseval's identity gives
\[
  \E_t[|V(t)\conj{U(t)}|^2]
  \leq q\norm{\widehat U}_2^2\E_t[|V(t)|^2]
  =q\norm{\widehat U}_2^2\norm{\widehat V}_2^2. \qedhere
\]
\end{proof}

\subsection{Computational model}
\label{sec:model}

The input $\X$ is fixed before the algorithm draws its randomness, and the
algorithm accesses $\X$ only through point queries.
Throughout, a \emph{query set} is a multiset of query locations with fixed
indices.  Repeated locations are counted once per occurrence, and the
decoder accesses these indexed samples in the specified order.  Thus a
query set need not consist of distinct locations.

We work in the exact arithmetic model.  Arithmetic on individual
coordinates in $\Z/2^a\Z$, exact complex arithmetic, exact real
comparisons, and square roots of nonnegative reals have unit cost.
A group element is represented by its $d$ coordinates.  Addition or
equality testing of two group elements takes $O(d)$ coordinate
operations.
Each independent fair random bit also has unit cost.  A uniform point
of $G$ is assembled from $ad=\log N$ such bits.  A uniform point of
the subgroup $K_j$ defined in \Cref{sec:projections} uses at most $ad$
bits, by fixing the appropriate coordinates or binary digits to zero.
These constructions take $O(\log N)$ operations on every execution.
\Cref{sec:fixed-base} specifies the corresponding random-digit primitive
for the extension to other fixed bases.

A primitive $2^a$-th root of unity is supplied.  Computing
$\langle f,t\rangle$ modulo $2^a$ takes $O(d)$ coordinate operations;
repeated squaring then evaluates $\chi_f(t)$ in $O(a)$ complex
operations.  Thus character evaluation costs $O(d+a)=O(\log N)$.

Rounded logarithms, powers of two, and ceilings of nonnegative reals, which
appear in the parameters of the algorithm, are computed by repeated doubling
and binary search in $O(\log(2x))$ operations for an argument $x\geq1$.

The independence from coefficient range in our bounds concerns samples,
rounds, and exact operations.  We do not prove a bit-complexity or
finite-precision analogue.  On the good event in \Cref{lem:ls}, the
Gram matrices used for fitting have eigenvalues between $1/2$ and $3/2$,
and hence condition number at most three.  This controls the
least-squares systems, but does not by itself control errors in all
residual computations and threshold decisions.  In particular, the
rules testing $B=0$, $M_S=0$, or exact singularity, and the exact-recovery
conclusion for a zero tail, are interpreted in this model.  A
finite-precision version would require explicit input and error
representations and a separate analysis of those decisions.

\subsection{Probability tools}

Several estimators in this paper take a coordinatewise median of independent
group means.  We analyze them with the following lemma.

\begin{lemma}[Median of means]\label{lem:median}
Let $r\geq1$ be odd.
\begin{enumerate}
\item \label{item:median-majority}
  Let $0\leq p\leq1$, let $\mathcal E_1,\dots,\mathcal E_r$ be independent
  events with
  $\Prob[\mathcal E_i]\leq p$ for every $i$, and let $N$ be the number of
  events that
  occur.  Then
  \[
    \Prob\Bigl[N\geq\frac{r+1}{2}\Bigr]
    \leq\binom{r}{(r+1)/2}p^{(r+1)/2}
    \leq\frac14(4p)^{(r+1)/2}.
  \]
\item \label{item:median-means}
  Let $x\in\C$, $\nu\geq0$, $\omega>0$, and let $v\geq1$ be an integer.
  Let $\xi_{i,b}$, for $i\in[r]$ and $b\in[v]$, be independent complex
  random variables with
  $\E[\xi_{i,b}]=x$ and $\E[|\xi_{i,b}-x|^2]\leq\nu$.  Let
  $\mu_i=\frac1v\sum_{b=1}^v\xi_{i,b}$ be the group means, and let $\lambda$ be
  their coordinatewise median, that is,
  $\mathrm{Re}(\lambda)=\operatorname{med}_i\mathrm{Re}(\mu_i)$ and
  $\mathrm{Im}(\lambda)=\operatorname{med}_i\mathrm{Im}(\mu_i)$.  Then, with
  $p=\min\{1,\nu/(v\omega^2)\}$,
  \[
    \Prob\bigl[|\lambda-x|\geq\sqrt2\,\omega\bigr]
    \leq\binom{r}{(r+1)/2}p^{(r+1)/2}
    \leq\frac14(4p)^{(r+1)/2}.
  \]
\end{enumerate}
\end{lemma}

\begin{proof}
For item~\ref{item:median-majority}, consider any $I \subseteq [r]$ such that
$|I| = (r + 1)/2$.
By independence of the events, we have
\[
  \Prob\left[\bigwedge_{i \in I} \mathcal E_i\right] = \prod_{i \in I}
  \Prob[\mathcal E_i] \leq p^{(r+1)/2}.
\]
Since there are $\binom{r}{(r+1)/2}$ such subsets $I$, a union bound gives
\[
  \Prob\Bigl[N \ge \frac{r+1}{2}\Bigr] \le \binom{r}{(r+1)/2} p^{(r+1)/2} \le
  \frac{1}{4}(4p)^{(r+1)/2}.
\]
The second inequality follows from $\binom{r}{(r+1)/2}\leq2^{r-1}$, since the
two equal middle coefficients $\binom{r}{(r-1)/2}$ and $\binom{r}{(r+1)/2}$ sum
to at most $2^r$.

For item~\ref{item:median-means}, $\xi_{i,b}-x$ are independent variables with
zero mean, so $\E[|\mu_i-x|^2]\leq\nu/v$.
By Chebyshev's inequality, the event $\mathcal E_i=\{|\mu_i-x|\geq\omega\}$ has
probability at most $p$.
$\mathcal E_1,\dots,\mathcal E_r$ are independent events because the groups are
disjoint.
If fewer than $(r+1)/2$ of the events occur, then more than half of the values
of $\mathrm{Re}(\mu_i)$ lie in $(\mathrm{Re}(x)-\omega,\mathrm{Re}(x)+\omega)$.
Thus $\operatorname{med}_i\mathrm{Re}(\mu_i) \in
(\mathrm{Re}(x)-\omega,\mathrm{Re}(x)+\omega)$.
A similar argument holds for the imaginary parts, so
$|\lambda-x|<\sqrt2\,\omega$.
Therefore $\Prob[|\lambda-x|\geq\sqrt2\,\omega]\leq\Prob[N\geq(r+1)/2]$, and
both bounds follow from item~\ref{item:median-majority}.
\end{proof}

\Cref{lem:median} holds on any probability space; we will apply it under
conditional probabilities.

For the tree search, we use Ville's maximal inequality, which extends
Markov's inequality to the running maximum of a nonnegative
supermartingale.

\begin{lemma}[Ville's maximal inequality]\label{lem:ville}
Let $(Z_j)_{j=0}^r$ be nonnegative integrable random variables adapted to
a filtration $(\mathcal F_j)_{j=0}^r$ and satisfying
$\E[Z_{j+1}\mid\mathcal F_j]\leq Z_j$ for $0\leq j<r$.
Then, for every $\lambda>0$,
\[
  \Prob\left[\max_{0\leq j\leq r}Z_j\geq\lambda\right]
  \leq\frac{\E[Z_0]}{\lambda}.
\]
In particular, if $Z_0=1$, the bound is $1/\lambda$.
\end{lemma}

\begin{proof}
Let $\tau$ be the first index at which $Z_j\geq\lambda$, with
$\tau=r$ if there is no such index.  Bounded optional stopping gives
$\E[Z_\tau]\leq\E[Z_0]$: apply the conditional supermartingale inequality
at each step to the stopped process $Z_{\min\{j,\tau\}}$ and take
expectations.  On the event of a crossing, $Z_\tau\geq\lambda$;
on its complement, $Z_\tau\geq0$.  Hence
\[
  \lambda\,\Prob\left[\max_{0\leq j\leq r}Z_j\geq\lambda\right]
  \leq\E[Z_\tau]\leq\E[Z_0].
\]
\end{proof}

We use the following standard concentration bound in the analysis;
the cited source gives its proof.

\begin{proposition}[Matrix Chernoff bound
\citeat{ref:Tro12}{Corollary~5.2}]\label{prop:Chernoff}
  Let $\Psi_1, \ldots, \Psi_n$ be independent $q \times q$ random Hermitian
  matrices which almost surely satisfy $0 \preceq \Psi_b \preceq \Lambda I_q$
  for constant $\Lambda > 0$.
  Take
  \[
    \Psi = \sum_{b = 1}^n \Psi_b, \qquad \mu_{\rm min} = \lambda_{\rm
    min}(\E[\Psi]), \qquad \mu_{\rm max} = \lambda_{\rm max}(\E[\Psi]).
  \]
  For $0 < \zeta < 1$ we have
  \[
    \Prob[\lambda_{\rm min}(\Psi) \le (1 - \zeta)\mu_{\rm min}] \le q
    \left(\frac{e^{-\zeta}}{(1 - \zeta)^{1 - \zeta}}\right)^{\mu_{\rm min} /
    \Lambda},
  \]
  and for $\zeta > 0$ we have
  \[
    \Prob[\lambda_{\rm max}(\Psi) \ge (1 + \zeta)\mu_{\rm max}] \le q
    \left(\frac{e^{\zeta}}{(1 + \zeta)^{1 + \zeta}}\right)^{\mu_{\rm max} /
    \Lambda}.
  \]
\end{proposition}

\section{Subgroup projections and the average-magnitude statistic}
\label{sec:projections}

We define the frequency search tree and the statistics used to test its
nodes in \Cref{sec:locator}.

We locate candidate frequencies by searching a binary tree of depth
$ad=\log N$, rooted at $G$, with leaves indexed by single frequencies.
Specifically, for integer $0\leq j=\ell a+c<ad$, where $0\le c<a$, define
\[
  H_j=\{0\}^\ell\times(2^c\Z/2^a\Z)\times(\Z/2^a\Z)^{d-\ell-1},
\]
and let $H_{ad}=\{0\}^d$ and $H_0=G$.
The $2^{j}$ nodes at depth $j$ are all the cosets of $H_j$.
For each $j<ad$, choose $u_j\in H_j\setminus H_{j+1}$.
The two children of a node $v+H_j$ are $v+H_{j+1}$ and $v+u_j+H_{j+1}$, which
partition the parent coset into two equal parts.

We use the following statistics to decide which nodes to explore.
For a residual signal $\R$ with spectrum $\Rhat$, define
\[
  h_{v, j, \R}(t)=\sum_{f\in v+H_j}\Rhat(f)\chi_f(t), \qquad
  F_{v, j, \R}=\E_t|h_{v, j, \R}(t)|,\qquad W_{v, j, \R}=\sum_{f\in
  v+H_j}|\Rhat(f)|^2.
\]
When $\R$ is clear from context, or when a statement holds for an arbitrary
residual, we omit it from the subscripts.
The function $h_{v, j, \R}$ is the Fourier projection of $\R$ onto the
frequencies in $v+H_j$.
Thus $F_{v, j, \R}$ is the average magnitude of this projected signal, while
$W_{v, j, \R}$ is its energy.

The subgroup structure makes the tree searchable with point queries.
In particular, the following proposition shows that the projection of $\R$ onto
$v+H_j$ at time $t$ is a character-weighted average of $\R$ over a coset of the
annihilator $K_j$.

\begin{proposition}
\label{prop:orthogonality}
  For any $v \in G, 0\le j\le (\log N)$ and $\R$,
  \begin{equation}
\label{eqn:coset-projection}
    h_{v, j, \R}(t) = \E_{z \in K_j} \R(t + z)\conj{\chi_v(z)}.
  \end{equation}
  Here $K_j$ is the annihilator of $H_j$ under the character pairing.
  For $j=\ell a+c<ad$, it is given by
  \[
    K_j
    :=\{z\in G:\chi_h(z)=1 \text{ for every }h\in H_j\}
    = (\Z/2^a\Z)^\ell \times (2^{a-c}\Z/2^a\Z) \times \{0\}^{d-\ell-1}.
  \]
  At the final depth, $K_{ad}=G$.
\end{proposition}
\begin{proof}
  Expanding $\R$ in the character basis and using
  $\chi_f(t+z)\overline{\chi_v(z)} = \chi_f(t)\chi_{f-v}(z)$, we obtain
  \begin{equation}
\label{eqn:ortho-proof}
    \E_{z\in K_j}
      \bigl[\R(t+z)\overline{\chi_v(z)}\bigr] = \sum_{f\in G}\Rhat(f)\chi_f(t)
      \E_{z\in K_j}\chi_{f-v}(z).
  \end{equation}

  The evaluation of the expectation term splits into two cases.
  If $f - v \in H_j$, then $\E_{z \in K_j}\chi_{f - v}(z) = 1$ by the
  definition of $K_j$.
  If $f - v \notin H_j$, we claim that there exists $z_0 \in K_j$ such that
  $\chi_{f - v}(z_0) \neq 1$.
  Write $g := f - v = (g_1, \ldots, g_d) \in (\Z/2^a\Z)^d$ and $j = \ell a +
  c$.
  If $j=ad$, then $K_j=G$, and any nonzero coordinate of $g$
  gives such a $z_0$ by choosing the corresponding coordinate vector.
  Otherwise $j<ad$, and since $g \notin H_j$, one of the following holds:
  \begin{itemize}
    \item $\exists i \le \ell : g_i \neq 0$.
    Then let $z_0 = (0, \ldots, 0, 1, 0, \ldots, 0)$ where the only $1$ is in
    the $i$-th coordinate.
    This gives $\chi_g(z_0) = \exp(2\pi\iu\langle g, z_0\rangle/2^a) =
    \exp(2\pi\iu g_i/2^a) \neq 1$, since $g_i \neq 0$ in $\Z/2^a\Z$;
    \item $g_{\ell + 1} \notin 2^c\Z$.
    Write $g_{\ell + 1} = p2^c + q$, where $p \in \Z/2^a\Z, q \in \Z/2^c\Z, q
    \neq 0$.
    Set $z_0 = (0, \ldots, 0, 2^{a - c}, 0, \ldots, 0)$ where the only nonzero
    coordinate is the $(\ell + 1)$-th.
    This gives $\chi_g(z_0) = \exp(2\pi\iu\, q2^{a-c}/2^a) = \exp(2\pi\iu\,
    q/2^c) \neq 1$, since $0 < q < 2^c$.
  \end{itemize}
  For this choice of $z_0$, the translation $z \mapsto z + z_0$ is a bijection
  on $K_j$.
  This gives
  \[
    \E_{z \in K_j}\chi_{f - v}(z) = \E_{z \in K_j}\chi_{f - v}(z + z_0) =
    \chi_{f - v}(z_0)\E_{z \in K_j}\chi_{f - v}(z)
  \]
  and therefore $\E_{z \in K_j}\chi_{f - v}(z) = 0$ since $\chi_{f - v}(z_0)
  \neq 1$.

  Thus only the frequencies in $v+H_j$ contribute to the summation in
  \Cref{eqn:ortho-proof}, yielding
  \[
    \E_{z\in K_j}
      \bigl[\R(t+z)\overline{\chi_v(z)}\bigr]
    =\sum_{f\in v+H_j}\Rhat(f)\chi_f(t)
    =h_{v, j, \R}(t). \qedhere
  \]
\end{proof}

The next proposition relates the average magnitude and energy of a
projection to its Fourier coefficients.
\begin{proposition}
\label{prop:statistics}
  The following holds for any $v, j, \R$:
  \begin{enumerate}
    \item $\E_t[|h_{v, j, \R}(t)|^2]=W_{v, j, \R}$; \label{prop:statistics-1}
    \item For any $f\in v+H_j$, we have $|\Rhat(f)|\le F_{v, j,
    \R}\le\sqrt{W_{v, j, \R}}$. \label{prop:statistics-2}
  \end{enumerate}
  In particular, a node $v$ at depth $j$ containing a coefficient of magnitude
  at least $T$ has $F_{v, j, \R}\ge T$.
\end{proposition}
\begin{proof}
  The equality follows from \eqref{eqn:parseval} by definition.

  Taking the Fourier coefficient of $h_{v, j, \R}$ gives
  \[
    \E_t[h_{v, j, \R}(t)\conj{\chi_{f}(t)}] = \sum_{g \in v + H_j}
    \Rhat(g)\E_t[\chi_g(t)\conj{\chi_f(t)}] = \Rhat(f),
  \]
  so
  \[
    |\Rhat(f)| = \left|\E_t[h_{v, j, \R}(t)\conj{\chi_{f}(t)}]\right| \le
    \E_t[|h_{v, j, \R}(t)\conj{\chi_{f}(t)}|] = \E_t[|h_{v, j, \R}(t)|] = F_{v,
    j, \R}.
  \]
  This establishes the first inequality.  The second inequality follows from
  Cauchy--Schwarz and \Cref{prop:statistics-1}.
\end{proof}

\Cref{prop:statistics} gives the two statistics complementary roles in the
coset search.
A heavy coefficient forces $F_v$ to be large, which underlies the
completeness of the locator.
Meanwhile, the energy $W_v$ controls the ``width'' of the retained search space
at each depth,
ensuring that no more than $O(\norm{\Rhat}_2^2/T^2)$ nodes at each depth have
$F_v = \Omega(T)$ by \Cref{prop:statistics}.
We use this bound in the traversal analysis.

The test is built on $F_v$ rather than $W_v$ for a statistical reason.
Classical coset-search algorithms, such as those of
Kushilevitz--Mansour~\cite{ref:KM93} and
Goldreich--Levin, estimate the energy $W_v$ directly, from products of two
point
values.
They are designed for bounded functions, for which such estimates are well
behaved.
For a general complex signal, however, the variance of a product of two point
values is governed by fourth moments of $\R$, and $\E_t|\R(t)|^4$ can exceed
$\norm{\Rhat}_2^4$ by a factor as large as $N$.
This dependence is unacceptable since we require the sample complexity to be
polynomial
in $\log N$.

By contrast, the test of \Cref{sec:locator} estimates $h_v(t)$ directly, from
samples whose second moment is $\norm{\Rhat}_2^2$.
The next section shows that this second-moment bound suffices for a test with
the required sample count.

\section{Multiscale bucket test and error-tolerant locator}
\label{sec:locator}

This section turns the statistics of \Cref{sec:projections} into the
locator of \Cref{prop:locator}.  Given a residual with
$\norm{\Rhat}_2\leq M$ and a threshold $T$, the locator returns, with
probability at least $1-\gamma$, every frequency with $|\Rhat(f)|\geq T$
and only frequencies with $|\Rhat(f)|>T/2$, where $\kappa=M^2/T^2$
governs its cost.
We first construct a node test (\Cref{lem:multiscale}) that separates
$F_v\geq T$ from $W_v\leq\theta T^2$ with constant error probability.
We then use it in a breadth-first traversal governed by a node potential,
and verify the retained leaves separately to remove false positives.

\subsection{Multiscale bucket test with second moment assumption}

We first state the test for a function $V$ of two independent random
inputs $U$ and $Z$.  In the locator, we take $U=t$ uniform on $G$,
$Z=z$ uniform on $K_j$, and $V_v(t,z)=\R(t+z)\conj{\chi_v(z)}$.
Then each pair $(t,z)$ costs one point query at $t+z$,
and the quantities $h$, $F$, and $W$ of \Cref{lem:multiscale} become $h_v$,
$F_v$, and $W_v$ by \Cref{prop:orthogonality,prop:statistics}.
\begin{lemma}[Decision test]\label{lem:multiscale}
  Let $\mathcal{U}$ and $\mathcal{Z}$ be finite sets with probability
  distributions $\mathcal{D}_U$ and $\mathcal{D}_Z$, and let
  $V:\mathcal{U}\times\mathcal{Z}\to\C$ be a function.
  Let $U\sim\mathcal{D}_U$ and $Z\sim\mathcal{D}_Z$ be independent, and suppose
  that $\E[|V(U, Z)|^2]\le\kappa T^2$, where $T>0$ and $\kappa\ge1$.
  Define
\[
  h(u)=\E[V(u, Z)], \qquad
  F=\E[|h(U)|],\qquad W=\E[|h(U)|^2],
\]
  and set $\delta_{\rm test} = 2^{-20}, A_0=128$, $C_0=2^{30}$,
  $\theta=2^{-28}$, and $\kappa'=2^{\ceil{\log\kappa}}$.
  There is a test $\mathcal{A}$ returning \emph{positive} or \emph{negative},
  using exactly $15A_0C_0\kappa'(1+\log\kappa')$ pairs of samples $(u, z)$ with
  $u\sim\mathcal{D}_U$ and $z\sim\mathcal{D}_Z$,
  such that
  \[
    F\ge T\ \Longrightarrow\ \Prob\{\mathcal{A} \text{ outputs
    \emph{negative}}\}<\delta_{\rm test},
    \qquad
    W\le\theta T^2\ \Longrightarrow\ \Prob\{\mathcal{A} \text{ outputs
    \emph{positive}}\}<\delta_{\rm test}.
  \]
\end{lemma}

The test draws samples $u$ of $U$, which we call \emph{anchors}, and
estimates $h(u)$ at each anchor by averaging $V(u,z)$ over samples $z$ of $Z$.
It does so at every power-of-two resolution $w\in\mathcal{W}_{\kappa'}$.
Resolution $w$ uses about $\kappa'/w$ anchors, about $w$ samples of $Z$ per
anchor, and the threshold $\vartheta_w=\kappa'T/(16w)$.
Thus a small $w$ gives many anchors with rough estimates against a high
threshold, and a large $w$ gives few anchors with accurate estimates against a
low threshold.
Each resolution costs $O(\kappa)$ samples.
One repetition accepts if and only if some anchor, at some resolution,
has an estimate above its threshold.  The test takes a majority vote
over five independent repetitions.
In particular, the test never estimates $F$: when $F\geq T$, a single anchor
whose estimate exceeds its threshold witnesses acceptance in one repetition,
and when $W\leq\theta T^2$ such anchors are rare.

We state the rejection case with $W\leq\theta T^2$ because energy also
controls how many nodes can require further exploration.  The quadratic
tail bound $\Prob[|h(U)|\geq\lambda]\leq W/\lambda^2$ gives a convergent
geometric sum over resolutions, with the stated constant $\theta$.
The same test also has a constant-gap soundness guarantee in terms of $F$.
For that guarantee, we sum the dyadic threshold indicators before taking
expectation, as explained after the proof.

\begin{proof}[Proof of \Cref{lem:multiscale}]
  Consider the following algorithm with slightly weaker soundness and
  completeness guarantees.
  At the end of the proof, we amplify the guarantees by taking a majority vote
  over five independent repetitions.
  Throughout this proof, sums and products over $w$ range over
  $w\in\mathcal{W}_{\kappa'}$.

  For every $w\in\mathcal{W}_{\kappa'}$, draw $n_w = A_0\kappa'/w$ independent
  samples $u_{w, 1}, \ldots, u_{w, n_w}$ from $\mathcal{D}_U$, which we call
  \emph{anchors}.
  At each anchor $u_{w, i}$, draw three independent blocks of $C_0w$ samples
  from $\mathcal{D}_{Z}$, denoted by $z_{w, i, \ell, b}$ for $\ell = 1, 2, 3$
  and $b = 1, \ldots, C_0w$.
  Take
  \[
      \mu_{w, i, \ell} = \frac{1}{C_0w} \sum_{b=1}^{C_0w} V(u_{w, i}, z_{w, i,
      \ell, b})
  \]
  and let $\lambda_{w,i}$ be the coordinatewise median of the group means:
  \[
      \mathrm{Re}(\lambda_{w, i}) = \operatorname{med}_{\ell\in\{1,2,3\}}
      \mathrm{Re}(\mu_{w, i, \ell}), \mathrm{Im}(\lambda_{w, i}) =
      \operatorname{med}_{\ell\in\{1,2,3\}} \mathrm{Im}(\mu_{w, i, \ell}).
  \]
  Put thresholds $\vartheta_w=\kappa' T/(16w)$, return positive if at least one
  such $\lambda_{w, i}$ has $|\lambda_{w, i}|\ge \vartheta_w$ among all $(w,
  i)$, and negative otherwise.
  At resolution $w$, an anchor's estimate is accurate to within $\vartheta_w/2$
  except with small probability (as shown below), and thus resolves values of
  $|h|$ at the scale $\vartheta_w$;
  the test is \emph{multiscale} in the sense that it uses all resolutions
  $w\in\mathcal{W}_{\kappa'}$ simultaneously.
  $3n_wC_0w = 3A_0C_0\kappa'$ samples are used at each resolution $w$, so the
  sample count for one repetition, $3A_0C_0\kappa'(1+\log\kappa')$, follows by
  summing over the $1+\log\kappa'$ resolutions $w\in\mathcal{W}_{\kappa'}$.

  \emph{Bounding estimation errors of $\lambda$.}
  We denote the conditional second moment by $\nu(u) = \E[|V(u, Z)|^2]$.
  Conditional on $u_{w,i}=u$, the $3C_0w$ values $V(u, z_{w,i,\ell,b})$ are
  independent.
  Each of these values has mean $h(u)$ and variance
  $\E[|V(u,Z)-h(u)|^2]\leq\nu(u)$.
  Apply item~\ref{item:median-means} of \Cref{lem:median} with
  \[
    r=3,\quad v=C_0w,\quad x=h(u),\quad \nu=\nu(u),\quad
    \omega=\frac{\vartheta_w}{2\sqrt2},
  \]
  so that $\sqrt2\,\omega=\vartheta_w/2$ and
  $p=\min\{1,8\nu(u)/(C_0w\vartheta_w^2)\}$.
  Since $\binom32\cdot8^2=192$ we obtain
  \begin{equation}
\label{eqn:conditional-noise}
    \Prob_{\mathcal{A}}\left[|\lambda_{w, i}-h(u_{w, i})|>\frac{\vartheta_w}{2}
    \ \middle| \ u_{w, i} = u\right]
    \le\min\left\{1,192\left(\frac{\nu(u)}{C_0w\vartheta_w^2}\right)^2\right\}.
  \end{equation}
  The following bound lets us sum these errors over all resolutions.
  \begin{claim}
\label{clm:dyadic-sum}
      For every $u\ge0$, we have
      \begin{equation}
\label{eqn:dyadic-sum}
      \sum_{w\in\mathcal{W}_{\kappa'}} \frac{\kappa'}{w}\min\left\{1,
      \left(\frac{wu}{\kappa'}\right)^2\right\} \le 4u.
      \end{equation}
  \end{claim}
  To see this inequality, we may assume $u>0$.
  Split $\mathcal{W}_{\kappa'}$ into $\mathcal{W}_-=\{w: wu\le\kappa'\}$ and
  $\mathcal{W}_+=\{w: wu>\kappa'\}$, according to the choice of the minimum
  function.
  If $\mathcal{W}_-$ is nonempty, its largest element $w_-$ satisfies
  $w_-\le\kappa'/u$, and since the elements of $\mathcal{W}_-$ are distinct
  powers of $2$,
  \[
\sum_{w\in\mathcal{W}_-}\frac{\kappa'}{w}\left(\frac{wu}{\kappa'}\right)^2
      = \frac{u^2}{\kappa'}\sum_{w\in\mathcal{W}_-} w
      < \frac{2w_-u^2}{\kappa'} \le 2u.
  \]
  If $\mathcal{W}_+$ is nonempty, its smallest element $w_+$ satisfies
  $w_+>\kappa'/u$, and similarly
  \[
      \sum_{w\in\mathcal{W}_+}\frac{\kappa'}{w} < \frac{2\kappa'}{w_+} < 2u.
  \]
  Adding the two parts proves \eqref{eqn:dyadic-sum}.

  Define $\nu'(u)=\nu(u)/(\kappa' T^2)$ and $\eta_0=256\sqrt{192}/C_0$.
  A union bound over all $(w, i)$ gives the following, using
  \eqref{eqn:conditional-noise} and then \eqref{eqn:dyadic-sum}:
  \begin{align*}
      \Prob_{\mathcal{A}}\left[\exists (w, i):\ |\lambda_{w, i}-h(u_{w,
      i})|>\frac{\vartheta_w}{2}\right]
      & \le \sum_{w} \frac{A_0\kappa'}{w}
\E_{\mathcal{D}_U}\left[\min\left\{1,
  192\left(\frac{\nu(U)}{C_0w\vartheta_w^2}\right)^2\right\}\right]
            \\
      & = A_0 \E_{\mathcal{D}_U} \left[\sum_{w} \frac{\kappa'}{w}
            \min\left\{1,\left( \frac{\eta_0 \nu'(U)w}{\kappa'}
            \right)^2\right\}\right] \\
      & \le A_0 \E_{\mathcal{D}_U} \left[4\eta_0 \nu'(U)\right] \\
      & \le 512\eta_0 < \frac1{512}.
  \end{align*}
  On the second line, the expectation of the sum uses the common marginal
  anchor distribution at every resolution;
  the fourth line follows from the law of total expectation
  \[
      \E[\nu'(U)] = \frac{1}{\kappa' T^2} \E[\nu(U)] = \frac{\E[|V(U,
      Z)|^2]}{\kappa' T^2} \le 1.
  \]

  \emph{Completeness.}
  For completeness, suppose $F = \E_{\mathcal{D}_U} [|h(U)|] \ge T$.
  For fixed $w$, each anchor $u_{w, i}$ has the same probability of being
  ``strong'', meaning that the magnitude of $h(u_{w, i})$ is not too small.
  Specifically, let $p_w=\Prob_{\mathcal{D}_U}\{|h(U)| \ge 2\vartheta_w\}$.
  Using independence of the anchors, the probability of sampling no strong
  anchor is at most
  \begin{equation}
    \prod_w(1-p_w)^{A_0\kappa'/w} \le \exp\left(-A_0 \sum_w
    \frac{\kappa'p_w}{w}\right).
  \end{equation}

  The dyadic resolutions give enough weight to every magnitude above the
  smallest threshold.  Values below that threshold contribute little to
  $F$, while the second moment controls the loss from truncating very large
  values.  To make this precise, put $Y=|h(U)|/T$ and define
  \[
    g_{\rm c}(y)
    =\sum_w\frac{\kappa'}w
       \ind_{\{y\geq\kappa'/(8w)\}}.
  \]
  By definition $\vartheta_w = \kappa'T/(16w)$ we have
  \begin{equation}
\label{eqn:no-strong-anchor}
    \prod_w(1-p_w)^{A_0\kappa'/w} \le \exp\left(-A_0 \sum_w \frac{\kappa'
    \E[\ind_{Y \ge \kappa'/(8w)}]}{w}\right)
    \leq\exp\bigl(-A_0\E[g_{\rm c}(Y)]\bigr).
  \end{equation}

  For $y\geq1/8$, let $w^*$ be the smallest contributing resolution.
  If $w^*=1$, its contribution is $\kappa'$.  Otherwise dyadic rounding
  gives $w^*\leq\kappa'/(4y)$, so its contribution is at least $4y$.
  In either case,
  \[
    g_{\rm c}(y)\geq\min\{y,\kappa'\}
    \qquad(y\geq1/8).
  \]
  It follows that
  \begin{align*}
    \E[g_{\rm c}(Y)]
    & \geq\E[\min\{Y,\kappa'\}]-\frac18 \\
    & = \E[Y] - \E[\max\{Y - \kappa', 0\}] - \frac18
    \ge \E[Y]-\frac{\E[Y^2]}{4\kappa'}-\frac18
  \end{align*}
  The last inequality above follows from $\max\{y - \kappa', 0\} \le
  y^2/(4\kappa')$.
  Jensen's inequality gives $\E[Y^2]\leq\kappa'$, since
  $\E|\E[V(U,Z)\mid U]|^2\leq\E|V(U,Z)|^2$, so we obtain
  \[
  \E[g_{\rm c}(Y)] \ge 1 - \frac14 - \frac18 = \frac58.
  \]
  Substituting this bound into \eqref{eqn:no-strong-anchor} gives
  \[
    \prod_w(1-p_w)^{A_0\kappa'/w}
    \leq\exp\bigl(-A_0\E[g_{\rm c}(Y)]\bigr)
    \leq e^{-80}.
  \]
  A strong anchor with estimation error at most $\vartheta_w/2$ yields a
  positive answer.
  Completeness failure is therefore less than $1/512+e^{-80}<1/256$.

  \emph{Soundness.}
  For soundness, a positive answer with all estimation errors at most
  $\vartheta_w/2$ requires an anchor $u_{w, i}$ with $|h(u_{w, i})| \ge
  \vartheta_w/2$.
  By the union bound and Chebyshev's inequality $\Prob[|h(U)|\geq\lambda]\leq
  W/\lambda^2$, the probability of sampling any such anchor is at most
  \begin{equation}
\label{eqn:soundness-error}
    \sum_w \frac{A_0\kappa'}{w} \Prob \left[|h(U)| \ge \frac{\kappa'T}{32w}
    \right]
    \le \sum_w \frac{A_0\kappa'}{w}\cdot\frac{(32w)^2W}{\kappa'^2T^2}
    = \frac{1024A_0W}{T^2}\sum_{w\in\mathcal{W}_{\kappa'}} \frac{w}{\kappa'}
    < \frac{2048A_0W}{T^2},
  \end{equation}
  where the last step uses $\sum_{w\in\mathcal{W}_{\kappa'}}w=2\kappa'-1$.
  Incorporating the estimation error and substituting $W \le \theta T^2 =
  2^{-28} T^2$, the total soundness failure probability is bounded above by
  \[
    \frac{2048A_0W}{T^2} + \frac{1}{512} = \frac{2^{18}W}{T^2} + \frac{1}{512}
    \le \frac{1}{1024} + \frac{1}{512} < \frac{1}{256}.
  \]

  \emph{Success probability amplification.}
  Finally, we boost the success probability in both promised cases by repeating
  the test five times and taking a majority vote.
  The total number of samples needed is $5 \times 3A_0C_0\kappa'(1+\log\kappa')
  = 15A_0C_0\kappa'(1+\log\kappa')$.
  The five repetitions use disjoint samples, so in each promised case their
  failure events are independent, each of probability less than $1/256$.
  By item~\ref{item:median-majority} of \Cref{lem:median} with $r=5$, the
  completeness and soundness failure probabilities are thus bounded by
  \[
    \binom53\left(\frac{1}{256}\right)^3 = 10\cdot2^{-24} < 2^{-20}. \qedhere
  \]
\end{proof}

\begin{remark}[Soundness from a first-moment bound]
The same test also rejects with probability greater than
$1-\delta_{\rm test}$ when $F\leq2^{-23}T$.  To see this, consider one
repetition and put
\[
  g_{\rm s}(x)
  =\sum_{w\in\mathcal W_{\kappa'}}\frac{\kappa'}w
     \ind_{\{x\geq\kappa'T/(32w)\}}.
\]
If $w_0$ is the smallest contributing resolution, the sum is at most
$2\kappa'/w_0\leq64x/T$; if no resolution contributes, it is zero.
Outside the estimation-error event, whose probability is less than
$1/512$, acceptance requires an anchor with magnitude at least
$\vartheta_w/2=\kappa'T/(32w)$.  A union bound over anchors therefore gives
\begin{align*}
  \Prob[\text{one repetition accepts}]
  &<\frac1{512}+A_0\E[g_{\rm s}(|h(U)|)]\\
  &\leq\frac1{512}+\frac{64A_0F}{T}
   \leq\frac1{512}+\frac1{1024}<\frac1{256}.
\end{align*}
The five independent repetitions reduce this probability below
$\delta_{\rm test}$, as in the proof above.  Summing the dyadic indicators
before taking expectation avoids the logarithmic loss that would result
from applying Markov's inequality separately at every resolution.
\end{remark}

\subsection{Potential-based traversal and supermartingale analysis}
\label{sec:potential-traversal}

We now present the locator.  It explores the tree breadth-first, using a
positive node potential to tolerate occasional incorrect tests.  Positive
answers increase the potential up to a fixed cap, and negative answers
multiply it by a smaller factor; a branch is pruned when its potential
falls below one.  An exponential supermartingale protects heavy paths, while a
drift bound
for the sum of all retained node potentials controls the total search.  All
nodes at one depth share
their samples, so these arguments condition on whole depths and do not
assume independence between nodes at the same depth.  Independent leaf
verification then removes false positives.

\begin{proposition}[Complete norm-bounded locator]\label{prop:locator}
  Let $\norm{\Rhat}_2\le M$, $0<T\le M$, $\kappa=M^2/T^2$, and $0<\gamma<1/2$.
  There exists a random algorithm that, with probability at least $1-\gamma$,
  returns a list containing every $f$ with $|\Rhat(f)|\ge T$ \emph{(which we
  call heavy frequencies)} and containing only frequencies with
  $|\Rhat(f)|>T/2$.
  The algorithm makes
  \[
    O\bigl(
      \kappa (\log N)\log(2\kappa)
      +\kappa\log(\kappa (\log N)/\gamma)
    \bigr)
  \]
  nonadaptive point queries, counting repetitions.
  On every execution, including executions that terminate with an
  abort, its running time and space usage are bounded by a fixed
  polynomial in $\kappa$, $\log N$, and $\gamma^{-1}$.
\end{proposition}

\begin{proof}

\emph{The node potential and traversal.}
Set
\begin{equation}
\label{eqn:potential-parameters}
  \phi_{\max}=\left(\frac{3\kappa\log N}{\gamma}\right)^{1/4},
  \qquad
  D=\frac{42\kappa\log N}{\theta\gamma}\,\phi_{\max}.
\end{equation}
Here $\theta=2^{-28}$ is the constant in \Cref{lem:multiscale}.
The root has potential $\phi_{\max}$.  For a child $v$ of a retained node,
run the multiscale test with
$V_v(t,z)=\R(t+z)\conj{\chi_v(z)}$, where $t$ is uniform over $G$ and
$z$ is uniform over $K_j$ at depth $j$, and set
\begin{equation}
\label{eqn:node-potential-update}
  \phi(v)=
  \begin{cases}
    \min\{\phi_{\max},2^{1/4}\phi(\mathrm{par}(v))\},
      &\text{positive},\\
    \phi(\mathrm{par}(v))/4,&\text{negative}.
  \end{cases}
\end{equation}
Retain $v$ precisely when $\phi(v)\geq1$, and expand retained nodes in
breadth-first order.  Abort if the cumulative number of retained nodes,
including the root, would exceed $D$.
The cap bounds the potential at entry into a low-energy branch, regardless
of the preceding positive answers.  These formulas require no root
extraction in the algorithm: store $\phi(v)^4$, starting at
$3\kappa\log N/\gamma$, and update it by doubling (subject to the cap
$\phi_{\max}^4$)
or dividing by $256$.  Likewise, $D^4$ is computed by ordinary arithmetic;
comparisons with $D$ and the rounded logarithms below can be evaluated by
comparing fourth powers.  The cutoff $D$ need not be an integer.

The test at $v$ uses point queries to $\R$ to evaluate $V_v$.
All nodes at depth $j$ use the same batch of pairs $(t,z)$, with only the
characters $\chi_v$ changing.  Batches for different depths are independent
and can be chosen in advance.

If the algorithm does not abort, let $\mathcal{C}$ be its list of retained
leaves.  We verify these candidates to remove those with small
$|\Rhat(f)|$.  Independently allocate $q_v$ groups of $v_0$ uniform
points $t_{i, \ell}$, where
\begin{equation}
\label{eqn:validation}
  v_0=1024\ceil\kappa,\qquad q_v=8\ceil{\log(3D/\gamma)}+1.
\end{equation}
On each retained leaf $f \in \mathcal{C}$, compute
\[
  \mu_i(f) = \frac{1}{v_0} \sum_{\ell=1}^{v_0} \R(t_{i,
  \ell})\conj{\chi_f(t_{i, \ell})}
\]
for every $i \in [q_v]$.
As in the multiscale test, compute $\lambda(f)$ as the coordinatewise real and
imaginary median of $\{\mu_i(f)\}_{i=1}^{q_v}$, and keep $f$ if $|\lambda(f)|
\ge 3T/4$.

For the traversal analysis, first omit the abort rule.  Let
$\mathcal F_j$ be the $\sigma$-field generated by the query batches through
depth $j$, with $\mathcal F_0$ trivial.  The residual and the locator
parameters are fixed.  The retained nodes and their potentials through
depth $j$ are $\mathcal F_j$-measurable.  The next depth uses a fresh batch,
so the bounds in \Cref{lem:multiscale} hold conditionally on $\mathcal F_j$.
For analysis, a test at an unvisited node is also defined by its depth's
batch; this requires no additional queries by the algorithm.

\emph{Heavy paths: an exponential supermartingale.}
Fix a frequency $f$ with $|\Rhat(f)|\geq T$ and its root-to-leaf path.
Write $A_j=2^{1/4}$ for a positive test on this path at depth $j$, and
$A_j=1/4$ for a negative test.  These are the multipliers before capping.
Every node on the path satisfies $F_v\geq T$, so
\Cref{prop:statistics,lem:multiscale} give
\begin{equation}
\label{eqn:heavy-potential-drift}
  \E[A_j^{-4}\mid\mathcal F_{j-1}]
  \leq\frac{1-\delta_{\rm test}}{2}+256\delta_{\rm test}<\frac34.
\end{equation}
For each fixed starting depth $a\in\{0,\ldots,\log N-1\}$, define
\[
  Z_{a,a}=1,\qquad
  Z_{a,j}=\prod_{i=a+1}^{j}A_i^{-4}
  \quad(a<j\leq\log N).
\]
Thus $(Z_{a,j})_{j\geq a}$ is a nonnegative supermartingale with respect to
$(\mathcal F_j)$: its conditional expectation at the next depth is at most
three quarters of its current value.  Ville's maximal inequality
(\Cref{lem:ville}), which extends Markov's inequality to the running maximum of
a nonnegative
supermartingale, gives
\begin{equation}
\label{eqn:heavy-potential-crossing}
  \Prob\left[\max_{a\leq j\leq\log N}Z_{a,j}
    >\phi_{\max}^{4}\right]
  \leq\phi_{\max}^{-4}.
\end{equation}

If the path is pruned, let $a$ be its last depth at the cap before pruning.
Such a depth exists because the root starts at the cap.  No capping occurs
after $a$ and before pruning, so at the pruning depth $j$,
$\phi_{\max}\prod_{i=a+1}^{j}A_i<1$; hence
$Z_{a,j}>\phi_{\max}^{4}$.  We apply
\eqref{eqn:heavy-potential-crossing} separately to every deterministic
starting depth $a$ and take a union bound.  The last visit to the cap need
not itself be a stopping time.  There are at most $\kappa$ heavy frequencies,
since $\norm{\Rhat}_2^2\leq\kappa T^2$.  Consequently,
\[
  \Prob[\text{some heavy path is pruned}]
  \leq\kappa(\log N)\phi_{\max}^{-4}
  =\frac{\gamma}{3}.
\]

\emph{Search size: a sum of node potentials.}
We bound the search by summing the potentials of all retained nodes.
Low-energy children contribute less potential on average than their
parents, while the energy bound limits the number of high-energy children
at each depth.  The cap bounds the contribution of each such child.

Let $\mathcal V_j$ be the retained nodes at depth $j$, and put
\begin{equation}
\label{eqn:search-potential}
  \Phi_j=\sum_{v\in\mathcal V_j}\phi(v).
\end{equation}
Each summand is at least one, so $\Phi_j$ bounds the number of retained
nodes at depth $j$, and $\Phi_0=\phi_{\max}$.

Condition on $\mathcal F_j$.  If a child $v$ of a retained node $u$ has
$W_v<\theta T^2$, its positive-test probability is at most
$\delta_{\rm test}$.  Its uncapped multiplier therefore has conditional
mean at most
\[
  \delta_{\rm test}2^{1/4}+(1-\delta_{\rm test})/4<\frac13.
\]
Capping and discarding a node can only decrease its contribution.
Since every parent has two children, all low-energy children together
contribute at most $2\Phi_j/3$ in conditional expectation.  This uses
linearity of expectation, with no independence assumption between children.
At depth $j+1$, at most $\kappa/\theta$ nodes have
$W_v\geq\theta T^2$, because their energies sum to at most $M^2$.
Each contributes at most $\phi_{\max}$, so these nodes contribute at most
$\kappa\phi_{\max}/\theta$.  Hence
\begin{equation}
\label{eqn:search-potential-drift}
  \E[\Phi_{j+1}\mid\mathcal F_j]
  \leq\frac23\Phi_j+\frac{\kappa\phi_{\max}}{\theta}.
\end{equation}

To control the accumulated search, reserve
$\kappa\phi_{\max}/\theta$ for each remaining depth and keep track of
the potential already spent.  Specifically, define
\begin{equation}
\label{eqn:search-supermartingale}
  \Psi_j=\Phi_j+\frac13\sum_{i=0}^{j-1}\Phi_i
    +\frac{\kappa\phi_{\max}}{\theta}(\log N-j),
  \qquad 0\leq j\leq\log N.
\end{equation}
These variables are nonnegative and adapted to $(\mathcal F_j)$.
They are integrable because the tree is finite and every potential is
capped.  The drift bound \eqref{eqn:search-potential-drift} gives
\[
  \E[\Psi_{j+1}\mid\mathcal F_j]\leq\Psi_j,
  \qquad
  \Psi_0=\phi_{\max}\left(1+\frac{\kappa\log N}{\theta}\right).
\]
The decrease in the current potential pays for its accumulated contribution,
and the reserve pays for the next depth's high-energy nodes.
Moreover, $\Psi_j\geq\frac13\sum_{i=0}^j\Phi_i$, so it bounds one third
of the cumulative number of retained nodes through depth $j$.

The actual traversal agrees with the untruncated traversal until an abort.
If it aborts while processing depth $j$, the untruncated traversal retains
more than $D$ nodes through that depth, and hence $\Psi_j>D/3$.
Ville's maximal inequality (\Cref{lem:ville}) now gives
\begin{align*}
  \Prob[\text{abort}]
  &\leq\Prob\left[\max_{0\leq j\leq\log N}\Psi_j>D/3\right]\\
  &\leq\frac{3\phi_{\max}(1+\kappa(\log N)/\theta)}{D}
   \leq\frac{6\kappa\phi_{\max}\log N}{\theta D}
    =\frac{\gamma}{7}<\frac{\gamma}{3},
\end{align*}
where we use $\kappa(\log N)/\theta\geq1$ and the definition of $D$.
Each retained nonleaf has only two children to test; bounding retained
nodes therefore also bounds all visited nodes.

\emph{Accuracy of the verification.}
The verification points are independent of the traversal, so we condition on
$\mathcal{C}$.
Fix $f\in\mathcal{C}$.
The $q_vv_0$ values $\R(t_{i,\ell})\conj{\chi_f(t_{i,\ell})}$ are independent,
and each has mean $\Rhat(f)$ by \Cref{prop:statistics}.
By \eqref{eqn:one-sample}, each has
\[
  \E_t\bigl[|\R(t)\conj{\chi_f(t)}-\Rhat(f)|^2\bigr] = \|\Rhat\|_2^2 -
  |\Rhat(f)|^2 \le M^2 = \kappa T^2.
\]
Since $q_v$ is odd, we apply item~\ref{item:median-means} of \Cref{lem:median}
with
\[
  r=q_v,\quad v=v_0,\quad x=\Rhat(f),\quad \nu=\kappa T^2,\quad \omega=\frac
  T8.
\]
Since $v_0 = 1024\ceil{\kappa}$, we have $p\leq64\kappa/v_0\leq1/16$, and hence
$(4p)^{(q_v+1)/2}\leq4^{-(q_v+1)/2}$.
Therefore, we obtain
\[
  \Prob\left[|\lambda(f) - \Rhat(f)| \geq \frac{\sqrt{2}T}{8} \right] \leq
  \frac14\cdot4^{-(q_v+1)/2} = 2^{-q_v-3}.
\]
The abort check ensures $|\mathcal{C}| \le D$, and $q_v\geq\log(3D/\gamma)$, so
\[
  \Prob\left[\exists f \in \mathcal{C} : |\lambda(f) - \Rhat(f)| \geq
  \frac{\sqrt{2}T}{8}\right] \leq D\cdot2^{-q_v-3} < \frac{\gamma}{3}.
\]
Because $T/2 + \sqrt{2}T/8 < 3T/4 < T - \sqrt{2}T/8$, with probability at least
$1-\gamma/3$, all heavy leaves are kept, and every kept frequency has magnitude
greater than $T/2$.

The failure probabilities of the three parts sum to less than $\gamma$, which
completes the proof of correctness.

The search-tree size limit permits at most $2D$ node tests and $D$ leaf
verifications on every run.  Each takes polynomial time, so the running
time is bounded by
\begin{equation}
\label{eqn:D-growth}
  O(D)\text{poly}(\kappa, (\log N), \gamma^{-1}) = O\bigl(\kappa^{5/4}(\log
  N)^{5/4}\gamma^{-5/4}\bigr)\text{poly}(\kappa, (\log N), \gamma^{-1}) =
  \text{poly}(\kappa, (\log N), \gamma^{-1}).
\end{equation}

\Cref{prop:shared-runtime} gives the sharper operation bound used in
the recovery theorems.

By \Cref{lem:multiscale}, the sample complexity for traversal at each depth is
$15A_0C_0\kappa'(1+\log\kappa')$ (recall that samples are reused within the
same depth), and the number of additional samples for final verification is
\[
  q_v v_0 = 1024(8\ceil{\log(3D/\gamma)} + 1)\ceil{\kappa} =
  O(\kappa\log(\kappa (\log N)/\gamma)),
\]
so the total sample complexity is given by $O(\kappa (\log N) \log(2\kappa) +
\kappa\log(\kappa (\log N)/\gamma))$.

\end{proof}

\section{Least-squares residual norm estimate for residual spectra}
\label{sec:least-squares}

We now study the input signal $\X$ and its spectrum $\Xhat$, rather than
an arbitrary residual.

For the residual norm estimate in this section, we assume the tail-norm bound
\eqref{eqn:upper-promise}, i.e., $\sigma_s(\Xhat)_2^2 \le s\tau^2$, where
$\tau\geq0$.
The case $\tau=0$ means that $\Xhat$ is $s$-sparse; then $\bar\sigma=0$ and
$\Omega=\supp(\Xhat)$ below.

\begin{lemma}[Simultaneous residual norm bounds from least
squares]\label{lem:ls}
    Define $\Omega = \{f: |\Xhat(f)| > \tau/4\}, \bar\sigma = \sqrt{s}\tau$.
    Use $n_{\rm ls}=170s\ceil{\log(2040s)}$ samples $t_b \in G$.
    Define $A \in \C^{n_{\rm ls} \times G}$, $\mathbf{b} \in \C^{n_{\rm ls}}$
    by
    $A_{bf}=\chi_f(t_b)/\sqrt{n_{\rm ls}}$ and
    $\mathbf{b}_b=\X(t_b)/\sqrt{n_{\rm ls}}$.
    Let $b_{\rm ls} = \sqrt{255}/2, \delta_{\rm ls} = 1/30$ be universal
    constants.

    Define $C_S = \arg\min_{c \in \C^S} \norm{\mathbf{b} - A_Sc}_2$ as the
    least-squares fit on the columns $A_S$, extended by zero outside $S$; we
    specifically designate $C_\varnothing = 0$.
    Let $z_S = \mathbf{b}-A_SC_S$, $M_S = 2\norm{z_S}_2+3b_{\rm ls}\bar\sigma$.
    Then there exists a polynomial-time algorithm, which either aborts or
    computes $C_S, M_S$ from input samples $t_1, \cdots, t_{n_{\rm ls}}$, set
    $S \subseteq G$ and $\bar\sigma$.

    Furthermore, with probability at least $1 - \delta_{\rm ls}$ over uniform
    and independent samples, the following holds for all $S \subseteq \Omega$
    simultaneously:
    \begin{equation}
        \norm{\Xhat-C_S}_2 \leq \sqrt2(\norm{z_S}_2+b_{\rm ls}\bar\sigma) +
        \norm{\Xhat_{G\setminus\Omega}}_2, \qquad \norm{z_S}_2 \leq
        \sqrt{3/2}\norm{\Xhat_{\Omega\setminus S}}_2 +b_{\rm ls}\bar\sigma;
        \label{eqn:ls-raw}
    \end{equation}
    \begin{equation}
        \norm{\Xhat-C_S}_2\leq M_S, \qquad M_S^2 \leq
        12\norm{\Xhat_{\Omega\setminus S}}_2^2+50b_{\rm ls}^2\bar\sigma^2.
        \label{eqn:ls-certificate}
    \end{equation}
\end{lemma}

The set $\Omega$ must be fixed independently of the least-squares
sample.  The supports $S\subseteq\Omega$ may depend on that sample;
the event in the lemma holds for all of them simultaneously.

\begin{proof}

For $S \neq \emptyset$, construct $A_S$ and $\mathbf{b} \in \C^{n_{\rm ls}}$
from the supplied sample locations and outcomes in polynomial time.
If $A_S^*A_S$ is singular, the algorithm aborts;
otherwise, let $C_S = (A_S^*A_S)^{-1}(A_S^*\mathbf{b})$ be the unique
least-squares fit.
Then the algorithm computes $z_S=\mathbf{b}-A_SC_S, M_S=2\norm{z_S}_2+3b_{\rm
ls}\bar\sigma$.

For analysis, consider $t_1, \ldots, t_{n_{\rm ls}}$ as samples uniformly and
independently drawn from $G$.
Define two ``good events'' $\mathcal{E}_1, \mathcal{E}_2$, which separately
control the distortion of this random restriction inside and outside $\Omega$:
\begin{align*}
  \mathcal{E}_1 & := \text{``}\frac12I_\Omega \preceq A_\Omega^*A_\Omega
  \preceq \frac32I_\Omega\text{''}, \\
  \mathcal{E}_2 & := \text{``}\norm{A_{G \setminus \Omega}\Xhat_{G \setminus
  \Omega}}_2 \le b_{\rm ls}\bar\sigma\text{''}.
\end{align*}
We first prove the desired property on the intersection of $\mathcal{E}_1,
\mathcal{E}_2$, and postpone the probabilistic analysis.

\emph{Correctness on $\mathcal{E}_1 \cap \mathcal{E}_2$.}
Define $n_\Omega = |\Omega|$.
For $S\subseteq \Omega$, put $v_S=(\Xhat-C_S)_\Omega \in \C^{n_\Omega}$.
If $n_\Omega=0$, then $S=\varnothing$ is the only admissible support, and $v_S$
is the empty vector, so the bound on $\norm{v_S}_2$ holds trivially;
hence we assume $n_\Omega\geq1$ whenever $A_\Omega$ or \Cref{prop:Chernoff} is
used.

On $\mathcal{E}_1$, for any $S \subseteq \Omega$, the matrix $A_S^*A_S$ is
nonsingular, so the algorithm never aborts for any such $S$.

We can split the residual $z_S$ as
\begin{align}
  z_S & = A(\Xhat - C_S) \notag\\
  & = A_\Omega(\Xhat - C_S)_\Omega + A_{G \setminus \Omega}(\Xhat - C_S)_{G
  \setminus \Omega} \notag \\
  & = A_\Omega v_S + A_{G \setminus \Omega} \Xhat_{G \setminus \Omega}.
  \label{eqn:split-residual}
\end{align}
The last equality uses $(C_S)_{G \setminus \Omega}=0$, since $S \subseteq
\Omega$.
On $\mathcal{E}_1 \cap \mathcal{E}_2$, \eqref{eqn:split-residual} gives
\[
  \norm{v_S}_2 \leq \frac{1}{\sigma_{\rm min}(A_\Omega)}\norm{z_S - A_{G
  \setminus \Omega}\Xhat_{G \setminus \Omega}}_2 \le \sqrt{2}(\norm{z_S}_2 +
  b_{\rm ls}\bar\sigma),
\]
and therefore
\begin{equation*}
  \norm{\Xhat - C_S}_2 \le \norm{v_S}_2 + \norm{\Xhat_{G \setminus \Omega}}_2
  \le \sqrt{2}(\norm{z_S}_2 + b_{\rm ls}\bar\sigma) + \norm{\Xhat_{G \setminus
  \Omega}}_2.
\end{equation*}

Least-squares optimality of $C_S$, using $\Xhat_S$ as a competitor, gives
\begin{align*}
  \norm{z_S}_2 & \leq \norm{\mathbf{b} - A\Xhat_S}_2 = \norm{A_{\Omega
  \setminus S}\Xhat_{\Omega \setminus S} + A_{G \setminus \Omega}\Xhat_{G
  \setminus \Omega}}_2 \\
  & \le \sigma_{\rm max}(A_{\Omega \setminus S}) \norm{\Xhat_{\Omega \setminus
  S}}_2 + b_{\rm ls}\bar\sigma \le
  \sqrt{\frac{3}{2}}\norm{\Xhat_{\Omega\setminus S}}_2 + b_{\rm ls}\bar\sigma,
\end{align*}
where the last inequality uses the two events $\mathcal{E}_1, \mathcal{E}_2$.
This completes the proof of \eqref{eqn:ls-raw}.

Fix the best-$s$ support $S^*_s = \supp(\HT_s(\Xhat))$.
By definition of $\Omega$ and $S^*_s$, coordinates in $S^*_s \setminus \Omega$
contribute Fourier energy of at most $s\tau^2/16$.
Hence we conclude that
\begin{align}
  \norm{\Xhat_{G \setminus \Omega}}_2^2 & = \norm{\Xhat_{S^*_s \setminus
  \Omega}}_2^2 + \norm{\Xhat_{(G \setminus S^*_s) \setminus \Omega}}_2^2
  \notag\\
  & \le \norm{\Xhat_{S^*_s \setminus \Omega}}_2^2 + \norm{\Xhat_{G \setminus
  S^*_s}}_2^2
  = \norm{\Xhat_{S^*_s \setminus \Omega}}_2^2 + \sigma_s(\Xhat)_2^2 \notag\\
  & \le \frac{17}{16} s\tau^2 = \frac{17}{16} \bar{\sigma}^2.
  \label{eqn:H-effective}
\end{align}

Combining \eqref{eqn:ls-raw} with the QM--AM inequality gives
\[
    M_S^2 = (2\|z_S\|_2 + 3b_{\rm ls}\bar\sigma)^2 \le
    \left(\sqrt{6}\norm{\Xhat_{\Omega\setminus S}}_2 + 5b_{\rm ls}\bar\sigma
    \right)^2 \le 12\norm{\Xhat_{\Omega\setminus S}}_2^2 + 50b_{\rm
    ls}^2\bar\sigma^2,
\]
and combining \eqref{eqn:ls-raw} with \eqref{eqn:H-effective}  gives
\[
  \norm{\Xhat - C_S}_2 \le \sqrt{2}\norm{z_S}_2 + \left(\sqrt{2}b_{\rm ls} +
  \frac{\sqrt{17}}{4}\right)\bar\sigma \le 2\norm{z_S}_2 + 3b_{\rm
  ls}\bar\sigma = M_S.
\]
This proves \eqref{eqn:ls-certificate}.

\emph{Probabilistic analysis of $\mathcal{E}_1$ and $\mathcal{E}_2$.}
Write $\Omega = \{f_1, \ldots, f_{n_\Omega}\}$.
If $\tau>0$, \eqref{eqn:upper-promise} implies that fewer than $16s$
coordinates in $G\setminus S^*_s$ have magnitude greater than $\tau/4$.
If $\tau=0$, then $\Omega=\supp(\Xhat)$ has at most $s$ elements.
Thus $n_\Omega<17s$ in both cases.

For $b \in [n_{\rm ls}]$, let $a_b^*$ be the $b$-th row of $A_\Omega$, i.e.,
\[
  a_b = \frac{1}{\sqrt{n_{\rm ls}}}\overline{(\chi_{f_1}(t_b), \ldots,
  \chi_{f_{n_\Omega}}(t_b))},
\]
and put $\Psi_b = a_ba_b^*$.
Then $\Psi := \sum_b \Psi_b = A_\Omega^*A_\Omega$.

To bound the failure probability of $\mathcal{E}_1$, we apply
\Cref{prop:Chernoff} to $\Psi$.  Its hypotheses hold as follows:
\begin{itemize}
  \item All samples $t_b$ are independent;
  \item Each $\Psi_b$ is positive semidefinite and has maximum eigenvalue
  $\norm{a_b}_2^2 = n_\Omega/n_{\rm ls}$, so $0 \preceq \Psi_b \preceq \Lambda
  I_{n_\Omega}$ almost surely holds for $\Lambda = n_\Omega/n_{\rm ls}$;
  \item For uniform $t_b$ over $G$, we can obtain $\E_{t_b} [\Psi_b] =
  I_{n_\Omega}/n_{\rm ls}$ and $\E [\Psi] = I_{n_\Omega}$ by
  \eqref{eqn:char-orthogonality}, so $\mu_{\rm min} = \mu_{\rm max} = 1$.
\end{itemize}
We therefore apply the upper-tail and lower-tail bounds in \Cref{prop:Chernoff}
with $\zeta = 1/2$, which gives
\begin{equation}
\label{eqn:matrix-Chernoff}
  \Prob\left[ \overline{\mathcal{E}_1} \right] \leq
  2n_\Omega\exp\left(-\frac{c_*n_{\rm ls}}{n_\Omega}\right),
\end{equation}
where
\[
c_*=\min\left\{\frac{1-\ln2}{2},
  \frac{3}{2}\ln\frac{3}{2}-\frac{1}{2}\right\}>\frac{1}{10}
\]
is a universal constant.
Notice that
\[
  n_{\rm ls} \ge 170s\ln(2040s) \ge \frac{17s}{c_*}\ln\frac{68s}{\delta_{\rm
  ls}}, \qquad n_\Omega < 17s,
\]
so we have
\[
  \Prob\left[ \overline{\mathcal{E}_1} \right] \leq
  2n_\Omega\exp\left(-\frac{c_* n_{\rm ls}}{n_\Omega}\right) \le
  34s\exp\left(-\frac{c_* n_{\rm ls}}{17s}\right) \le \frac{\delta_{\rm
  ls}}{2}.
\]

Regarding $\mathcal{E}_2$,
if $\tau=0$, then $\Xhat_{G\setminus\Omega}=0$ and $\mathcal{E}_2$ holds
surely.
Otherwise, applying \eqref{eqn:parseval} to $G\setminus\Omega$ gives
\[
  \E\left[\norm{A_{G \setminus \Omega}\Xhat_{G \setminus \Omega}}_2^2\right]
  = \sum_{b = 1}^{n_{\rm ls}} \E\left[\frac{1}{n_{\rm ls}}\left|\sum_{f \in G
  \setminus \Omega} \Xhat(f)\chi_f(t_b)\right|^2\right]
  = \frac{1}{n_{\rm ls}} \cdot n_{\rm ls} \norm{\Xhat_{G \setminus \Omega}}_2^2
  = \norm{\Xhat_{G \setminus \Omega}}_2^2.
\]
By Markov's inequality,
\[
  \Prob\left[ \overline{\mathcal{E}_2} \right] \leq \frac{\norm{\Xhat_{G
  \setminus \Omega}}_2^2}{b_{\rm ls}^2\bar\sigma^2} \le \frac{17}{16b_{\rm
  ls}^2} = \frac{\delta_{\rm ls}}{2}.
\]

Finally, a union bound gives $\Prob\left[ \mathcal{E}_1 \cap \mathcal{E}_2
\right] \ge 1 - \delta_{\rm ls}$, which finishes the proof.
\end{proof}

\section{Coordinatewise recovery from a tail-norm upper bound}
\label{sec:schedule}

This section solves the \emph{coordinatewise recovery problem}, whose inputs
are a sparsity parameter $s$ and a scale $\tau\geq0$.
It assumes only a one-sided bound on the tail $\sigma_s(\Xhat)_2$, and it
guarantees an $\ell_\infty$ error of order $\tau$ for an output with $O(s)$
nonzero coordinates.
Our goal is the following result, which serves as an intermediate step toward
\Cref{thm:main-dyadic-intro}.

\begin{theorem}[Recovery at a supplied tail-norm upper bound]
\label{thm:upper-tail}
Let $2\leq s<N$ and $\tau\geq0$, and suppose
\begin{equation}
\label{eqn:upper-promise}
  \sigma_s(\Xhat)_2^2\leq s\tau^2.
\end{equation}
There is a randomized algorithm with nonadaptive sampling which, taking $s$ and
$\tau$ as inputs, with
probability at least $9/10$, returns a vector $\Xtilde$ satisfying
\begin{equation}
\label{eqn:upper-conclusion}
  |\supp(\Xtilde)|<17s,\qquad \norm{\Xhat-\Xtilde}_\infty\leq\tau/2.
\end{equation}
It uses at most
\begin{equation}
\label{eqn:upper-samples}
  \min\{N,Cs\log N\log s\}
\end{equation}
point queries, for an absolute constant $C$.
The algorithm takes at most $O\bigl(s^{7/2}(\log N)^{9/4}\log(s\log N)\bigr)$
exact operations.
\end{theorem}

\Cref{sec:resources} derives \Cref{thm:main-dyadic-intro} from
\Cref{thm:upper-tail}.
For $B>0$, it takes $s=\ceil{25c^2k/\eps^2}$ and
$\tau=\eps B/(5c\sqrt k)$, using a full Fourier transform if $s\geq N$.
The case $B=0$ is handled separately there.
When $s<N$, we have $s\geq k$, so the lower bound
$\sigma_k(\Xhat)_2\leq B$ implies the promise \eqref{eqn:upper-promise}.
Keeping the $k$ largest coordinates of the output turns the $\ell_\infty$
guarantee into the $\ell_2/\ell_2$ guarantee \eqref{eqn:main-error}; this step
uses \Cref{lem:hard-thresholding} and the upper bound $B\leq
c\,\sigma_k(\Xhat)_2$.
The variants in \Cref{sec:tail-free,sec:relative-additive} also build on the
algorithm of this section.

The parameter $s$ controls every sample size and every work budget below,
whereas $\tau$ enters only the thresholds and the residual norm estimate.
This section develops the main algorithmic ideas for \Cref{thm:upper-tail}; the
proof is completed in \Cref{sec:resources}, after we construct the nonadaptive
query set and establish the resource bounds.
We first analyze an algorithm with \emph{adaptive} sampling, using fresh
randomness for each requested batch.
In \Cref{sec:tapes}, we show that all query locations can be fixed before any
answer is revealed, making the sampling nonadaptive.
We then verify the sample and operation bounds in \Cref{sec:resources}.

Throughout this section, we implicitly assume the tail-norm bound
\eqref{eqn:upper-promise} with $\tau\geq0$.
Let $\delta_{\rm ls} = 1/30$ and $b_{\rm ls} = \sqrt{255}/2$ be the constants
from \Cref{lem:ls}, and fix
\begin{equation}
\label{eqn:schedule-parameters}
  \delta_{\rm loc}=\delta_{\rm est}=\frac1{30},
  \qquad \bar\sigma=\sqrt s\,\tau,
  \qquad T_{\min}=\frac\tau2,
  \qquad K=31000s,
  \qquad \gamma=\frac{\delta_{\rm loc}}{K}.
\end{equation}
Here $K$ bounds the total ratio $\sum_i\kappa_i$ of all locator calls.
Every call uses the same failure probability parameter $\gamma$.

\subsection{The recovery algorithm}
\label{sec:recovery-algo}

The recovery algorithm first combines the locator from \Cref{sec:locator} with
the residual norm estimate from \Cref{sec:least-squares} to find the heavy
frequencies.
It then estimates the coefficients on the identified set of frequencies in a
final coefficient estimation step.

Each locator call searches the current residual.  We simulate a point
query to its time signal $\R$ using one query to $\X$.  For residual
spectrum $\Rhat = \Xhat - C_S$, we use
\begin{equation}
\label{eqn:residual-synthesis}
  \R(t)=\X(t)-\sum_{f\in S}C_S(f)\chi_f(t),
\end{equation}
which does not introduce any additional sample complexity.

The algorithm performs the following steps.

\begin{enumerate}
\item Draw $n_{\rm ls}=170s\ceil{\log(2040s)}$ independent uniform points $U =
\{t_1, \dots, t_{n_{\rm ls}}\}$ of $G$, as in \Cref{lem:ls}.
  Initialize support $S=\varnothing$, current fit $C_S=0$, and cumulative cost
  parameter $\kappa_{\rm tot}=0$, and compute $M_\varnothing$ by the
  least-squares routine in \Cref{lem:ls}.
  If $M_\varnothing=0$, return the zero vector; otherwise set the threshold
  $T=M_\varnothing$.
\item Compute $\kappa=M_S^2/T^2$.
  Abort if $\kappa<1$, if $T<T_{\min}$, or if $\kappa_{\rm tot}+\kappa>K$.
  Otherwise the call is considered \emph{accepted};
  replace $\kappa_{\rm tot}$ by $\kappa_{\rm tot}+\kappa$.
\item Independently draw a new batch of samples, according to the sample size
prescribed by \Cref{prop:locator} for norm-to-threshold ratio $\kappa$.
  Using this batch and \eqref{eqn:residual-synthesis}, run the locator on
  $\Rhat = \Xhat - C_S$ with norm bound $M_S$, threshold $T$, and failure
  probability parameter $\gamma$.
  Abort if the locator aborts.
  If the locator terminates successfully, it returns a set of frequencies
  $S_{\rm new}$.
  Abort if $|S \cup S_{\rm new}| \geq 17s$, before constructing any new
  columns;
  otherwise replace $S$ by $S \cup S_{\rm new}$.
\item
  Invoke the least-squares routine in \Cref{lem:ls} with input $U, S$.
  If the least-squares routine aborts, abort the entire algorithm.
  Otherwise, update $C_S$ and $M_S$ based on the output of the least-squares
  routine.
  If $M_S=0$, return $C_S$ without entering step 5.
  If $T=T_{\min}$, jump to step 5 and leave the loop.
  Otherwise, update the threshold $T$ using
  \begin{equation}
\label{eqn:threshold-update}
    T\leftarrow\max\left\{T_{\min},\min\left\{\frac{T}{2},M_S\right\}\right\}
  \end{equation}
  and return to step 2.
\item
  Draw the final coefficient estimation sample and estimate the coefficients on
  $S$ as in \Cref{sec:cleanup}.
  Return the estimated values for coefficients in $S$, and zero outside $S$.
\end{enumerate}

The stopping rule for a zero residual estimate in steps 1 and 4 can apply only
when $\tau=0$, since $M_S\geq3b_{\rm ls}\bar\sigma>0$ whenever $\tau>0$.
Conversely, when $\tau=0$ we have $T_{\min}=0$, so the loop is left only
through this rule, and the final coefficient estimation step is never reached.

We may assume that an abort returns the zero vector.
The check in step 2 ensures $\kappa_i \ge 1$, and the sum of all accepted
$\kappa_i$ is at most $K$.
Likewise, the support-size limit ensures that fewer than $17s$ least-squares
columns are constructed on every execution.
These bounds hold on every execution, without conditioning on a good event.
They make the failure-probability bound in \Cref{prop:adaptive-upper-tail}, the
query allocation analysis in \Cref{sec:tapes}, and
the resource consumption analysis in \Cref{sec:resources} valid on every
execution.

\subsection{Analysis of support identification}
\label{sec:discovery}

For the analysis, index the passes through step 2 as rounds $i=1,2,\ldots$;
round $i$, if not aborted, makes call $i$ to the locator.
Let $S_i$ denote the support \emph{before} call $i$, and let $T_i$, $M_{S_i}$,
and $\kappa_i=M_{S_i}^2/T_i^2$ be the corresponding quantities.
Let $\Rhat_i = \Xhat - C_{S_i}$ be the residual considered by call $i$.
Call $i$ \emph{succeeds} if its locator does not abort and returns a list
containing every $f$ with $|\Rhat_{i}(f)|\geq T_i$ and only frequencies with
$|\Rhat_{i}(f)|>T_i/2$.
We also use the heavy subset $\Omega = \{f : |\Xhat(f)| > \tau/4\}$ from
\Cref{lem:ls}.

The next lemma guarantees that, as long as no locator has failed, the checks
stay inactive and the next locator receives a valid input.
\begin{lemma}[Validity before the first failure]\label{lem:charging}
Suppose the least-squares sample $U$ falls into the event of \Cref{lem:ls}.
Let $n\geq1$, and suppose that
the algorithm reaches round $n$ and that calls $1,\ldots,n-1$ all succeeded.
Then:
\begin{enumerate}
\item $S_1\subseteq S_2\subseteq\cdots\subseteq S_n\subseteq \Omega$;
\label{item:charging-1}
\item $T_n > 0$, $T_{\min}\leq T_n\leq M_{S_n}$, and
$\norm{\Xhat-C_{S_n}}_2\leq M_{S_n}$, so call $n$
  receives a valid norm promise; \label{item:charging-2}
\item the ratios of the first $n$ rounds satisfy \label{item:charging-3}
  \begin{equation}
\label{eqn:charging}
    \sum_{i=1}^n\kappa_i<30838s<K.
  \end{equation}
\end{enumerate}
\end{lemma}

\Cref{lem:charging} implies that if the least-squares sample is good and
every locator call succeeds, then the algorithm terminates with the
following support guarantee.
\begin{corollary}[Successful support identification]\label{cor:discovery}
Suppose that the event of \Cref{lem:ls} holds and that every call made by the
algorithm succeeds.
Then the algorithm terminates without aborting, and the loop ends after fewer
than $30838s$ rounds.
If $\tau>0$, the final support satisfies $S\subseteq \Omega$ and contains every
$f$ with $|\Xhat(f)|\geq T_{\min}$.
If $\tau=0$, the algorithm returns $\Xhat$.
\end{corollary}

\begin{proof}
We prove the non-abort property by induction on the number of rounds $n$.
Suppose that the algorithm starts round $n$ and that calls $1,\ldots,n-1$ all
succeeded without aborting.
Thus \Cref{lem:charging} applies.

In step 2 of round $n$, item~\ref{item:charging-2} of \Cref{lem:charging} gives
$\kappa_n \geq 1$ and $T_n \geq T_{\min}$.
The earlier rounds were accepted, so the cumulative cost parameter is
$\sum_{i\leq n}\kappa_i$, which is below $K$ by \eqref{eqn:charging};
hence round $n$ is accepted.
Call $n$ succeeds by hypothesis, so its locator does not abort.
By soundness of the locator, a returned frequency $f$ outside $S_n$ satisfies
$|\Xhat(f)|>T_n/2\geq\tau/4$, so it lies in $\Omega$;
together with item~\ref{item:charging-1} of \Cref{lem:charging}, the union $S_n
\cup S_{\rm new}$ formed in step 3 is contained in $\Omega$, and therefore has
size less than $17s$.
Since the good event of \Cref{lem:ls} holds, the least-squares call in step 4
does not abort.
Thus the algorithm completes round $n$ without aborting.

We next bound the number of rounds and prove the correctness of $S$.
We have $S \subseteq \Omega$ by the preceding paragraph.
Since every $\kappa_i\geq1$, \eqref{eqn:charging} shows that the total number
of rounds $n^*$ satisfies $n^*<30838s$.
If the algorithm stops already in step 1, then $M_\varnothing=0$, and
\eqref{eqn:ls-certificate} with $S=\varnothing$ gives $\norm{\Xhat}_2\leq0$, so
the returned zero vector equals $\Xhat$; this can happen only if $\tau=0$.
Otherwise, without aborts, the loop can be left only in step 4, either by the
stopping rule for a zero residual estimate or because $T_{n^*}=T_{\min}$.

If $\tau=0$, then $T_{\min}=0<T_{n^*}$ by item~\ref{item:charging-2} of
\Cref{lem:charging}, so the algorithm stops with $M_S=0$ for the current
support $S\subseteq\Omega$.
By \eqref{eqn:ls-certificate}, $\norm{\Xhat-C_S}_2\leq M_S=0$, so the returned
vector $C_S$ equals $\Xhat$.

If $\tau>0$, the stopping rule for a zero residual estimate never applies, so
$T_{n^*}=T_{\min}$.
Suppose for contradiction that there exists $f$ with $|\Xhat(f)|\geq T_{\min}$
but $f\notin S$.
By completeness of locator call $n^*$, $|(\Xhat-C_{S_{n^*}})(f)| < T_{\min}$
must hold.
However, since $f \notin S$, we have $f \notin S_{n^*}$ and thus
$|(\Xhat-C_{S_{n^*}})(f)| = |\Xhat(f)| \geq T_{\min}$, which is a
contradiction.
Hence the final support contains every $f$ with $|\Xhat(f)|\geq T_{\min}$.
\end{proof}

Now we prove \Cref{lem:charging}.
\begin{proof}[Proof of \Cref{lem:charging}]
\emph{\Cref{item:charging-1} and \Cref{item:charging-2}.}
We first note that $M_{S_i}>T_{\min}$ for all $i\leq n$.
If $\tau>0$, this follows from $M_{S_i}\geq3b_{\rm ls}\bar\sigma>T_{\min}$ by
definition of $M_S$; if $\tau=0$, round $i$ is reached only if
$M_{S_i}>0=T_{\min}$, by the stopping rule for a zero residual estimate.

Since $T_1 = M_{S_1} = M_\varnothing$ and round $1$ is reached, we have $T_1 >
0$ and $T_{\min} \le T_1 \le M_{S_1}$.
The update \eqref{eqn:threshold-update} in round $i$ ensures $T_{i+1} > 0$ and
$T_{\min} \le T_{i + 1} \le M_{S_{i + 1}}$, so we have $T_i > 0$ and
$T_{\min}\leq T_i\leq M_{S_i}$ for every round $i\leq n$.
This completes the first half of \Cref{item:charging-2}.

The nested inclusions in \Cref{item:charging-1} follow from the support
update in step 3.
For any $i \le n - 1$ and frequency $f \in S_{i + 1} \setminus S_{i}$,
$|\Xhat(f)| > T_{i}/2 \ge T_{\min}/2 = \tau/4$, so $f \in \Omega$.
The first inequality here is due to the soundness of the locator, given that
calls $1, \ldots, n - 1$ are successful.
By induction, $S_i\subseteq \Omega$ for every $i\leq n$, which completes
\Cref{item:charging-1}.

For the remaining assertion in \Cref{item:charging-2}, the norm bound
$\norm{\Xhat-C_{S_n}}_2\leq M_{S_n}$ follows from $S_n \subseteq \Omega$ and
\eqref{eqn:ls-certificate} in \Cref{lem:ls}.

\emph{\Cref{item:charging-3}.}
We first record the geometric decrease of the thresholds:
\begin{claim}
\label{clm:threshold-geometric}
  For $n \ge 2$, we have $T_n < T_{n - 1}$, and $T_i \le T_{i - 1} / 2$ for
  every $i = 2, \ldots, n - 1$.
\end{claim}
Indeed, $T_i \le T_{i - 1} / 2$ whenever $T_i > T_{\min}$, as holds for
$i = 2, \ldots, n - 1$ by the loop-exit condition in step 4.
For the strict inequality $T_n<T_{n-1}$, note that round $n$ is reached only if
$T_{n-1}>T_{\min}$, by the same loop-exit condition;
since also $T_{n-1}>0$, \eqref{eqn:threshold-update} gives
$T_n\leq\max\{T_{\min},T_{n-1}/2\}<T_{n-1}$.

Another claim controls $|\Xhat(f)|$ by thresholds in the rounds where $f$ is
not yet identified.
\begin{claim}
\label{clm:coordinate-charge}
  Fix any $f\in \Omega$ and consider the rounds $i\leq n$ for which $f\notin
  S_i$.
  These rounds form a nonempty initial segment $1, \ldots, \ell_f$.
  Then $|\Xhat(f)| < 2T_{\ell_f}$, and if $\ell_f \ge 2$, we also have
  $|\Xhat(f)| < T_{\ell_f - 1}$.
\end{claim}
If $\ell_f=1$, the residual certificate gives
$|\Xhat(f)|\leq M_\varnothing=T_1<2T_1$.
Suppose $\ell_f\geq2$.  Since $f$ is still missing, completeness of the
preceding successful call gives $|\Xhat(f)|<T_{\ell_f-1}$.
The current residual certificate also gives
$|\Xhat(f)|\leq M_{S_{\ell_f}}<2M_{S_{\ell_f}}$.
Combining these two strict upper bounds and using the threshold update,
\[
  |\Xhat(f)|
  <2\min\{T_{\ell_f-1}/2,M_{S_{\ell_f}}\}
  \leq2T_{\ell_f}.
\]
This proves both assertions of \Cref{clm:coordinate-charge}.

Now we bound the sum of the ratios $\kappa_i$ using \eqref{eqn:ls-certificate}
in \Cref{lem:ls}:
\begin{align}
  \sum_{i=1}^n\kappa_i & = \sum_{i=1}^n\frac{M_{S_i}^2}{T_i^2} \le \sum_{i =
  1}^n \frac{12 \norm{\Xhat_{\Omega \setminus S_i}}_2^2 + 50b_{\rm
  ls}^2\bar{\sigma}^2}{T_i^2} \notag \\
  & = 12\sum_{i = 1}^n \sum_{f \in \Omega \setminus S_i}
  \frac{|\Xhat(f)|^2}{T_i^2} + 50b_{\rm ls}^2\bar{\sigma}^2 \sum_{i = 1}^n
  T_i^{-2} \notag \\
  & = 12\sum_{f \in \Omega} \sum_{i = 1}^{\ell_f} \frac{|\Xhat(f)|^2}{T_i^2} +
  50b_{\rm ls}^2\bar{\sigma}^2 \sum_{i = 1}^n T_i^{-2}.
  \label{eqn:coordinate-charge-sum}
\end{align}
The first summation in \eqref{eqn:coordinate-charge-sum} can be bounded by a
geometric series:
\begin{align}
  \sum_{i = 1}^{\ell_f} \frac{|\Xhat(f)|^2}{T_i^2}
  < 4 + \sum_{i = 1}^{\ell_f - 1} \frac{T_{\ell_f - 1}^2}{T_i^2}
  \le 4 + \sum_{i = 1}^{\ell_f - 1} 4^{-(\ell_f - 1 - i)} < \frac{16}{3}.
  \label{eqn:coordinate-charge-bound}
\end{align}
Here the first inequality uses \Cref{clm:coordinate-charge} and the second
inequality uses \Cref{clm:threshold-geometric}.

If $\tau=0$, then $\bar\sigma=0$, the second summation in
\eqref{eqn:coordinate-charge-sum} does not contribute,
and \eqref{eqn:coordinate-charge-bound} gives
$\sum_{i=1}^n\kappa_i<64\cdot17s<30838s$ using $|\Omega|<17s$.
Suppose now that $\tau>0$.
A similar argument then applies to the second summation in
\eqref{eqn:coordinate-charge-sum}:
\begin{align}
  \sum_{i = 1}^n T_i^{-2}
  \le \frac{1}{T_{\min}^2} + \sum_{i = 1}^{n - 1} T_i^{-2}
  \le \frac{1}{T_{\min}^2} \left(1 + \sum_{i = 1}^{n - 1} 4^{-(n - 1 -
  i)}\right)
  < \frac{7}{3T_{\min}^2}. \label{eqn:inverse-threshold-sum-bound}
\end{align}

Combining \eqref{eqn:coordinate-charge-sum},
\eqref{eqn:coordinate-charge-bound}, and
\eqref{eqn:inverse-threshold-sum-bound} gives
\begin{align*}
  \sum_{i=1}^n\kappa_i
  &< 12\sum_{f \in \Omega} \frac{16}{3} + 50b_{\rm ls}^2\bar{\sigma}^2
  \frac{7}{3T_{\min}^2}
  = 64|\Omega| + \frac{350b_{\rm ls}^2\bar{\sigma}^2}{3T_{\min}^2}.
\end{align*}
Substituting $|\Omega| < 17s$, $\bar{\sigma}^2 = s\tau^2$, $b_{\rm ls} =
\sqrt{255}/2$, and $T_{\min} = \tau/2$ gives
\begin{align*}
  \sum_{i=1}^n\kappa_i
  < 64 \cdot 17s + \frac{350(\sqrt{255}/2)^2 s \tau^2}{3(\tau/2)^2}
  = 1088s + \frac{350 \cdot 255 s}{3}
  =30838s,
\end{align*}
which finishes the proof of \Cref{item:charging-3}.
\end{proof}

\subsection{Final coefficient estimation and analysis}
\label{sec:cleanup}

The final coefficient estimation sample consists of $r_c$ groups of $v_c$
independent uniform points $t_{i, j}$ from $G$, where
\begin{equation}
\label{eqn:cleanup-bank}
  v_c=\ceil{16384b_{\rm ls}^2s},
  \qquad
  r_c=2\ceil{\log(17s/\delta_{\rm est})}+1,
\end{equation}
and $i\in[r_c]$, $j\in[v_c]$.
The samples are drawn after the support identification phase has fixed the
final support $S$ and fit $C_S$, and the same groups are used for every $f\in
S$.
This step is reached only when $\tau>0$, which we assume throughout this
subsection.

We set $\Xtilde(f) = 0$ for every $f\notin S$, and estimate the coefficients on
$S$ by a median-of-means procedure similar to the verification in
\Cref{prop:locator} and the testing in \Cref{lem:multiscale}.
For each $f\in S$ and each group $i\in[r_c]$, compute the group mean
\[
  \mu_i(f) = \frac{1}{v_c}\sum_{j=1}^{v_c} \R(t_{i,j})\conj{\chi_f(t_{i,j})},
\]
where $\R$ is the residual signal computed by \eqref{eqn:residual-synthesis}.
Then take the coordinatewise median $\lambda(f)$ over the real and imaginary
parts of the $r_c$ group means.
Finally, take $\Xtilde(f) = \lambda(f) + C_S(f)$.

\begin{lemma}[Final coefficient estimation]\label{lem:cleanup}
Fix an execution of the support identification phase on which the event of
\Cref{lem:ls}
holds and the loop ends without an abort, with final support $S\subseteq
\Omega$
containing every $f$ with $|\Xhat(f)|\geq T_{\min}$.  Conditional on these
observations and decisions, the
output $\Xtilde$ satisfies
\[
  |\supp(\Xtilde)|<17s,
  \qquad
  \norm{\Xhat-\Xtilde}_\infty<\frac\tau2,
\]
with probability at least $1 - \delta_{\rm est}/8$ over the final coefficient
estimation sample.
\end{lemma}

\begin{proof}
The support bound is immediate: $\supp(\Xtilde)\subseteq S\subseteq \Omega$ and
$|\Omega|<17s$.  We proceed to show the $\ell_\infty$ bound.

Let $S^*_s = \supp(\HT_s(\Xhat))$ be the best-$s$ support.
At most $s$ coordinates of $G \setminus S$ lie in $S^*_s$, and due to the
assumption on the support identification result, each contributes energy less
than $\tau^2/4$.
Meanwhile, by the tail-norm bound \eqref{eqn:upper-promise}, the coordinates of
$G \setminus S$ outside $S^*_s$ contribute at most $\sigma_s(\Xhat)_2^2\leq
s\tau^2$.
Hence
\begin{equation}
\label{eqn:final-tail}
  \norm{\Xhat_{G \setminus S}}_2^2 \leq \frac{s\tau^2}{4} + s\tau^2 =
  \frac54\bar\sigma^2.
\end{equation}
On the event of \Cref{lem:ls}, \Cref{eqn:ls-certificate} gives
\begin{equation}
  \norm{\Xhat-C_S}_2^2\leq M_S^2
  \leq12 \norm{\Xhat_{G \setminus S}}_2^2 + 50b_{\rm ls}^2\bar\sigma^2
  \leq15 \bar\sigma^2 + 50b_{\rm ls}^2\bar\sigma^2 \le 64b_{\rm
  ls}^2\bar\sigma^2.
\end{equation}
Here, the $\norm{\Xhat_{\Omega \setminus S}}_2^2$ term in the original
statement of \Cref{eqn:ls-certificate} is absorbed into $\norm{\Xhat_{G
\setminus S}}_2^2$ since $\Omega \setminus S \subseteq G \setminus S$.

Conditional on the preceding observations and decisions before the final
coefficient estimation sample, fix $f\in S$.
The $r_cv_c$ values $\R(t_{i,j})\conj{\chi_f(t_{i,j})}$ are independent, each
with mean $(\Xhat-C_S)(f)$, and by \eqref{eqn:one-sample} each has
\[
  \E_t\bigl[|\R(t)\conj{\chi_f(t)}-(\Xhat-C_S)(f)|^2\bigr] \leq
  \norm{\Xhat-C_S}_2^2 \leq 64b_{\rm ls}^2\bar\sigma^2 = 64b_{\rm ls}^2s\tau^2.
\]
Since $r_c$ is odd, we apply item~\ref{item:median-means} of \Cref{lem:median}
with
\[
  r=r_c,\quad v=v_c,\quad x=(\Xhat-C_S)(f),\quad \nu=64b_{\rm
  ls}^2s\tau^2,\quad \omega=\frac{\tau}{4\sqrt2}.
\]
Given $v_c=\ceil{16384b_{\rm ls}^2s}$, this implies $\sqrt2\,\omega=\tau/4$ and
$p\leq 2048b_{\rm ls}^2s/v_c\leq1/8$.
Thus we have $(4p)^{(r_c+1)/2}\leq2^{-(r_c+1)/2}$ and
\begin{align}
  \Prob\left[|\Xtilde(f) - \Xhat(f)| \geq \frac{\tau}{4}\right] & =
  \Prob\left[|\lambda(f) - (\Xhat-C_S)(f)| \geq \frac{\tau}{4}\right] \notag\\
  & \leq \frac14\cdot2^{-(r_c+1)/2} = 2^{-\ceil{\log(17s/\delta_{\rm est})}-3}
  \leq \frac{\delta_{\rm est}}{8\cdot17s}. \label{eqn:cleanup-majority}
\end{align}

Applying a union bound over all $f \in S$ shows that, conditional on the good
event of \Cref{lem:ls} and successful support identification,
\[
  \Prob\left[\norm{\Xhat - \Xtilde}_\infty \geq \frac{\tau}{2}\right] \leq
  \Prob\left[\exists f \in S : |\Xtilde(f) - \Xhat(f)| \geq
  \frac{\tau}{4}\right] \leq \frac{\delta_{\rm est}}{8\cdot17s} \cdot |S| <
  \frac{\delta_{\rm est}}8,
\]
where the first inequality uses the fact that, for every $f \notin S$,
$\Xtilde(f) = 0$ and $|\Xhat(f)| < T_{\min} = \tau/2$.
\end{proof}

\subsection{Correctness of the recovery algorithm}

\begin{proposition}[Correctness of coordinatewise recovery with adaptive
sampling]
\label{prop:adaptive-upper-tail}
Assume \eqref{eqn:upper-promise} with $\tau\geq0$.
The recovery algorithm in \Cref{sec:recovery-algo} fails with probability at
most
\begin{equation}
\label{eqn:upper-failure-bound}
  \delta_{\rm ls}+\delta_{\rm loc}+\frac{\delta_{\rm est}}8
  =\frac{17}{240}<\frac1{10}.
\end{equation}
On success, it returns a vector $\Xtilde$ satisfying
\begin{equation}
\label{eqn:adaptive-conclusion}
  |\supp(\Xtilde)|<17s,
  \qquad
  \norm{\Xhat-\Xtilde}_\infty\leq\frac\tau2.
\end{equation}
\end{proposition}

\begin{proof}
Let $\mathcal E_{\rm ls}$ be the good event of \Cref{lem:ls}.
It depends only on the least-squares sample and fails with probability at most
$\delta_{\rm ls}$.

For $i\geq1$, let $\mathcal F_i$ be the event that
$\mathcal E_{\rm ls}$ holds, calls $1,\ldots,i-1$ are accepted and succeed,
and call $i$ is accepted and fails.  These first-failure events are
disjoint.

Condition on an execution before call $i$ in which the least-squares
sample is good, all preceding calls succeeded, and call $i$ is accepted.
By \Cref{lem:charging}, the residual satisfies
$\norm{\Xhat-C_{S_i}}_2\leq M_{S_i}$ and $0<T_i\leq M_{S_i}$.
The fresh batch therefore allows \Cref{prop:locator} to be applied with
failure parameter $\gamma$.  The conditional probability that call $i$
fails is at most $\gamma$.  Averaging over preceding histories gives
$\Prob[\mathcal F_i]\leq\gamma$.
Since at most $K$ calls are accepted on every execution,
$\mathcal F_i=\varnothing$ for $i>K$, and hence
\[
  \Prob\left[\bigcup_i\mathcal F_i\right]
  \leq K\gamma=\delta_{\rm loc}.
\]

On $\mathcal E_{\rm ls}\setminus\bigcup_i \mathcal{F}_i$, every invoked locator
succeeds, and \Cref{cor:discovery} applies.
If $\tau=0$, the algorithm returns $\Xhat$, and $|\supp(\Xhat)|\leq s<17s$; the
failure probability is at most $\delta_{\rm ls}+\delta_{\rm loc}<1/10$.
If $\tau>0$, no check is activated, and the loop terminates with $S\subseteq
\Omega$ containing every $f$ with $|\Xhat(f)|\geq T_{\min}$.
In particular we have $|S|\leq|\Omega|<17s$.
Conditional on such an execution, the final coefficient estimation sample is
fresh, so by \Cref{lem:cleanup} the output satisfies
\eqref{eqn:adaptive-conclusion} except with probability less than $\delta_{\rm
est}/8$.
The union bound now gives \eqref{eqn:upper-failure-bound}.
\end{proof}

\begin{remark}
   On every execution, the recovery algorithm draws $O(s\log N\log s)$ samples:
   by \Cref{prop:locator}, call $i$ uses $O(\kappa_i(\log
   N)\log(2\kappa_i)+\kappa_i\log(\kappa_i(\log N)/\gamma))$ samples,
   and the budget check gives $\sum_i\kappa_i\leq K=O(s)$.
   \Cref{sec:tapes} shows that fixing all samples in advance costs only a
   constant factor, so nonadaptivity follows directly once the adaptive sample
   bound is established.
\end{remark}

\section{Constructing the nonadaptive query set}
\label{sec:tapes}

\Cref{sec:schedule} analyzed the recovery algorithm with fresh samples.
At each locator call, the algorithm chooses the batch size from preceding
answers and then draws a fresh batch with the prescribed distribution.
Points within a tree search block share an anchor and need not be
independent.  The ratio $\kappa_i$ determines the batch size before any
of its locations or answers is revealed.

To remove this dependence, we choose all query locations before decoding
and store the answers in indexed query sets.  As in \Cref{sec:model},
these are multisets: the same location may occur more than once.
Each sample request reads the next entries of the corresponding query
set.  The decoding decisions remain adaptive, but they determine which
stored entries to \emph{read}; all locations have already been \emph{queried}.

The same query sets also serve the two variants in
\Cref{sec:tail-free,sec:relative-additive}.
These variants change the thresholds, the residual norm estimate, and the final
step.
However, they find frequencies by the same combination of least-squares fits
and locator calls with resource checks, as in \Cref{sec:schedule}.
We therefore define a class of procedures that contains all three algorithms,
and we prove the query set and cost bounds once for the whole class.

Throughout this section, $s$ is a fixed integer with $2\leq s<N$.
In \Cref{sec:schedule}, it is the sparsity parameter of \Cref{thm:upper-tail}.
The tail-free variants choose it as a function of sparsity $k$ and the accuracy
parameters.
The parameters determining the query allocation are functions of $s$ and $\log
N$ only; none of them depends on $\tau$ or on the signal.
Specifically, we use the parameters of \eqref{eqn:schedule-parameters} that do
not involve $\tau$, so the same allocation also applies to the tail-free
variants:
\[
  K=31000s,\qquad \gamma=\frac{\delta_{\rm loc}}{K}=\frac1{30K},\qquad n_{\rm
  ls}=170s\ceil{\log(2040s)},
\]
together with $r_c$ and $v_c$ from \eqref{eqn:cleanup-bank}.

\subsection{Query sets}
\label{sec:tapes-def}

We first specify the batch used in step 3 of the recovery algorithm.
For $\kappa\geq1$, put $\kappa'=2^{\ceil{\log\kappa}}$.
Recall that the binary tree has depth $ad=\log N$.
Set $D(\kappa,N,\gamma)$, $v_0(\kappa)$, and $q_v(\kappa)$ to be the quantities
$D$, $v_0$, and $q_v$ of \eqref{eqn:potential-parameters} and
\eqref{eqn:validation} for this $\kappa$.

A \emph{tree search block of depth $j$ and resolution $w$} is built by taking
an anchor $t\in G$ together with three groups of $C_0w$ uniform offsets $z\in
K_j$.
The anchor and all offsets are independent and uniformly distributed.
The block contains $3C_0w$ query points in the form $t+z$, with $z$ ranging
over all offsets in the three groups.

By \Cref{lem:multiscale} and \Cref{prop:locator}, the batch for launching the
locator at ratio $\kappa$ consists of the following independent components:
\begin{itemize}
  \item tree search blocks: for every repetition $\ell\in\{1,\ldots,5\}$, depth
  $j\in\{1,\ldots,\log N\}$, and resolution $w\in\mathcal{W}_{\kappa'}$,
    there should be exactly $A_0\kappa'/w$ tree search blocks of depth $j$ and
    resolution $w$;
  \item verification points: $q_v(\kappa)$ groups of $v_0(\kappa)$ uniform
  points from $G$.
\end{itemize}
At each repetition and depth, the multiscale test uses the corresponding tree
search blocks across all candidate nodes at the same depth.
The final verification uses the verification points.

Put $K'=2^{\ceil{\log K}}$ and
\begin{equation}
\label{eqn:q-star}
  q_*=q_v(K)=8\ceil{\log(3D(K,N,\gamma)/\gamma)}+1.
\end{equation}
Since $D(\kappa,N,\gamma)$ is nondecreasing in $\kappa$, we have
$q_v(\kappa)\leq q_*$ whenever $\kappa\leq K$.

The query sets are defined as the following independent sequences:
\begin{itemize}
\item a \emph{least-squares query set} of $n_{\rm ls}$ uniform points of $G$;
\item a \emph{tree search query set} $\mathcal N_{j,\ell,w}$ for every
$j\in\{1,\ldots,\log N\}$, $\ell\in\{1,\ldots,5\}$, and $w\in\mathcal{W}_{K'}$,
  each consisting of $\ceil{2A_0K/w}$ tree search blocks of depth $j$ and
  resolution $w$;
\item a \emph{verification query set} of $2048Kq_*$ uniform points of $G$;
\item a \emph{final-stage query set} of $r_cv_c$ uniform points of $G$, with
$r_c$ and
  $v_c$ as in \eqref{eqn:cleanup-bank}.
\end{itemize}
The ceilings in this allocation round the total capacity up to an
integer.  Each individual locator call still uses the exact block
counts in \Cref{lem:multiscale}; \Cref{lem:capacity} shows that the
allocated capacity suffices for all calls.
All these query counts are functions of $s$ and $\log N$ only.
In particular, the same query sets can be used with any value of $\tau$, or
without any $\tau$.
All query locations are generated and queried before decoding begins.
For simplicity, we assume that duplicate locations are queried repeatedly
instead of being merged.

The total number of query occurrences is
\begin{equation}
\label{eqn:Q-upper}
  Q_{\rm total}
  =n_{\rm ls}+Q_{\rm node}+2048Kq_*+r_cv_c,
  \qquad
  Q_{\rm node}=15C_0\,(\log
  N)\sum_{w\in\mathcal{W}_{K'}}w\ceil{\frac{2A_0K}{w}},
\end{equation}
where $Q_{\rm node}$ accounts for $\log N$ depths, five repetitions, and tree
search blocks
of $3C_0w$ points at resolution $w$.

\begin{lemma}[Total query count]\label{lem:Q-upper}
$Q_{\rm total}=O(s\log N\log s)$.
\end{lemma}

\begin{proof}
Since $\ceil{x}<x+1$, $|\mathcal{W}_{K'}|=1+\log K'$, and $K'<2K$,
\begin{align*}
  Q_{\rm node}
  &\leq15C_0\,(\log N)\Bigl(2A_0K(1+\log
  K')+\sum_{w\in\mathcal{W}_{K'}}w\Bigr)\\
  &\leq15C_0\,(\log N)\bigl(2A_0K(1+\log K')+4K\bigr)
  =O(K\log N\log K),
\end{align*}
which is $O(s\log N\log s)$ because $K=31000s$.
By \eqref{eqn:potential-parameters}, $D(K,N,\gamma)$ is bounded by a fixed
polynomial in $K$ and $\log N$;
and $\gamma^{-1}=30K$.
Therefore $q_*=O(\log(s\log N))$ and $2048Kq_*=O(s\log N\log s)$, using
$\log\log N\leq (\log N)$.
By definition, $n_{\rm ls}=O(s\log s)$.
Finally, \eqref{eqn:cleanup-bank} gives $r_cv_c=O(s\log s)$.
\end{proof}

\begin{remark}[Repeated locations]\label{rem:repeated-locations}
We count repeated locations once per occurrence, so that the number of
point queries is the deterministic quantity $Q_{\rm total}$, computed from
input parameters before collecting the samples.  Since queries are exact, an
implementation may instead query each distinct location once and reuse the
answer.  This changes neither the answers nor the output distribution,
provided the decoder accesses answers only through the entries it reads and
does not use information about repeated locations.
In particular, the proof of \Cref{lem:simulation} is unaffected.
The bound $Q_{\rm total}$ then remains an upper bound on the number of distinct
queries.
\end{remark}

\subsection{Admissible recovery procedures}
\label{sec:admissible}

All three algorithms, i.e.,
the recovery algorithm in \Cref{sec:schedule} and the two variants in
\Cref{sec:tail-free,sec:relative-additive},
maintain a support $S$, the least-squares fit $C_S$ of \Cref{lem:ls}, and a
residual norm estimate $M_S$.
They search the residual $\Xhat-C_S$ by locator calls, and they enlarge $S$ by
the returned frequencies.
They differ only in how they choose thresholds, when they stop, and what they
do after the last call.
The following definition captures what they have in common.

\begin{definition}[Admissible recovery procedure]\label{def:admissible}
Let $\bar\sigma\geq0$ be an arbitrary fixed constant.
An \emph{admissible recovery procedure with parameter $s$} is a procedure that
reads the query sets of \Cref{sec:tapes-def} and has the following properties.
\begin{enumerate}
\item \label{item:adm-ls}
  At initialization, it reads the entire least-squares query set as $U$, and it
  starts with $S=\varnothing$ and $C_\varnothing=0$.
  It computes least-squares fits $C_S$ on $U$ as in \Cref{lem:ls}, at most once
  after each accepted call, and it stops if a Gram matrix is singular.
  The residual norm estimate of a fit is $M_S=2\norm{z_S}_2+3b_{\rm
  ls}\bar\sigma$, where $z_S=\mathbf{b}-A_SC_S$.
\item \label{item:adm-support}
  It enlarges $S$ only by frequencies returned by locator calls.
  It never constructs least-squares columns for a support of $17s$ or more
  frequencies; the size of an enlargement is checked before any of its columns
  is constructed.
\item \label{item:adm-guards}
  Before each locator call on $\Rhat=\Xhat-C_S$ with norm bound $M_S$ and
  threshold $T$, it checks that $T>0$, that $\kappa=M_S^2/T^2\geq1$, and that
  the cumulative cost parameter including $\kappa$ is at most $K$.
  If a check fails, the call is not made.
  A call that passes these checks is \emph{accepted}, and it uses the
  fixed failure probability parameter $\gamma$.
\item \label{item:adm-reading}
  Each accepted call $i$ reads its batch by the following \emph{sequential
  access rule}.
  For every $j$, every $\ell$, and every $w\in\mathcal{W}_{\kappa_i'}$, it
  reads the next $A_0\kappa_i'/w$ tree search blocks from $\mathcal
  N_{j,\ell,w}$.
  It also reads the next $v_0(\kappa_i)q_v(\kappa_i)$ points of the
  verification query set.
  Entries that are read but not used, for instance because the locator aborts,
  are discarded and never used again.
\item \label{item:adm-final}
  It reads the final-stage query set only after its last locator call, in a
  single final step.
\item \label{item:adm-causal}
  Every decision, including the residual, the parameters of each call, and
  whether to stop, depends only on entries already read.
  Unread locations, anchors, offsets, and answers cannot influence it.
\end{enumerate}
The procedure may stop earlier for other reasons, and it aborts if a query set
runs out.
\end{definition}

\begin{remark}[Instances]\label{rem:admissible-instances}
The recovery algorithm of \Cref{sec:recovery-algo}, with each fresh draw
replaced by the corresponding accesses to stored entries, is admissible with
$\bar\sigma=\sqrt s\,\tau$.
Its steps 1--4 satisfy items~\ref{item:adm-ls}--\ref{item:adm-reading}, and its
final coefficient estimation in step 5 is the final step.
Its threshold floor and its stopping rule for a zero residual estimate only
stop it earlier.
The candidate construction in \Cref{sec:tail-free} is admissible with $s=k$ and
$\bar\sigma=0$.
In \Cref{sec:relative-additive}, every decoding run is an instance of the
recovery algorithm.
\end{remark}

\begin{lemma}[Sufficiency of the query allocation]\label{lem:capacity}
The allocated query entries suffice for every admissible recovery
procedure on every execution.
\end{lemma}

\begin{proof}
The least-squares and final-stage query sets are read once, in full.
Their allocated query counts suffice by definition.
By item~\ref{item:adm-guards} of \Cref{def:admissible}, every accepted call
satisfies $\kappa_i\geq1$, and summing over accepted calls gives
$\sum_i\kappa_i\leq K$ on every execution.
Therefore, $\kappa_i'=2^{\ceil{\log\kappa_i}}\leq K'$, so
$\mathcal{W}_{\kappa_i'}\subseteq\mathcal{W}_{K'}$ and every tree search query
set that call $i$ reads is well-defined.

Fix any $j \in \{1,2,\dots,\log N\}$, $\ell \in \{1, \dots, 5\}$, and $w \in
\mathcal{W}_{K'}$.
Call $i$ reads from $\mathcal N_{j,\ell,w}$ only if $w\leq\kappa_i'$, in which
case it reads $A_0\kappa_i'/w$ blocks.
Since $\kappa_i'\leq2\kappa_i$, the total number of blocks read from $\mathcal
N_{j,\ell,w}$ is at most
\[
  \frac{A_0}{w}\sum_i\kappa_i'\leq\frac{2A_0}{w}\sum_i\kappa_i
  \leq\frac{2A_0K}{w}\leq\ceil{\frac{2A_0K}{w}}.
\]
For the verification query set,
$v_0(\kappa_i)=1024\ceil{\kappa_i}\leq2048\kappa_i$ and
$q_v(\kappa_i)\leq q_*$, so the total is at most
$2048q_*\sum_i\kappa_i\leq2048Kq_*$.
\end{proof}

\begin{lemma}[Equivalence to fresh sampling]\label{lem:simulation}
Every admissible recovery procedure has the same joint distribution
of revealed batches and output as with fresh sampling.
In particular, conditional on the observations and decisions before a batch is
read, that batch has its prescribed fresh distribution.
\end{lemma}

\begin{proof}
Compare a procedure run on the query sets with the same procedure
drawing each batch when it is requested.  Its decisions and output are
functions of the revealed samples; since $\X$ is fixed, each answer is a
deterministic function of its location.  Call the least-squares sample,
the sample for one locator call, or the sample for the final stage a
\emph{batch}.
By construction (\Cref{sec:tapes-def} and \Cref{def:admissible}), the entries
read in each step have the same count, types, and roles as the samples drawn in
that step with fresh sampling.
By induction on the number of batches, it suffices to show that, conditionally
on the first $n-1$ batches, the $n$-th batch read from the query sets has the
law of a fresh batch of the same composition.

Condition on the contents of the first $n-1$ batches.
Reading is sequential, and the composition of each batch is a function of the
earlier batches (through the ratios $\kappa_i$).
Hence the conditioning fixes the indexed entries that have been read, and which
entries form the $n$-th batch.
Furthermore, \Cref{lem:capacity} guarantees that the entries forming the $n$-th
batch do exist.

The conditioning event therefore only prescribes the contents of a fixed set of
entries.
Since entries are accessed in their fixed order and no entry is read twice
within a run, the $n$-th batch consists of entries outside this set.
Because the indexed blocks and the individual uniform samples are
mutually independent, and each has exactly the distribution used in fresh
sampling (\Cref{sec:tapes-def}),
the $n$-th batch is conditionally distributed as a fresh batch.
\end{proof}

\subsection{Cost of an admissible recovery procedure}

\begin{proposition}[Cost of an admissible recovery procedure]
\label{prop:shared-runtime}
For an admissible recovery procedure, generating all query sets and
performing its locator calls, residual computations, support enlargements,
least-squares fits, and residual norm estimates can be implemented using
\begin{equation}
\label{eqn:shared-runtime}
  O\bigl(s^{7/2}(\log N)^{9/4}\log(s\log N)\bigr)
\end{equation}
exact operations on every execution.
This count covers these common operations.  It excludes any additional
computation used to choose thresholds or decide when to stop, and the
final step; each algorithm accounts for these costs separately.
It assumes neither a tail-norm bound nor the validity of any
locator norm bound.
\end{proposition}

\begin{proof}
Recall $\gamma=1/(30K)$.
Throughout, $\kappa_i$ ranges over the accepted calls of an execution.
By item~\ref{item:adm-guards} of \Cref{def:admissible}, $\kappa_i\geq1$ and
$\sum_i\kappa_i\leq K$ on every execution,
including executions on which some locator fails.

\emph{Locator calls.}
By \eqref{eqn:potential-parameters}, the search-tree size limit satisfies
\begin{equation}
\label{eqn:D-bound}
  D(\kappa,N,\gamma)
  =\frac{42\kappa\log N}{\theta\gamma}
    \left(\frac{3\kappa\log N}{\gamma}\right)^{1/4}
  =O\bigl(\kappa^{5/4}(\log N)^{5/4}\gamma^{-5/4}\bigr).
\end{equation}
Let $D_i=D(\kappa_i,N,\gamma)$.
Since $\gamma^{-1}=30K$, $K = 31000s$, and
$\sum_i\kappa_i^p\leq(\sum_i\kappa_i)^p\leq K^p$ for $p\geq1$, we have
\begin{equation}
\label{eqn:amortized-visits}
\begin{aligned}
  \sum_iD_i
    &=O\bigl(K^{5/2}(\log N)^{5/4}\bigr)
     =O\bigl(s^{5/2}(\log N)^{5/4}\bigr),\\
  \sum_i\kappa_iD_i
    &=O\bigl(K^{7/2}(\log N)^{5/4}\bigr)
     =O\bigl(s^{7/2}(\log N)^{5/4}\bigr).
\end{aligned}
\end{equation}

According to the search-tree size limit of \Cref{prop:locator}, call $i$
performs at most $2D_i$ node tests and $D_i$ leaf verifications.
In call $i$, a node test at depth $j$ processes the tree search blocks of depth
$j$ (\Cref{sec:tapes-def}), which contain
\[
  15A_0C_0\,\kappa_i'(1+\log\kappa_i') \leq 30A_0C_0\,\kappa_i(2+\log K) =
  O(\kappa_i\log s)
\]
points.
A leaf verification processes
$v_0(\kappa_i)q_v(\kappa_i)\leq2048\kappa_iq_*=O(\kappa_i\log(s\log N))$
points,
with $q_*$ as in \eqref{eqn:q-star}.
For each point $(t, z)$, the remaining work is to compute $\R(t +
z)\conj{\chi_v(z)}$ and accumulate the group mean.
Using the supplied primitive $2^a$-th root of unity, a single character
evaluation costs $O(\log N)$.
By reusing residual values from \Cref{eqn:residual-synthesis}, as discussed
below, the computation at each point costs $O(\log N)$.
Group means and three-way medians add only constant work per anchor,
and the majority vote over five repetitions adds only constant work per test.

Therefore, call $i$ uses $O((\log N)D_i\kappa_i\log s)$ operations for node
tests.
The median in each leaf verification can be computed in linear time,
so verification also has amortized cost $O(\log N)$ per point.
Hence call $i$ spends $O((\log N)D_i\kappa_i\log(s\log N))$ for leaf
verifications.

By \eqref{eqn:amortized-visits}, all calls together spend
\begin{equation}
\label{eqn:node-time}
  O\Bigl((\log N)\log(s\log N)\sum_i\kappa_iD_i\Bigr)=O\bigl(s^{7/2}(\log
  N)^{9/4}\log(s\log N)\bigr)
\end{equation}
on node tests and leaf verifications.

Represent each node by its depth $j$ and a representative $v$ of
$v+H_j$.  For $j=\ell a+c<ad$, its children have representatives $v$
and $v+2^ce_{\ell+1}$, where $e_{\ell+1}$ is a coordinate vector.
Generate children only from retained nodes, and enforce the search-tree size
limit before appending a node.  Constructing a representative costs
$O(d+a)=O(\log N)$ operations, so traversal and materializing the leaf and
output lists cost $O((\log N)\sum_iD_i)$.
On an arbitrary execution, a returned list can contain $D_i$ frequencies.
Form its union with $S$ by scanning against the fewer than $17s$ frequencies
retained so far, stopping before the union reaches $17s$ and before forming
any new columns.  Each comparison costs $O(\log N)$, so all scans cost
$O(s(\log N)\sum_iD_i)=O(s^{7/2}(\log N)^{9/4})$.
Together with \eqref{eqn:node-time}, this is within
\eqref{eqn:shared-runtime}.

\emph{Computing and reusing residual values.}
Call $i$ reads its batch according to item~\ref{item:adm-reading} of
\Cref{def:admissible}.
At that moment, compute the residual value \eqref{eqn:residual-synthesis} once
for every point occurrence in the batch, with the fit $C_S = C_{S_i}$ plugged
in.
This costs $O(s\log N)$ per occurrence, since $|S_i|<17s$ and each character
evaluation costs $O(\log N)$.
Every node test and leaf verification uses these stored residual values.

Each query set occurrence belongs to at most one call's batch, so the total
residual-computation time is
\begin{equation}
\label{eqn:residual-cache-time}
  O(s(\log N)Q_{\rm total})=O(s^2(\log N)^2\log s).
\end{equation}

\emph{Incremental least squares.}
We implement the least-squares computation of \Cref{lem:ls} with the standard
bordered Cholesky update for least squares with an increasing set of
columns~\cite{ref:GVL13,ref:Bjo96}.
Adding a column to a support of size $p$ costs $O(n_{\rm ls}(\log N))$ to form
the column,
$O(n_{\rm ls}(p+1))$ to maintain its inner products with $\mathbf{b}$ and the
existing columns, and
$O(p^2)$ to append a row to the Cholesky factor.

In exact arithmetic, the bordered Cholesky update succeeds and yields an
invertible factor exactly
when the Gram matrix $A_{S'}^*A_{S'}$ of the new support $S'=S\cup\{f\}$ is
nonsingular~\cite{ref:GVL13};
otherwise we abort, as the least-squares routine of \Cref{lem:ls} does.

The support $S$ grows monotonically, and its size is checked before
constructing columns for an enlargement.  Hence fewer than $17s$ columns
are constructed in total, including an attempted column at which
singularity is detected.
Therefore, adding all columns costs
\begin{equation}
\label{eqn:column-cost}
  O(n_{\rm ls}s(\log N)+n_{\rm ls}s^2+s^3)=O(s^2(\log N)\log s+s^3\log s).
\end{equation}

Each invocation of the least-squares routine recovers $C_S$ by solving two
triangular systems in $O(s^2)$ operations.
After initialization, at most $K$ fits are computed,
so the total cost for solving linear systems is $O(Ks^2) = O(s^3)$, which is
asymptotically dominated by the column addition cost.
Each invocation also outputs the residual norm estimate $M_S$.
Since $A_S^*\mathbf{b}$ is maintained and $\norm{\mathbf{b}}_2^2$ is computed
once in $O(n_{\rm ls})$ operations, the identity
$\norm{z_S}_2^2=\norm{\mathbf{b}}_2^2-\mathbf{b}^*A_SC_S$ yields $M_S$ in
$O(|S|)$ further operations, for a total of $O(Ks)=O(s^2)$.
Taking the square root and adding $3b_{\rm ls}\bar\sigma$ require
constant additional work.

\emph{Constructing the query set.}
A uniform point of $G$, or a uniform element of $K_j$, is generated from
at most $ad$ independent fair bits in $O(\log N)$ operations, as in
\Cref{sec:model}.
Hence the query sets are generated in $O((\log N)Q_{\rm total})$ operations.
The input parameters, including $Q_{\rm total}$, are computed at a cost
absorbed by the other bounds.

All displayed costs are bounded by \eqref{eqn:shared-runtime}.
They use only the properties in \Cref{def:admissible}, so the bound continues
to hold after a locator failure or when $M_S$ is not a valid norm bound.
\end{proof}

\section{Proof of the main theorem}
\label{sec:resources}

We prove \Cref{thm:main-dyadic-intro} using the locator,
least-squares fits, and query sets developed above.
We first complete the proof of \Cref{thm:upper-tail}, including the case
$\tau=0$,
by running the recovery algorithm on the query sets and applying the
cost bound of \Cref{prop:shared-runtime}.
Hard thresholding then proves \Cref{thm:main-dyadic-intro}.

\subsection{Proof of \texorpdfstring{\Cref{thm:upper-tail}}{the coordinatewise
recovery theorem}}

We run the recovery algorithm of \Cref{sec:recovery-algo} on the query sets of
\Cref{sec:tapes-def}, with the parameter $s$ of \Cref{thm:upper-tail}.
By \Cref{rem:admissible-instances}, it is then an admissible recovery procedure
with $\bar\sigma=\sqrt s\,\tau$, and its final coefficient estimation is its
final step.
The following proposition adds a full-FFT case for small $N$.

\begin{proposition}[Nonadaptive coordinatewise recovery]
\label{prop:nonadaptive-upper-tail}
Assume \eqref{eqn:upper-promise}.
Consider the following algorithm.
Compute $Q_{\rm total}$ from the input parameters.
If $N = |G| \leq Q_{\rm total}$, query every point of $G$, compute $\Xhat$
exactly,
and return the vector $\Xtilde$ defined by
\[
  \Xtilde(f) = \begin{cases}
    \Xhat(f) & \text{if } |\Xhat(f)|\geq\tau/2,\\
    0 & \text{otherwise}.
  \end{cases}
\]
Otherwise generate and query the query sets, and run the recovery algorithm on
them.

This algorithm fixes all queries before any answer is used, and it makes
$\min\{N,Q_{\rm total}\}$ point queries, counted with multiplicity.
With probability at least $9/10$, it returns $\Xtilde$ satisfying
\eqref{eqn:adaptive-conclusion}.
\end{proposition}

\begin{proof}
The branch is chosen from input parameters before any query.
Let $S^*_s = \supp(\HT_s(\Xhat))$ denote the best-$s$ support.
In the branch with $N \leq Q_{\rm total}$, we have
\[
  |\supp(\Xtilde)| \leq |S^*_s| + |\supp(\Xtilde) \setminus S^*_s| \leq s + 4s
  = 5s < 17s.
\]
The second inequality holds because, for $\tau > 0$, every coordinate in
$\supp(\Xtilde)\setminus S^*_s$
has $|\Xhat(f)|^2\geq\tau^2/4$, while $\sigma_s(\Xhat)_2^2 \le s\tau^2$;
and if $\tau=0$, this set is empty.
Every omitted coefficient has magnitude less than $\tau/2$, so
$\norm{\Xhat-\Xtilde}_\infty\leq\tau/2$ holds deterministically.

In the other branch, all query sets are generated and queried before the
recovery algorithm starts, with total length $Q_{\rm total}$.
By \Cref{lem:simulation}, the output has the same distribution as in
\Cref{prop:adaptive-upper-tail}, which proves the claim.
\end{proof}

\begin{proof}[Proof of \Cref{thm:upper-tail}]
Use the algorithm of \Cref{prop:nonadaptive-upper-tail}.
It fixes every query location and multiplicity before any answer is used, so
the algorithm is nonadaptive.
With probability at least $9/10$ it returns $\Xtilde$ satisfying
\eqref{eqn:upper-conclusion}.
It makes $\min\{N,Q_{\rm total}\}$ point queries, and $Q_{\rm total}\leq Cs\log
N\log s$
for an absolute constant $C$ by \Cref{lem:Q-upper}, which proves
\eqref{eqn:upper-samples}.

In the sparse-recovery case, the algorithm is an admissible recovery procedure,
and its final step is the final coefficient estimation.
The threshold and cost parameter updates take constant work per round.
Hence \Cref{prop:shared-runtime} bounds all work before the final coefficient
estimation by $O(s^{7/2}(\log N)^{9/4}\log(s\log N))$ on every execution,
including executions that abort.
It remains to account for the final coefficient estimation and for the full-FFT
case.

\emph{Final coefficient estimation.}
Compute the residual \eqref{eqn:residual-synthesis} with the final fit $C_S$
once at each of the $r_cv_c$ points of the final-stage query set.
This costs $O(s\log N)$ per point, since $|S|<17s$ and each character
evaluation costs $O(\log N)$, and hence $O(s^2(\log N)\log s)$ in total, since
$r_cv_c=O(s\log s)$.
With these values, each group mean $\mu_i(f)$ costs $O((\log N)v_c)=O(s\log
N)$.
Since $r_c=O(\log s)$ and $|S|=O(s)$, all group means cost $O(s^2(\log N)\log
s)$, and the medians add linear work.
Therefore the final coefficient estimation takes
\begin{equation}
\label{eqn:cleanup-cost}
  O(s^2(\log N)\log s)
\end{equation}
operations.

\emph{Full-FFT case.}
In the branch where all points in $G$ are queried, apply radix-two Fourier
transforms coordinate by coordinate using the supplied root of unity.
On each of the $d$ coordinates, $N/2^a$ radix-two transforms of length $2^a$
are applied, so the total cost is $O(Nad)=O(N\log N)$.
This branch occurs only when $N\leq Q_{\rm total}=O(s\log N\log s)$, so the
total cost is $O(s(\log N)^2\log s)$, including the final thresholding.

Both costs are within $O(s^{7/2}(\log N)^{9/4}\log(s\log N))$, which proves the
operation bound.
\end{proof}

\subsection{Hard thresholding and proof of
\texorpdfstring{\Cref{thm:main-dyadic-intro}}{Theorem 1}}

The following lemma compares two $k$-sparse approximations of $\Xhat$:
the optimal one, $\HT_k(\Xhat)$, and the one obtained by keeping the $k$
largest coordinates of an intermediate spectral estimate $\Xtilde^{\rm pre}$.
In particular, with fixed $\Xhat$ and $k$, the extra approximation error
vanishes linearly as the $\ell_\infty$ perturbation $\Delta \to 0$.

\begin{lemma}[Hard-thresholding stability]\label{lem:hard-thresholding}
For any signal $\X$ on $G$, any $\Xtilde^{\rm pre}\in\C^G$, and $1\leq k<N$,
with $\Delta=\norm{\Xhat-\Xtilde^{\rm pre}}_\infty$, we have
\begin{equation}
\label{eqn:hard-thresholding}
  \norm{\Xhat-\HT_k(\Xtilde^{\rm pre})}_2^2
  \leq\sigma_k(\Xhat)_2^2+4\sqrt k\,\sigma_k(\Xhat)_2\,\Delta+5k\Delta^2.
\end{equation}
\end{lemma}

\begin{proof}
Let $S_*$ and $S_{\rm pre}$ be the selected top-$k$ coordinate sets of
$\Xhat$ and $\Xtilde^{\rm pre}$, respectively.  Set
\[
  A=S_*\setminus S_{\rm pre},\qquad
  B=S_{\rm pre}\setminus S_*,\qquad |A|=|B|=r.
\]
The definition of $S_{\rm pre}$ gives
$\norm{\Xtilde^{\rm pre}_A}_2\leq\norm{\Xtilde^{\rm pre}_B}_2$.
The coordinatewise error on each of these sets is at most $\Delta$, so
\[
  \norm{\Xhat_A}_2
  \leq\norm{\Xhat_B}_2+2\sqrt r\,\Delta.
\]
Write $\sigma=\sigma_k(\Xhat)_2$.
The estimate has squared error at most $k\Delta^2$ on $S_{\rm pre}$.
Outside that set, its error differs from the optimal tail by exchanging
$B$ for $A$.  Consequently,
\begin{align}
  \norm{\Xhat-\HT_k(\Xtilde^{\rm pre})}_2^2
  &\leq\sigma^2+k\Delta^2
       +\norm{\Xhat_A}_2^2-\norm{\Xhat_B}_2^2
       \label{eqn:hard-thresholding-intermediate}\\
  &\leq\sigma^2+k\Delta^2
       +4\sqrt r\,\Delta\norm{\Xhat_B}_2+4r\Delta^2\notag\\
  &\leq\sigma^2+4\sqrt k\,\sigma\Delta+5k\Delta^2.\notag
\end{align}
The last step uses $r\leq k$ and $B\subseteq G\setminus S_*$, so
$\norm{\Xhat_B}_2\leq\sigma$.  The same argument includes $r=0$ and
any ties resolved by the fixed coordinate order.
\end{proof}

We now restate and prove the guarantee with a supplied tail estimate.

\begin{mainresultrestated}[Restated]
Fix a constant $c\geq1$.
Let $a,d\geq1$ be integers, $G=(\Z/2^a\Z)^d$, $N=|G|=2^{ad}$, $\eps \in
(0,1/2]$, and
$2 \leq k < N$.
There is a randomized algorithm which, given point-query access to any fixed
$\X \in \C^G$ and a tail estimate $B$ satisfying
\[
  \sigma_k(\Xhat)_2 \leq B \leq c\,\sigma_k(\Xhat)_2,
\]
uses non-adaptive queries, computes an estimate $\Xtilde^{\rm pre}$
with $O_c(k/\eps^2)$ nonzero coordinates, and returns the $k$-sparse vector
$\Xtilde=\HT_k(\Xtilde^{\rm pre})$.  With probability at least $9/10$,
both
\[
  \norm{\Xhat-\Xtilde^{\rm pre}}_\infty
  \leq\frac{\eps}{\sqrt{k}}\,\sigma_k(\Xhat)_2
\]
and
\[
  \norm{\Xhat-\Xtilde}_2^2
  \leq(1+\eps)\sigma_k(\Xhat)_2^2,
\]
where $\Xhat \in \C^G$ denotes the Fourier transform of $\X$ over $G$
and $\sigma_k(\Xhat)_2$ denotes its best $k$-sparse approximation error in
$\ell_2$.
Moreover, the algorithm makes at most
\[
  \min\!\left\{N,
    O\left(\frac{k}{\eps^2} (\log N) \log(k/\eps)\right)\right\}
\]
queries.
The runtime of the algorithm (assuming exact operations) is at most
\[
  O\!\left(
    \frac{k^{7/2}}{\eps^7}(\log N)^{9/4}
    \log\!\left(\frac{k\log N}{\eps}\right)
  \right) = \mathsf{poly}((k/\eps) \log N).
\]
The hidden constants may depend on $c$.
\end{mainresultrestated}

\begin{proof}[Proof of \Cref{thm:main-dyadic-intro}]
Write $\sigma=\sigma_k(\Xhat)_2$ for analysis only; the algorithm receives $B$.
All implicit constants below may depend on the fixed constant $c$.

If $B=0$, the promise implies $\sigma=0$, so $\Xhat$ is $k$-sparse.
Run the algorithm of \Cref{thm:upper-tail} with $s=k$ and $\tau=0$, denote its
output by $\Xtilde^{\rm pre}$, and return $\Xtilde=\HT_k(\Xtilde^{\rm pre})$.
The intermediate estimate has fewer than $17k$ nonzero coordinates.
With probability at least $9/10$, $\Xtilde^{\rm pre}=\Xhat$ and
$\Xtilde=\Xhat$, since $|\supp(\Xhat)|\leq k$.
Both error bounds therefore hold on this event, with error zero.
With $s=k$, \Cref{thm:upper-tail} uses $\min\{N,O(k\log N\log k)\}$ queries and
$O\bigl(k^{7/2}(\log N)^{9/4}\log(k\log N)\bigr)$ operations, which are within
the bounds of \Cref{thm:main-dyadic-intro} since $\eps\leq1$.
Henceforth suppose $B>0$.

Set
\begin{equation}
\label{eqn:wrapper-parameters}
  s=\ceil{25c^2k/\eps^2},
  \qquad
  \tau=\frac{\eps B}{5c\sqrt k}.
\end{equation}

If $s\geq N$, query every point of $G$ and compute $\Xhat$ exactly
as in the full-FFT case of \Cref{thm:upper-tail}.
Set $\Xtilde^{\rm pre}=\Xhat$ and return
$\Xtilde=\HT_k(\Xtilde^{\rm pre})$.
The intermediate estimate has at most $N\leq s=O_c(k/\eps^2)$
nonzero coordinates and coordinatewise error zero, while
$\norm{\Xhat-\Xtilde}_2^2=\sigma^2$.  Thus both error bounds hold
deterministically in this branch.

Otherwise $s<N$.
Since $\eps \leq 1$, we have $s\geq k$, and therefore
\[
  \sigma_s(\Xhat)_2^2\leq\sigma^2\leq B^2=\frac{25c^2k}{\eps^2}\tau^2
  \leq s\tau^2.
\]
Thus the promise \eqref{eqn:upper-promise} holds.
Run the algorithm of \Cref{thm:upper-tail} once with parameters $s$ and $\tau$,
denote its output by $\Xtilde^{\rm pre}$, and return
$\Xtilde=\HT_k(\Xtilde^{\rm pre})$.
The intermediate estimate has fewer than $17s=O_c(k/\eps^2)$
nonzero coordinates.  With probability at least $9/10$,
\[
  \norm{\Xhat-\Xtilde^{\rm pre}}_\infty
  \leq\frac\tau2
  =\frac{\eps B}{10c\sqrt k}
  \leq\frac{\eps\sigma}{10\sqrt k}
  \leq\frac{\eps\sigma}{\sqrt k},
\]
where we used $B\leq c\sigma$.
This proves the coordinatewise guarantee.  On the same event,
\Cref{lem:hard-thresholding} with $\Delta\leq\tau/2<\tau$ gives
\begin{align*}
  \norm{\Xhat-\Xtilde}_2^2
  &\leq\sigma^2+4\sqrt k\,\sigma\tau+5k\tau^2\\
  &=\sigma^2+\frac{4\eps}{5c}\sigma B+\frac{\eps^2}{5c^2}B^2\\
  &\leq\left(1+\frac{4\eps}{5}+\frac{\eps^2}{5}\right)\sigma^2
  \leq(1+\eps)\sigma^2,
\end{align*}
where we used $B\leq c\sigma$ and $0<\eps\leq1$.
The vector $\Xtilde$ has at most $k$ nonzero coordinates.

Both branches are chosen from input parameters, and the algorithm of
\Cref{thm:upper-tail} makes nonadaptive queries,
so the algorithm for $\ell_2/\ell_2$ recovery is also nonadaptive.
It remains to account for the resources.
In the first branch, $N\leq s=O(k/\eps^2)$, so the $N$ queries and $O(N\log
N)=O((k/\eps^2)\log N)$ operations are within the bounds of
\Cref{thm:main-dyadic-intro}.
In the second branch, $s=\Theta(k/\eps^2)$ gives $\log s=O(\log(k/\eps))$, so
by \Cref{thm:upper-tail} the algorithm makes at most
\[
  \min\!\left\{N,O\left(\frac{k}{\eps^2}(\log N)\log(k/\eps)\right)\right\}
\]
queries.
Moreover, $\log s+\log\log N=O(\log(k/\eps)+\log\log N)=O(\log(k\log N/\eps))$,
so the operation bound of \Cref{thm:upper-tail} becomes
\[
  O\bigl(s^{7/2}(\log N)^{9/4}\log(s\log N)\bigr)
  =O\!\left(\frac{k^{7/2}}{\eps^7}(\log N)^{9/4}\log\!\left(\frac{k\log
  N}{\eps}\right)\right).
\]
Finally, the selection of the best $k$ coordinates can be implemented by
sorting the fewer than $17s$ nonzero entries of $\Xtilde^{\rm pre}$, which
takes $O(s\log N\log s)=O((k/\eps^2)(\log N)\log(k/\eps))$ operations and is
within the same bound.
\end{proof}

\section{Constant-factor recovery without a tail estimate}
\label{sec:tail-free}
The tail estimate in \Cref{thm:main-dyadic-intro} can be removed if we
replace the approximation factor $1+\eps$ by an absolute constant.
The resulting algorithm also covers a zero tail without being told that
the spectrum is exactly sparse.

\begin{theorem}[Recovery without a tail estimate; detailed form of
\Cref{thm:tail-free-intro}]\label{thm:tail-free}
Let $a,d\geq1$ be integers, $G=(\Z/2^a\Z)^d$, $N=2^{ad}$, and
$2\leq k<N$.  There are an absolute constant $C$ and a randomized
algorithm which, given point-query access to any fixed $\X\in\C^G$,
returns a $k$-sparse vector $\Xtilde\in\C^G$ such that, with
probability at least $9/10$,
\[
  \norm{\Xhat-\Xtilde}_2^2
  \leq C\,\sigma_k(\Xhat)_2^2.
\]
Here $\Xhat$ is the Fourier transform of $\X$ over $G$ and
$\sigma_k(\Xhat)_2$ is its best $k$-sparse approximation error
in $\ell_2$.  No signal norm or tail estimate is supplied.
The algorithm makes at most
\[
  \min\{N,O(k(\log N)\log k)\}
\]
nonadaptive point queries and uses at most
\[
  O\bigl(k^{7/2}(\log N)^{9/4}\log(k\log N)\bigr)
\]
exact operations on every execution, in the computational model of
\Cref{sec:model}.  All hidden constants are absolute.
In particular, if $\Xhat$ is $k$-sparse, the algorithm recovers it
exactly with probability at least $9/10$.
\end{theorem}

As before, $\X$ is the queried signal and $\Xhat$ its spectrum.
We impose no tail-norm bound in this section.  The algorithm saves a list of
least-squares fits and then selects a candidate using independent
samples.  Neither step receives a signal norm or a tail estimate.

\subsection{Saving candidates without a stopping scale}

The candidate construction is an admissible recovery procedure
(\Cref{def:admissible}) with $s=k$ and $\bar\sigma=0$.  It uses the
query sets of \Cref{sec:tapes} with $s=k$, the fits $C_S$ of
\Cref{lem:ls}, and the residual norm estimate $M_S=2\norm{z_S}_2$.  It differs
from
the recovery algorithm of \Cref{sec:recovery-algo} in four ways: it
saves every fit as a candidate, it has no threshold floor, it caps the support
at
$5k$, and its final step is candidate selection
(\Cref{lem:tail-free-selection}) instead of final coefficient estimation.
\begin{enumerate}
\item Start with $S=\varnothing$ and $C_\varnothing=0$, and save this
  fit.  If $M_\varnothing=0$, stop.  Otherwise set $T=M_\varnothing$.
\item Compute $\kappa=M_S^2/T^2$.  If the cumulative cost parameter plus
  $\kappa$ exceeds $K$, stop.  Otherwise run the locator on $\Xhat-C_S$
  with norm bound $M_S$, threshold $T$, and failure probability parameter
  $\gamma$,
  and add $\kappa$ to the cumulative cost parameter.
\item Enlarge $S$ by the returned list, and stop if its size would reach
  $5k$.  Otherwise refit on $S$ and append the new fit to the candidate list.
  If $M_S=0$, stop; otherwise update $T\leftarrow\min\{T/2,M_S\}$ and return to
  step 2.
\end{enumerate}
The support cap is compatible with \Cref{def:admissible}, since
$5k<17k$.  The other two checks always pass: the update
rule gives $T\leq M_S$, so $\kappa\geq1$, and $T>0$ because $M_S>0$
whenever step 2 is reached.
A singular fit or a locator abort also stops the candidate construction.
In every case, all previously saved fits are kept.

Without a threshold floor, the budget check is what makes candidate
generation finite on every execution.  Every accepted call has cost parameter
at least one, so there are at most $K$ accepted calls and at most $K+1$
saved fits, each supported on fewer than $5k$ frequencies.
\Cref{lem:tail-free-candidates} shows that, on a good event, an accurate
enough fit is saved before either the budget check or the support cap
stops the run.  Applying $\HT_k$ to every saved fit gives the
candidate list.

\begin{lemma}
\label{lem:tail-free-candidates}
For every fixed $\X$, the list contains a candidate $\Xtilde$ with
\[
  \norm{\Xhat-\Xtilde}_2\leq241\sigma_k(\Xhat)_2
\]
except with probability at most $\delta_{\rm ls}+\delta_{\rm loc}=1/15$.
In particular, when $\Xhat$ is $k$-sparse the list contains $\Xhat$ itself on
this event.
\end{lemma}

\begin{proof}
Let $\sigma=\sigma_k(\Xhat)_2\geq0$.  For analysis only, set
\[
  t_* = \frac{\sigma}{\sqrt k},\qquad
  \Omega=\{f:|\Xhat(f)|>t_*/2\}.
\]
The best-$k$ tail bound gives $|\Omega|<5k$ and
$\norm{\Xhat_{G\setminus\Omega}}_2^2\leq5\sigma^2/4$;
if $\sigma=0$, then $\Omega=\supp(\Xhat)$ and $\Xhat_{G\setminus\Omega}=0$.
Apply \Cref{lem:ls} with $s=k$ and the analysis-only choice
$\tau=2\sigma/\sqrt{k}$.  Its promise holds since
$\sigma_k(\Xhat)_2^2\leq k\tau^2$; its set $\Omega$ and sample
count agree with those above, and $\bar\sigma=2\sigma$.
The samples and ordinary fits do not depend on this choice of $\tau$,
so it supplies no information to the algorithm.
With probability at least $1-\delta_{\rm ls}$, every fit on $S\subseteq\Omega$
is nonsingular and the bounds \eqref{eqn:ls-raw} hold simultaneously.
Condition on this event.  Since $b_{\rm
ls}\bar\sigma=\sqrt{255}\sigma\leq20\sigma$,
these bounds give
\begin{align}
  \norm{\Xhat-C_S}_2
  &\leq\sqrt2(\norm{z_S}_2+20\sigma)+\sqrt{5/4}\,\sigma
   \leq\tfrac32\norm{z_S}_2+30\sigma,
   \label{eqn:tail-free-certificate}\\
  M_S^2
  &\leq12\norm{\Xhat_{\Omega\setminus S}}_2^2+3200\sigma^2.
   \label{eqn:tail-free-energy}
\end{align}
\emph{Good fits and required conditions.}
Call a saved fit $C_S$ \emph{good} if $\norm{\Xhat-C_S}_2\leq120\sigma$,
and \emph{bad} otherwise.
It suffices to show that some saved fit is good, because the last step
of this proof turns a good fit into a candidate with the claimed error.
Bad fits are still useful, because their residual norm estimates are valid norm
bounds.
To see this, let $S\subseteq\Omega$, and suppose that $C_S$ is bad.
Since $M_S=2\norm{z_S}_2$, \eqref{eqn:tail-free-certificate} gives
$\norm{\Xhat-C_S}_2\leq\tfrac34M_S+30\sigma.$
Since $C_S$ is bad, we have $30\sigma<\tfrac14\norm{\Xhat-C_S}_2$.
Substituting this bound gives $\tfrac34\norm{\Xhat-C_S}_2<\tfrac34M_S$.
Therefore
\begin{equation}
\label{eqn:tail-free-valid}
  M_S>\norm{\Xhat-C_S}_2>120\sigma
\end{equation}
holds whenever $S\subseteq\Omega$ and $C_S$ is bad.
In particular, $M_S>0$ for such a fit.
Thus the rule ``stop if $M_S=0$'' can only stop the run at a good fit.

\emph{The initial segment.}
Call a locator call \emph{early} if every fit saved before it is bad
and every earlier call succeeded, in the sense of \Cref{sec:discovery}.
The early calls form an initial segment of the run, which we call the
\emph{initial segment}.
The initial segment ends when a good fit is saved or when a call fails.
We show that every early call satisfies the locator's hypotheses and
that the initial segment is finite and cannot be stopped by a check.
It therefore ends with a good fit or a failed early call.  We then bound
the probability of the latter event.

\emph{Two invariants.}
In the algorithm with a supplied tail-norm bound of \Cref{sec:schedule}, the
threshold floor $T_{\min}$ had two roles.
It kept every new frequency inside the heavy set, and it bounded the sum
of $T_i^{-2}$ in the cost accounting argument.
Here the algorithm has no floor.
In the analysis, the level $t_*$ takes over both roles, although the
algorithm never uses it.
We prove the following invariants:
\begin{claim}
\label{clm:tail-free-invariants}
  For every early call, with support $S$ and threshold $T$, we have $
    S\subseteq\Omega \text{ and }
    T\geq t_*$.
\end{claim}
In particular, these invariants ensure the required conditions.
Indeed, the fit $C_S$ is bad, so \eqref{eqn:tail-free-valid} gives
$\norm{\Xhat-C_S}_2<M_S$.
The threshold also satisfies $0<T\leq M_S$, so $\kappa\geq1$.
For the first call, this holds because $T=M_\varnothing>0$.
Each later threshold is the minimum of $T/2$ and the current $M_S$,
and both numbers are positive.

We prove \Cref{clm:tail-free-invariants} by induction over the early calls.

\emph{The first call.}
If the zero approximation is good, there is no early call, and nothing needs
to be proved.
Otherwise the zero approximation is bad.
The first call then has $S=\varnothing\subseteq\Omega$ and
$T=M_\varnothing$.
By \eqref{eqn:tail-free-valid}, $M_\varnothing>120\sigma$.
Since $k\geq1$, we have $t_*\leq\sigma$, and therefore $T\geq t_*$.

\emph{The support after a successful call.}
Consider an early call with support $S$ and threshold $T$ satisfying
\Cref{clm:tail-free-invariants}, and suppose that it succeeds.
Let $S'$ be the enlarged support.
Write $\Rhat=\Xhat-C_S$ for the residual searched by the call.
By soundness, every returned frequency $f$ satisfies $|\Rhat(f)|>T/2$.
A returned frequency in $S$ does not change the support.
For a returned frequency $f\notin S$, we have $\Rhat(f)=\Xhat(f)$, since
$C_S$ vanishes outside $S$.
Hence $|\Xhat(f)|>T/2\geq t_*/2$, so $f\in\Omega$.
This shows $S'\subseteq\Omega$.
Consequently, the refit on $S'$ is nonsingular, and $|S'|\leq|\Omega|<5k$.
Neither the singularity check nor the support-size limit is triggered, and the
fit $C_{S'}$ is saved.

\emph{The threshold of the next call.}
If $C_{S'}$ is good, the initial segment ends here.
Otherwise $C_{S'}$ is bad, and the next call uses the threshold
$T'=\min\{T/2,M_{S'}\}$.
We show $T'\geq t_*$.
If $\sigma=0$, then $t_*=0$, and there is nothing to prove.
Assume $\sigma>0$.
We need $M_{S'}\geq t_*$ and $T/2\geq t_*$.

The first bound is direct:
since $C_{S'}$ is bad and $S'\subseteq\Omega$, \eqref{eqn:tail-free-valid}
gives $M_{S'}>120\sigma\geq t_*$.

The second bound uses completeness of the current call.
Suppose that $T<2t_*$, and let $f\notin S'$.
Then $f\notin S$, so $\Rhat(f)=\Xhat(f)$.
The call did not return $f$, so completeness gives
$|\Xhat(f)|<T<2t_*$.
Thus every frequency outside $S'$ has magnitude less than $2t_*$.
Let $S_k^*$ be the best-$k$ support of $\Xhat$.
We then obtain
\[
  \norm{\Xhat_{\Omega\setminus S'}}_2^2 =
  \norm{\Xhat_{(\Omega\setminus S')\cap S_k^*}}_2^2 +
  \norm{\Xhat_{\Omega\setminus (S'\cup S_k^*)}}_2^2
  \leq 4kt_*^2+\sigma^2=5\sigma^2,
\]
where the last equality uses $kt_*^2=\sigma^2$.
Now \eqref{eqn:tail-free-energy} gives
\[
  M_{S'}^2\leq12\cdot5\sigma^2+3200\sigma^2=3260\sigma^2.
\]
This contradicts $M_{S'}>120\sigma$.
Hence $T/2\geq t_*$, and therefore $T'\geq t_*$.
This completes the induction proving \Cref{clm:tail-free-invariants}.

It remains to consider the budget check.  Consider any initial segment
of the early calls, including the proposed next call before its cost parameter
check.  For each $f\in\Omega$, the rounds $i$ in which $f \notin S_i$ form an
initial segment.
At its last missing round $\ell_f$, we have $|\Xhat(f)| < 2T_{\ell_f}$.
For $\ell_f=1$, use $|\Xhat(f)|\leq M_\varnothing=T_1$.
For $\ell_f\geq2$, the valid residual bound and completeness of the
preceding call give $|\Xhat(f)|\leq M_{S_{\ell_f}}$ and
$|\Xhat(f)|<T_{\ell_f-1}$.  Since $M_{S_{\ell_f}}>0$, these imply
\[
  |\Xhat(f)|<2\min\{M_{S_{\ell_f}},T_{\ell_f-1}/2\}
  =2T_{\ell_f}.
\]
Since thresholds decrease by at least a factor of two, we obtain the following
geometric-series bounds, as in \Cref{lem:charging}:
\[
  \sum_{i:f\notin S_i}\frac{|\Xhat(f)|^2}{T_i^2}<\frac{16}{3},
  \qquad
  \sum_iT_i^{-2}\leq\frac{4}{3t_*^2}\ \text{ if }\sigma>0.
\]
The second bound uses $T_i\geq t_*$ from \Cref{clm:tail-free-invariants}.
Consequently, using \eqref{eqn:tail-free-energy}, an argument similar to that
in \Cref{lem:charging} gives
\[
  \sum_i\kappa_i
  <64|\Omega|+\frac{12800}{3}k<4587k<K;
\]
for $\sigma > 0$.
For $\sigma=0$, the term with $\sigma^2$ in \eqref{eqn:tail-free-energy}
vanishes and the second summand is absent.
Hence the budget check cannot stop the initial segment.

Together with the previous paragraphs, this shows that no check can stop
the initial segment.
Since each early call has $\kappa_i\geq1$, the initial segment also has fewer
than $K$ calls.
The initial segment therefore ends in one of two ways: a good fit is saved, or
an early call fails.
We next bound the probability of the latter case.

Fix a least-squares sample in the good event above.
For each $i\in[K]$, condition on the observations and decisions before
the $i$th accepted call.  They determine whether this call exists and is
early.  If it is early, its norm promise is valid, and
\Cref{lem:simulation} gives a fresh batch.  Thus \Cref{prop:locator}
bounds its conditional failure probability by $\gamma$.
Averaging over preceding histories and taking a union bound over at most
$K$ accepted calls bounds the probability of a failed early call by
$K\gamma=\delta_{\rm loc}=1/30$.
This proves that a good fit is saved except with probability at most
$\delta_{\rm ls}+\delta_{\rm loc}$.

Finally, let $C_S$ be a good fit.
Since $\HT_k(\Xhat)$ is $k$-sparse and $\HT_k(C_S)$ is a best $k$-sparse
approximation of $C_S$, we have
$\norm{C_S-\HT_k(C_S)}_2\leq\norm{C_S-\HT_k(\Xhat)}_2$.
Together with the triangle inequality, this gives
\begin{align*}
  \norm{\Xhat-\HT_k(C_S)}_2
  &\leq\norm{\Xhat-C_S}_2+\norm{C_S-\HT_k(C_S)}_2\\
  &\leq\norm{\Xhat-C_S}_2+\norm{C_S-\HT_k(\Xhat)}_2\\
  &\leq2\norm{\Xhat-C_S}_2+\norm{\Xhat-\HT_k(\Xhat)}_2\\
  &\leq2\cdot120\sigma+\sigma=241\sigma.
\end{align*}
The candidate $\HT_k(C_S)$ therefore satisfies the claimed bound.
\end{proof}

\subsection{Selecting a candidate without a norm estimate}

\begin{lemma}
\label{lem:tail-free-selection}
\label{lem:additive-selection}
Let $0<\eta<1$.
Given a list of $n_{\rm c}\geq1$ $k$-sparse vectors
$\Xtilde^{(1)},\ldots,\Xtilde^{(n_{\rm c})}\in\C^G$ independent of a fresh
sample query set, one can select a vector $\Xtilde$ from the list such that
\[
  \norm{\Xhat-\Xtilde}_2\leq\frac{1+\eta}{1-\eta}\min_{i\in[n_{\rm
  c}]}\norm{\Xhat-\Xtilde^{(i)}}_2
\]
with probability at least $99/100$, using $n_{\rm sel}=O(k\eta^{-2}\log(2n_{\rm
c}))$ point queries.
No signal norm or residual norm is supplied, and the sample size may be fixed
using any deterministic upper bound on $n_{\rm c}$.
The selection uses $O(n_{\rm c}k(\log N)n_{\rm sel}+n_{\rm c}^2n_{\rm
sel}+n_{\rm c}^2k(\log N))$ exact operations.
In particular, for $\eta=1/4$ the factor is $5/3$.
\end{lemma}

\begin{proof}
Discard duplicate vectors from the list.
If only one vector remains, return it.
From now on, assume that $n_{\rm c}\geq2$ and that the vectors are distinct.
Condition on the list; by assumption, the sample query set is independent of
it.
For analysis, let $\R_i$ be the time signal with spectrum
$\Rhat_i=\Xhat-\Xtilde^{(i)}$, and write
$e_i=\norm{\Rhat_i}_2$ for the error of the $i$th candidate.
For each pair of candidates, we estimate the difference $e_j^2-e_i^2$.

Fix $i\neq j$, and let $D_{ij}$ be the signal with spectrum
$\hat{D}_{ij}=\Xtilde^{(j)}-\Xtilde^{(i)}$.
Then $\Xhat-\Xtilde^{(j)}=\Rhat_i-\hat{D}_{ij}$.
Expanding the square gives
\[
  e_j^2=\norm{\Rhat_i-\hat{D}_{ij}}_2^2
  =e_i^2-2\,\mathrm{Re}\sum_{f\in G}\Rhat_{i}(f)\conj{\hat{D}_{ij}(f)}
   +\norm{\hat{D}_{ij}}_2^2 .
\]
By \Cref{lem:sparse-correlation}, the inner product in this expansion
is $\E_t[\R_i(t)\conj{D_{ij}(t)}]$.
Hence we define
\begin{equation}
\label{eqn:tail-free-difference}
  a_{ij}:=e_j^2-e_i^2
  =\norm{\hat{D}_{ij}}_2^2-2\,\E_t\bigl[\xi_{ij}(t)\bigr],
  \qquad
  \xi_{ij}(t)=\mathrm{Re}\bigl[\R_i(t)\conj{D_{ij}(t)}\bigr].
\end{equation}
The sign of $a_{ij}$ compares the two candidates: $a_{ij}<0$ means that
$\Xtilde^{(j)}$ has a smaller error than $\Xtilde^{(i)}$.
The first term in \eqref{eqn:tail-free-difference} is known, since
$\Xtilde^{(i)}$
and $\Xtilde^{(j)}$ are known.
Only the average of $\xi_{ij}$ must be estimated.
For a point $t$, one query to $\X$ gives $\R_i(t)$, as in
\eqref{eqn:residual-synthesis} with $\Xtilde^{(i)}$ in place of $C_S$.
The value $D_{ij}(t)$ is computed from the two candidates without any
query.

\emph{The variance of one sample.}
The spectrum $\hat{D}_{ij}$ has at most $2k$ nonzero coordinates.  Apply
\Cref{lem:sparse-correlation} with $U=D_{ij}$,
$V=\R_i$, and $q=2k$.  Taking real parts can only decrease the second
moment, so
\begin{equation}
\label{eqn:tail-free-variance}
  \E_t[\xi_{ij}(t)^2]
  \leq\E_t[|\R_i(t)\conj{D_{ij}(t)}|^2]
  \leq2k\norm{\hat{D}_{ij}}_2^2e_i^2.
\end{equation}
The variance of $\xi_{ij}(t)$ is at most this second moment.

\emph{The estimator.}
The candidate selection queries consist of $r$ groups of $v$ independent
uniform
points of $G$, where
\[
  v=\ceil{128k/\eta^2},
  \qquad
  r=2\ceil{\log(100n_{\rm c}^2)}+1.
\]
The same points serve every ordered pair $(i,j)$.
For each pair, compute the mean of $\xi_{ij}$ over each group.
Let $\lambda_{ij}$ be the median of these $r$ group means, and set $\widehat
a_{ij}=\norm{\hat{D}_{ij}}_2^2-2\lambda_{ij}$.
We claim that, with probability at least $99/100$,
\begin{equation}
\label{eqn:tail-free-pair-error}
  |\widehat a_{ij}-a_{ij}|
  \leq\eta\norm{\hat{D}_{ij}}_2\,e_i
  \qquad\text{for all }i\ne j.
\end{equation}

Fix a pair $(i,j)$.
If $e_i=0$, then $\R_i=0$, every sample is zero, and $\widehat a_{ij}=a_{ij}$.
Assume $e_i>0$.
We apply item~\ref{item:median-means} of \Cref{lem:median} to the real
samples $\xi_{ij}(t)$, viewed as complex numbers, with
$x=\E_t[\xi_{ij}(t)]$, $\nu=2k\norm{\hat{D}_{ij}}_2^2e_i^2$ from
\eqref{eqn:tail-free-variance}, $v=\ceil{128k/\eta^2}$, and
$\omega=\eta\norm{\hat{D}_{ij}}_2e_i/(2\sqrt2)$.
Then
\[
  p=\frac{\nu}{v\omega^2}
  \leq\frac{2k\norm{\hat{D}_{ij}}_2^2e_i^2\cdot8}
           {(128k/\eta^2)\cdot\eta^2\norm{\hat{D}_{ij}}_2^2e_i^2}
  =\frac18 .
\]
Hence $|\lambda_{ij}-x|<\sqrt2\,\omega=\eta\norm{\hat{D}_{ij}}_2e_i/2$,
except with probability at most
$\tfrac14\cdot2^{-(r+1)/2}\leq1/(800n_{\rm c}^2)$.
On this event,
$|\widehat a_{ij}-a_{ij}|=2|\lambda_{ij}-x|\leq\eta\norm{\hat{D}_{ij}}_2e_i$.
There are fewer than $n_{\rm c}^2$ ordered pairs, so a union bound
gives total failure probability less than $1/100$, which proves
\eqref{eqn:tail-free-pair-error}.
The query set has $n_{\rm sel}=rv=O(k\eta^{-2}\log(2n_{\rm c}))$ points.

\emph{The selection rule.}
Define the score
\[
  J_i=\max_{j\ne i}\frac{-\widehat
  a_{ij}}{\norm{\Xtilde^{(j)}-\Xtilde^{(i)}}_2}.
\]
The denominators are positive, since the candidates are distinct.
The algorithm returns a candidate $\Xtilde^{(\hat\imath)}$ with the smallest
score, with ties broken by index.

\emph{Analysis of the rule.}
Assume \eqref{eqn:tail-free-pair-error}, and let $i_*$ be an index that
minimizes $e_i$.
We compare two bounds: the best candidate has a small score, while a
candidate much worse than $\Xtilde^{(i_*)}$ has a large score because
$\Xtilde^{(i_*)}$ offers a large improvement.

We first bound the score of $\Xtilde^{(i_*)}$.
Fix any $j\ne i_*$.
By the choice of $i_*$, we have $a_{i_*j}=e_j^2-e_{i_*}^2\geq0$, so
\eqref{eqn:tail-free-pair-error} implies
\[
  -\widehat a_{i_*j}
  \leq\eta\norm{\Xtilde^{(j)}-\Xtilde^{(i_*)}}_2\,e_{i_*}
\]
for any $j \neq i_*$.
Dividing by $\norm{\Xtilde^{(j)}-\Xtilde^{(i_*)}}_2$ and taking the maximum
over $j$ gives
\begin{equation}
\label{eqn:tail-free-best-score}
  J_{i_*}\leq\eta e_{i_*}.
\end{equation}

Next, we bound the score of another candidate.
Fix any $i\ne i_*$.
\eqref{eqn:tail-free-pair-error} implies
\[
  -\widehat a_{ii_*}
  \geq-a_{ii_*}-\eta\norm{\Xtilde^{(i_*)}-\Xtilde^{(i)}}_2\,e_i
  =(e_i^2-e_{i_*}^2)-\eta\norm{\Xtilde^{(i_*)}-\Xtilde^{(i)}}_2\,e_i .
\]
Combining with $J_i \geq -\widehat a_{ii_*} /
\norm{\Xtilde^{(i_*)}-\Xtilde^{(i)}}_2$ then gives
\[
  J_i\geq\frac{e_i^2-e_{i_*}^2}{\norm{\Xtilde^{(i_*)}-\Xtilde^{(i)}}_2}-\eta
  e_i.
\]
The triangle inequality gives
\[
  \norm{\Xtilde^{(i_*)}-\Xtilde^{(i)}}_2
  \leq\norm{\Xhat-\Xtilde^{(i)}}_2+\norm{\Xhat-\Xtilde^{(i_*)}}_2
  =e_i+e_{i_*},
\]
so the first term is at least $e_i-e_{i_*}$, and
\begin{equation}
\label{eqn:tail-free-other-score}
  J_i\geq(e_i-e_{i_*})-\eta e_i=(1-\eta)e_i-e_{i_*}.
\end{equation}

Finally, we compare the two bounds.
If $\hat\imath=i_*$, the returned candidate has the smallest error.
Otherwise, $\hat\imath$ minimizes the score, so
$J_{\hat\imath}\leq J_{i_*}$.
We apply \eqref{eqn:tail-free-other-score} with $i=\hat\imath$ and then
\eqref{eqn:tail-free-best-score}, which gives
$(1-\eta)e_{\hat\imath}-e_{i_*}\leq J_{\hat\imath}\leq J_{i_*}\leq\eta
e_{i_*}$.
Rearranging gives $e_{\hat\imath}\leq\frac{1+\eta}{1-\eta}e_{i_*}$, as
required.

\emph{Cost.}
Compute the time signal of each candidate at every sample point in $O(k\log N)$
operations, for a total of $O(n_{\rm c}k(\log N)n_{\rm sel})$.
Then every value $\xi_{ij}(t)$ costs $O(1)$, and all group means and medians
cost $O(n_{\rm c}^2n_{\rm sel})$.
Sparse coordinate arithmetic computes all norms $\norm{\hat{D}_{ij}}_2$ in
$O(n_{\rm c}^2k(\log N))$ operations, with an additional logarithmic sorting
cost absorbed since $k<N$.
Discarding duplicates can also be done in $O(n_{\rm c}^2 k(\log N))$
operations.
\end{proof}

\subsection{Proof of the constant-factor theorem}

\begin{proof}[Proof of \Cref{thm:tail-free}]
Construct the candidate approximations and then apply
\Cref{lem:tail-free-selection} with $\eta=1/4$.
For candidate selection, use the final-stage query set, which the candidate
construction
never reads.
By \Cref{lem:simulation}, this query set is independent of the candidate
list, so \Cref{lem:tail-free-selection} applies once we check that the
query set is long enough.

The selection rule needs $r=2\ceil{\log(100n_{\rm c}^2)}+1$ groups of
$v=2048k$ points.
The final-stage query set has $r_cv_c$ points, where \eqref{eqn:cleanup-bank}
with $s=k$ and $\delta_{\rm est}=1/30$ gives
\[
  v_c=\ceil{16384b_{\rm ls}^2k}=1044480k=510v,
  \qquad
  r_c=2\ceil{\log(510k)}+1.
\]
Since $v_c=510v$, it suffices that $r\leq510r_c$.
We have $n_{\rm c}\leq K+1\leq31001k<(510k)^2$ and $100<510k$, hence
$100n_{\rm c}^2<(510k)^5$.
Therefore $r\leq10\ceil{\log(510k)}+1\leq5r_c$, and
$rv\leq5r_cv<r_cv_c$.

By \Cref{lem:tail-free-candidates,lem:tail-free-selection}, the output
satisfies
\[
  \norm{\Xhat-\Xtilde}_2\leq\frac53\cdot241\sigma_k(\Xhat)_2
  <402\sigma_k(\Xhat)_2
\]
except with probability at most
$\delta_{\rm ls}+\delta_{\rm loc}+1/100=23/300<1/10$.
Thus the squared-error statement holds, for example, with $C=10^6$.
When the tail vanishes, this same bound gives exact recovery.

The candidate construction is an admissible recovery procedure, so
\Cref{lem:capacity,lem:simulation} apply.  All samples are acquired
in advance, and \Cref{lem:Q-upper} with $s=k$ bounds their total number by
$Q_{\rm total}=O(k\log N\log k)$, including the final-stage query set used for
selection.  If $N\leq Q_{\rm total}$, instead query all of $G$, compute
$\Xhat$ by FFT, and return $\HT_k(\Xhat)$.

For the operation count, \Cref{prop:shared-runtime} with $s=k$ shows
that generating the query sets and constructing the candidates take
\[
  O\bigl(k^{7/2}(\log N)^{9/4}\log(k\log N)\bigr)
\]
operations on every execution.
Threshold updates and stopping checks use constant work per accepted call,
and there are at most $K=O(k)$ calls.
Saving the $O(k)$ fits and applying $\HT_k$ to them costs $O(k^2\log k)$
operations.
By \Cref{lem:tail-free-selection} with $n_{\rm c}=O(k)$, $\eta=1/4$, and
$n_{\rm sel}=O(k\log k)$, the selection costs $O(k^3(\log N)\log k)$
operations.
These costs, and the full-FFT case's $O(N\log N)$ cost when selected, are
within the displayed bound.
Combining these bounds gives the claimed running time.
\end{proof}

\section{Relative additive error without a tail estimate}
\label{sec:relative-additive}
Allowing an additive error relative to the total spectral energy
removes the need for a supplied tail estimate while retaining the
approximation factor $1+\eps$.  We obtain this guarantee by a reduction
to \Cref{thm:upper-tail}.

\begin{theorem}[Recovery with relative additive error; detailed form of
\Cref{thm:relative-additive-intro}]
\label{thm:relative-additive}
Let $a,d\geq1$ be integers, $G=(\Z/2^a\Z)^d$, $N=2^{ad}$,
$2\leq k<N$, $0<\eps\leq1/2$, and $0<\delta\leq1/4$.
There is a randomized algorithm which, given
point-query access to any fixed $\X\in\C^G$, returns a $k$-sparse
vector $\Xtilde$ such that, with probability at least $9/10$,
\begin{equation}
\label{eqn:relative-additive}
  \norm{\Xhat-\Xtilde}_2^2
  \leq(1+\eps)\sigma_k(\Xhat)_2^2+\delta\norm{\Xhat}_2^2.
\end{equation}
No signal norm or tail estimate is supplied.  The algorithm makes at most
\[
  \min\!\left\{N,
    O\!\left(\frac{k}{\eps^2}
      \left((\log N)\log\frac{k}{\eps}
        +\log\log\frac1{\eps\delta}\right)
    \right)\right\}
\]
nonadaptive point queries and uses at most
\begin{align*}
  O\!\Bigl(&
    \frac{k^{7/2}}{\eps^7}(\log N)^{9/4}
      \log\frac{k\log N}{\eps}\log\frac1{\eps\delta}
    \\[-2pt]
    &{}+\frac{k^2}{\eps^2}(\log N)
      \log^2\frac1{\eps\delta}\log\log\frac1{\eps\delta}
  \Bigr)
\end{align*}
exact operations on every execution, in the model of \Cref{sec:model}.
All hidden constants are absolute.
\end{theorem}

For fixed $\delta$, the query bound simplifies to
$O((k/\eps^2)(\log N)\log(k/\eps))$.
The additive term remains present when the spectrum is exactly
$k$-sparse.

We first take preliminary scale-estimation samples, called the pilot
sample below.  Either they give an interval containing a suitable
tail-norm bound or zero already satisfies the required error guarantee.
We then run coordinatewise recovery for a geometric list of candidate
tail bounds, using the corresponding accuracy thresholds and the same
stored query answers.  Independent samples select a candidate whose
error is within a prescribed multiplicative factor of the smallest
candidate error.  This factor can be arbitrarily close to one.

Throughout this section, write
\[
  E=\norm{\Xhat}_2,\qquad \sigma=\sigma_k(\Xhat)_2.
\]
Both quantities are used only in the analysis.  Our normalization gives
$\E_t|\X(t)|^2=E^2$ and $\norm \X_2=\sqrt N E$.

\subsection{Preliminary scale estimation and the zero approximation}

\begin{lemma}
\label{lem:additive-pilot}
Let $0<\psi\leq1$, let $n_{\rm p}=\ceil{4096k/\psi}$, and sample
$t_1,\ldots,t_{n_{\rm p}}$ independently and uniformly from $G$.  Set
\[
  E_{\rm p}^2=\frac1{n_{\rm p}}\sum_{b=1}^{n_{\rm p}}|\X(t_b)|^2.
\]
If $E>(1+\psi/4)\sigma$, then, with probability at least
$1-3/256$,
\begin{equation}
\label{eqn:additive-pilot}
  \frac{\sqrt\psi}{64}E\leq E_{\rm p}\leq16E.
\end{equation}
\end{lemma}

\begin{proof}
Assume $E>(1+\psi/4)\sigma$.  Define $\hat{H}=\HT_k(\Xhat)$,
and let $H$ be the signal with spectrum $\hat{H}$.
If $\norm{\hat{H}}_2^2\leq\psi E^2/4$, then
$\sigma^2=E^2-\norm{\hat{H}}_2^2\geq(1-\psi/4)E^2$.
Since $(1+\psi/4)^2(1-\psi/4)\geq1$, this contradicts the assumption.
Thus $\norm{\hat{H}}_2^2>\psi E^2/4>0$.  For analysis, define
\[
  \Xi_1=\frac1{n_{\rm p}}\sum_b \X(t_b)\conj{H(t_b)},\qquad
  \Xi_2=\frac1{n_{\rm p}}\sum_b|H(t_b)|^2.
\]
We bound $\Xi_1$ from below and $\Xi_2$ and $E_{\rm p}$ from above,
then use Cauchy--Schwarz to obtain the lower bound on the pilot estimate.

\emph{The mean and variance of $\Xi_1$.}
Apply \Cref{lem:sparse-correlation} with $U=H$, $V=\X$, and $q=k$.
Since $\hat{H}$ agrees with $\Xhat$ on its support, this gives
\[
  \E[\Xi_1]=\norm{\hat{H}}_2^2,
  \qquad
  \E\bigl[|\X(t)\conj{H(t)}|^2\bigr]
  \leq k\norm{\hat{H}}_2^2E^2.
\]
The $n_{\rm p}$ samples are independent, so
\[
  \E\Bigl[\bigl|\Xi_1-\norm{\hat{H}}_2^2\bigr|^2\Bigr]
  \leq\frac{k}{n_{\rm p}}\norm{\hat{H}}_2^2E^2.
\]
By Chebyshev's inequality, $\norm{\hat{H}}_2^2>\psi E^2/4$, and $n_{\rm
p}\geq4096k/\psi$,
\[
\Prob\Bigl[\bigl|\Xi_1-\norm{\hat{H}}_2^2\bigr|
  >\tfrac12\norm{\hat{H}}_2^2\Bigr]
  \leq\frac{4kE^2}{n_{\rm p}\norm{\hat{H}}_2^2}
  <\frac{16k}{n_{\rm p}\psi}
  \leq\frac{1}{256}.
\]
Outside this event, $|\Xi_1|\geq\norm{\hat{H}}_2^2/2$.

\emph{Upper bounds on $\Xi_2$ and $E_{\rm p}$.}
By Parseval, $\E[\Xi_2]=\E_t|H(t)|^2=\norm{\hat{H}}_2^2$ and $\E[E_{\rm
p}^2]=\E_t|\X(t)|^2=E^2$.
Markov's inequality therefore gives $\Xi_2\leq256\norm{\hat{H}}_2^2$ and
$E_{\rm p}^2\leq256E^2$, each except with probability at most $1/256$.
The second bound is the upper half of \eqref{eqn:additive-pilot}.

\emph{The lower bound on $E_{\rm p}$.}
The Cauchy--Schwarz inequality for the empirical averages gives
\[
  |\Xi_1|^2
  =\Bigl|\frac1{n_{\rm p}}\sum_b\X(t_b)\conj{H(t_b)}\Bigr|^2
  \leq\Bigl(\frac1{n_{\rm p}}\sum_b|\X(t_b)|^2\Bigr)
      \Bigl(\frac1{n_{\rm p}}\sum_b|H(t_b)|^2\Bigr)
  =E_{\rm p}^2\,\Xi_2 .
\]
Outside the three exceptional events above, this yields
\[
  \frac14\norm{\hat{H}}_2^4
  \leq|\Xi_1|^2
  \leq E_{\rm p}^2\,\Xi_2
  \leq256\norm{\hat{H}}_2^2E_{\rm p}^2 ,
\]
that is, $E_{\rm p}\geq\norm{\hat{H}}_2/32$.
Since $\norm{\hat{H}}_2>\sqrt\psi E/2$, we conclude that $E_{\rm p}>\sqrt\psi
E/64$.
A union bound over the three exceptional events completes the proof.
\end{proof}

\subsection{Reusing the coordinatewise recovery across scales}

We first prove the unsquared-error bound.  For any
$0<\psi,\varphi\leq1$, we construct a candidate satisfying
\begin{equation}
\label{eqn:additive-norm}
  \norm{\Xhat-\Xtilde}_2
  \leq(1+\psi)\sigma+\varphi E
\end{equation}
with probability at least $9/10$.
Set
\[
  \alpha=\psi/16,\qquad
  s=\ceil{25k/\alpha^2},\qquad
  n_\tau=\ceil{\log\frac{2048}{\varphi\sqrt\psi}}.
\]
If $s\geq N$, query the entire signal, compute its Fourier transform,
and return $\HT_k(\Xhat)$.  Otherwise, prepare the following
independent samples:
preliminary scale-estimation samples as in \Cref{lem:additive-pilot};
one collection of the query sets of \Cref{sec:tapes-def}
with parameter $s$;
and candidate selection samples as in \Cref{lem:tail-free-selection}, with
$\eta=\psi/16$ and the list bound $2n_\tau$.
All these query locations and multiplicities are fixed before reading
any answers.  If their total count is at least $N$, use the same full-FFT
case instead.

Compute $E_{\rm p}$ from the pilot sample.  If $E_{\rm p}=0$, return
zero.  Otherwise, for $0\leq\ell<n_\tau$ put
\[
  B_\ell=\frac{2^\ell\varphi E_{\rm p}}{16},\qquad
  \tau_\ell=\frac{\alpha B_\ell}{5\sqrt k}.
\]
Run the coordinatewise recovery algorithm, namely the one from
\Cref{sec:recovery-algo} with parameters $s$ and $\tau_\ell$, once on the
stored query sets.  Start every decoding run from its initial state and from
the beginning
of the stored query answers, retaining all the checks.  Replace
an aborted run by zero and apply $\HT_k$ to every output.  Include zero
as an additional candidate, and apply \Cref{lem:tail-free-selection} to
the resulting list, which has at most $2n_\tau$ entries.  This specifies
the algorithm even for scales at which the tail-norm bound fails.

\emph{Accuracy.}
If $E\leq(1+\psi/4)\sigma$, zero already achieves the required error.
Either the algorithm returns it directly, or the selection lemma
gives error at most
\[
  \frac{1+\psi/16}{1-\psi/16}(1+\psi/4)\sigma
  \leq(1+\psi/7)(1+\psi/4)\sigma
  \leq(1+\psi)\sigma.
\]

If $E>(1+\psi/4)\sigma$, condition on the
pilot event \eqref{eqn:additive-pilot}, whose failure probability is at
most $3/256$.  We then have $E_{\rm p}>0$, $B_0\leq\varphi E$, and
$B_{n_\tau-1}\geq E$.  Let $\Gamma=\max\{\sigma,\varphi E\}$, and let
$\ell_*$ be the first index with $B_{\ell_*}\geq\Gamma$.  Since $\Gamma\leq E$,
\[
  \sigma\leq B_{\ell_*}\leq2\Gamma.
\]
The index $\ell_*$ depends only on $\X$ and on the pilot sample, through
$E_{\rm p}$, and not on the query sets.
The algorithm does not know $\ell_*$; we use it only in the analysis.
Conditional on the pilot sample, $\ell_*$ and $\tau_{\ell_*}$ are fixed,
and the recovery query sets retain their prescribed distribution.
Moreover, the promise of \Cref{thm:upper-tail} holds at $\ell_*$:
\[
  \sigma_s(\Xhat)_2^2\leq\sigma^2\leq B_{\ell_*}^2
  \leq s\tau_{\ell_*}^2.
\]
The first inequality holds since $s\geq k$, the second by the choice of
$\ell_*$, and the third since $s\tau_{\ell_*}^2=s\alpha^2B_{\ell_*}^2/(25k)$
and $s\geq25k/\alpha^2$.
Thus the run at $\ell_*$ has fixed inputs satisfying the recovery
promise.  By \Cref{lem:simulation} and \eqref{eqn:upper-failure-bound},
its failure probability is at most $17/240$.
We need success only at this scale, not simultaneously at every scale.

For a successful output $\Xtilde^{\rm pre}$, use the weaker consequence
$\norm{\Xhat-\Xtilde^{\rm pre}}_\infty\leq\tau_{\ell_*}$ in
\Cref{lem:hard-thresholding}.  This gives
\[
  \norm{\Xhat-\HT_k(\Xtilde^{\rm pre})}_2^2
  \leq\sigma^2+\frac{4\alpha}{5}\sigma B_{\ell_*}
       +\frac{\alpha^2}{5}B_{\ell_*}^2
  \leq(\sigma+\alpha B_{\ell_*})^2.
\]
Recall $\alpha = \psi/16$ and $\Gamma \leq E$.  The list therefore contains a
candidate of error at most
\[
  \sigma+\alpha B_{\ell_*}\leq
  \sigma+2\alpha\Gamma
  \leq(1+\psi/8)\sigma+(\psi\varphi/8)E.
\]
The selection sample is independent of the complete candidate list.
With its additional failure probability at most $1/100$, the selected
error is at most
\[
  (1+\psi/7)
     \bigl((1+\psi/8)\sigma+(\psi\varphi/8)E\bigr)
  \leq(1+\psi)\sigma+\varphi E.
\]
The total failure probability is at most
\[
  \frac3{256}+\frac{17}{240}+\frac1{100}
  =\frac{1777}{19200}<\frac1{10},
\]
proving \eqref{eqn:additive-norm}.

\emph{Nonadaptivity and resources.}
The lengths of the query sets depend only on $s$ and $\log N$
(\Cref{sec:tapes-def}), and not on the thresholds $\tau_\ell$.
Every decoding run starts from the beginning of the same query sets,
and by \Cref{rem:admissible-instances} it is an admissible recovery
procedure with parameter $s$.  Hence \Cref{lem:capacity,lem:simulation}
apply to each run separately.  Runs at different scales are dependent,
but the accuracy argument uses only the run at $\ell_*$.  Its threshold
is fixed by the independent pilot sample, and it is decoded from its
initial state without using the observations or decisions of other runs.
Thus pilot answers determine thresholds for decoding but never query
locations.

Since $s=\Theta(k/\psi^2)$, the recovery query sets use
$O((k/\psi^2)\log N\log(k/\psi))$ samples by \Cref{lem:Q-upper}.
The pilot
uses $O(k/\psi)$, and the selection sample uses
$O((k/\psi^2)\log n_\tau)$.  The total sparse-recovery query count is
therefore
\begin{equation}
\label{eqn:additive-query-count}
  O\!\left(\frac{k}{\psi^2}
    \left((\log N)\log\frac{k}{\psi}+\log n_\tau\right)\right).
\end{equation}
There are $n_\tau$ decoding runs.  Threshold updates and stopping checks
take constant work per round and hence $O(s)$ operations per run.
By \Cref{prop:shared-runtime} and
\eqref{eqn:cleanup-cost}, each run, including its final coefficient estimation,
costs
$O((k^{7/2}/\psi^7)(\log N)^{9/4}\log(k(\log N)/\psi))$ operations on every
execution, whether or not the promise holds at its scale.  Hard
thresholding an output with fewer than $17s$ nonzero entries costs
$O(s\log N\log s)$, which is absorbed by this bound.
Adding the cost from \Cref{lem:tail-free-selection} with $n_{\rm c}\leq2n_\tau$
gives
\[
  O\!\left(\frac{n_\tau k^{7/2}}{\psi^7}(\log N)^{9/4}\log\frac{k(\log
  N)}{\psi}
    +\frac{n_\tau^2k^2}{\psi^2}(\log N)\log n_\tau\right).
\]
Computing the preliminary scale estimate is absorbed by these bounds.

In the full-FFT case, the cost is $O(N\log N)$, including hard thresholding.
If $s\geq N$, then $N=O(k/\psi^2)$.  Otherwise the branch is taken only
when $N$ is at most the planned query count in
\eqref{eqn:additive-query-count}.  In either case $O(N\log N)$ is absorbed
by the displayed operation bounds.  The full-FFT rule also proves the
minimum with $N$ in the query bound.

\begin{proof}[Proof of \Cref{thm:relative-additive}]
Apply the preceding construction with
$\psi=\eps/8$ and $\varphi=\sqrt{\eps\delta}/4$.
By \eqref{eqn:additive-norm}, with probability at least $9/10$ the
error is at most
$(1+\eps/8)\sigma+(\sqrt{\eps\delta}/4)E$.
The inequality $(a+b)^2\leq(1+\eps/4)a^2+(1+4/\eps)b^2$ gives
\[
  \norm{\Xhat-\Xtilde}_2^2
  \leq(1+\eps/4)(1+\eps/8)^2\sigma^2
       +\frac{\eps+4}{16}\delta E^2
  \leq(1+\eps)\sigma^2+\delta E^2.
\]
Here $n_\tau=\Theta(\log(1/(\eps\delta)))$ and
$\log n_\tau=\Theta(\log\log(1/(\eps\delta)))$.
Substituting these estimates and $\psi=\eps/8$ into
the query and operation counts proves the stated bounds.
\end{proof}

\section{Extension to other fixed bases}
\label{sec:fixed-base}

In this section, we describe the changes needed
to extend the results from $(\Z/2^a\Z)^d$ to
$G=(\Z/b^a\Z)^d$, where $b\geq2$ is any fixed integer and
$a,d\geq1$.  Write $N=b^{ad}$.
Then
\Cref{thm:main-dyadic-intro,thm:upper-tail,thm:tail-free,thm:relative-additive}
extend to $G$ with the same approximation guarantees, success
probabilities, and side information.
Their sample and operation bounds remain valid, with implicit constants
that may additionally depend on $b$, in the following exact sampling model.
Arithmetic on each coordinate in $\Z/b^a\Z$ has unit cost, and a
primitive $b^a$-th root of unity is supplied.  The other exact arithmetic
operations are as in \Cref{sec:model}.  We also count each independent
uniform draw from $\{0,\ldots,b-1\}$ as one operation.
For $b=2$, this is the fair-bit primitive used above.
For other bases, the model supplies these digits directly; it does not
assert a bounded-time exact simulation from fair bits.
The search depth is $ad=\log_bN$; logarithms in resource bounds remain to base
two.

We replace the binary frequency search tree of \Cref{sec:projections} by a
$b$-ary
frequency search tree.  For $j=\ell a+c<ad$, where $0\leq c<a$, set
\[
  H_j=\{0\}^{\ell}\times(b^c\Z/b^a\Z)
      \times(\Z/b^a\Z)^{d-\ell-1},
  \qquad H_{ad}=\{0\}^d.
\]
Writing $e_{\ell+1}$ for the $(\ell+1)$st coordinate vector, the children
of $v+H_j$ are
\[
  v+u b^c e_{\ell+1}+H_{j+1},\qquad 0\leq u<b.
\]
The projection identity of \Cref{prop:orthogonality} holds with
$K_j=H_j^{\perp}$, and \Cref{prop:statistics} is unchanged.
These facts use character orthogonality and the subgroup chain, so $b$
need not be prime.  The statistical resolutions and threshold schedule
may still use powers of two.

\begin{proposition}[Fixed-base locator]\label{prop:fixed-base-locator}
Fix an integer $b\geq2$.  Let $a,d\geq1$ be integers,
$G=(\Z/b^a\Z)^d$, and $N=b^{ad}$.  Suppose $\norm{\Rhat}_2\leq M$, $0<T\leq M$,
and
$0<\gamma<1/2$, and put $\kappa=M^2/T^2$.
In the exact arithmetic and sampling model above, a randomized
algorithm uses
\[
  O_b\bigl(\kappa(\log N)\log(2\kappa)
    +\kappa\log(\kappa(\log N)/\gamma)\bigr)
\]
nonadaptive point queries and, with probability at least $1-\gamma$,
returns every frequency with $|\Rhat(f)|\geq T$ and only frequencies
with $|\Rhat(f)|>T/2$.
On every execution, including aborts, its time and space are bounded
by a fixed polynomial in $\kappa$, $\log N$, and $\gamma^{-1}$, with
constants that may depend on $b$.
\end{proposition}

\begin{proof}
We use the locator of \Cref{prop:locator} on the $b$-ary tree above.
The larger branching factor requires a change to the potential updates.  Keep
the positive multiplier $2^{1/4}$, replace
the negative multiplier $1/4$ by $1/(4b)$, and amplify each node test
by a number of independent repetitions depending only on $b$, so that
its error probability is at most $p_b$, where
\[
  p_b\leq\frac{1}{16(4b)^4}.
\]
For the uncapped multiplier $A\in\{2^{1/4},1/(4b)\}$ these choices give
\[
  \E[A^{-4}\mid\text{earlier depths}]
  \leq \frac12+p_b(4b)^4<\frac34
  \quad\text{on a heavy path},
\]
and
\[
  \E[A\mid\text{earlier depths}]
  \leq p_b2^{1/4}+\frac{1}{4b}<\frac{1}{2b}
  \quad\text{at a low-energy node}.
\]
The first inequality preserves the heavy-path supermartingale
and its maximal bound, with
$\phi_{\max}=(3\kappa ad/\gamma)^{1/4}$.
For the search-size bound, let $\Phi_j$ sum the potentials of all retained
nodes at depth $j$.  Each retained parent has at most $b$ low-energy
children, whose total expected contribution is at most half its potential.
There are at most $\kappa/\theta$ high-energy nodes at the next depth,
each contributing at most $\phi_{\max}$.  Thus
\[
  \E[\Phi_{j+1}\mid\mathcal F_j]
  \leq\frac12\Phi_j+\frac{\kappa\phi_{\max}}\theta.
\]
As in \eqref{eqn:search-supermartingale}, adding half the accumulated
potential and the reserve $(ad-j)\kappa\phi_{\max}/\theta$ gives a
nonnegative supermartingale.  Ville's inequality bounds the total search
and permits a search-tree size limit
\[
  D=O_b\bigl(\kappa^{5/4}(ad)^{5/4}\gamma^{-5/4}\bigr).
\]
Choose the implicit constant in $D$ so that the abort probability is
at most $\gamma/3$.  The heavy-path argument also fails with probability
at most $\gamma/3$.  Each retained node has at most $b$ children, so
there are at most $bD$ node tests and $D$ retained leaves on every run.
Independent leaf verification uses the parameters in
\eqref{eqn:validation} with this value of $D$; its proof is unchanged
and contributes failure probability less than $\gamma/3$.
The union bound proves the recovery guarantee.
The amplified test costs $O_b(\kappa\log(2\kappa))$ samples at each of
$ad$ depths, and verification costs
$O_b(\kappa\log(\kappa(\log N)/\gamma))$ samples.
Their locations are chosen independently of the answers, as in
\Cref{prop:locator}.  Since $ad=\Theta_b(\log N)$ and the search is
bounded by $bD$ tests, the sample, time, and space bounds follow.
In particular, the exponent $1/4$ arising from the potential cap is
unchanged.
\end{proof}

We use the amplified tests and adjusted search-tree size limit in
the locator, with the corresponding query allocation.
Samples are still shared across all nodes at a given depth, and the
least-squares, recovery, and hard-thresholding arguments apply as before.
The supplied primitive $b^a$-th root of unity permits character evaluation
in $O(\log N)$ exact operations, and the full-FFT case uses a radix-$b$
FFT in $O_b(N\log N)$ operations.
Assembling each coordinate from $a$ independent base-$b$ digits gives
a uniform point of $G$ using $ad$ digits.  The displayed product
description of $K_j$ similarly gives a uniform point of $K_j$ using at
most $ad$ digits, by fixing the required coordinates or digits to zero.
Both constructions take $O_b(ad)=O_b(\log N)$ operations on every execution.
Since $ad=\Theta_b(\log N)$, the accounting of \Cref{prop:shared-runtime}
and \Cref{sec:resources} therefore gives the same exponents of $k$,
$\eps^{-1}$, and $\log N$,
with no additional dependence on $a$ or $d$ beyond $N=b^{ad}$.

The two results without a supplied tail estimate extend in the same way.
The candidate construction in \Cref{sec:tail-free} and every decoding run in
\Cref{sec:relative-additive} are admissible recovery procedures built
from the modified locator, so the query set and cost bounds of
\Cref{sec:tapes} apply with the constants above.
The proof of \Cref{lem:tail-free-candidates} uses only \Cref{lem:ls},
the locator guarantee, and the cost accounting argument, and none of these
depends on $b$ beyond the constants.
The pilot estimate of \Cref{lem:additive-pilot} and the selection rule
of \Cref{lem:tail-free-selection} use only character orthogonality and
the bound $|\sum_{f\in S}c(f)\chi_f(t)|\leq\sqrt{|S|}\,\norm{c}_2$, both of
which hold for every $G=(\Z/b^a\Z)^d$.

\section*{AI methodology}

The starting point of this project was the
first author's prior work \cite{ref:CI16} and
\cite{ref:CN20}.  The former provides a coset projection
approach to the sparse Fourier transform in
the special case of the Boolean hypercube (Walsh--Hadamard transform)
but with an $\ell_1/\ell_1$ guarantee (albeit
deterministic) rather than the stronger
$\ell_2/\ell_2$ or heavy hitters guarantees.  The latter
work uses a prefix tree akin to what we use
in the current work and achieves an
$\ell_2/\ell_2$ guarantee but requires arbitrary
linear measurements (or a Boolean group testing
setting), which Fourier queries do not provide.
The plan was therefore to combine the two
approaches and get the best of both worlds,
with the original hope of getting sublinear time
and a nonadaptive sample count (polynomial in
$k \log N$).  Another appeal of the prefix tree
approach was to provide a unified approach to
handle both the
Boolean cube and the cyclic
group (of power-of-two order), and all the groups in
between.

We used ChatGPT 6 and Claude 5.5 to optimize the
sample complexity, and the interactions led to
the current near-optimal sample bound while at
the same time maintaining polynomial time
reconstruction in the number of samples.
These tools were also used to assist with drafting, revising the
exposition, and \LaTeX{} typesetting.
The authors verified the correctness and originality of all content, including references. The authors thank OpenAI for the access to 
its ChatGPT for Academic Researchers program, and Anthropic
for providing us with the Claude Team plan for scientists
which we benefited from in this work.

\bibliographystyle{alpha}
\bibliography{references}
\end{document}